\documentclass[traditabstract]{aa}
\newcommand{\kms}{km~s$^{-1}$}
\usepackage{natbib}
\usepackage{graphicx}
\usepackage{txfonts}
\date{Received 27 Apr 2011 / 30 Aug 2011}
\usepackage{longtable}
\authorrunning{J.~K.~J{\o}rgensen et al.}
\titlerunning{A large SMA molecular line survey of IRAS~16293-2422}

\begin{document}

\title{Arcsecond resolution images of the chemical structure of the
  low-mass protostar IRAS~16293-2422}\subtitle{An overview of a large
  molecular line survey from the Submillimeter Array}

\author{J.~K. J{\o}rgensen\inst{1,2}, T.~L. Bourke\inst{3}, Q.~Nguy$\tilde{\hat{\rm e}}$n Lu{\hskip-0.65mm\small'{}\hskip-0.5mm}o{\hskip-0.65mm\small'{}\hskip-0.5mm}ng\inst{4,5},
  \and 
  S.~Takakuwa\inst{6}}
\institute{Centre for Star and Planet Formation, Niels Bohr Institute, University of Copenhagen, Juliane
  Maries Vej 30, DK-2100 Copenhagen {\O}., Denmark \and Natural History
  Museum of Denmark, University of Copenhagen, {\O}ster Voldgade 5-7,
  DK-1350 Copenhagen K., Denmark \and Harvard-Smithsonian Center for
  Astrophysics, 60 Garden Street MS42, Cambridge, MA 02138, USA \and
  Laboratoire AIM, CEA/DSM, IRFU/Service d'Astrophysique, 91191
  Gif-sur-Yvette Cedex, France \and Argelander Institute for
  Astronomy, University of Bonn, Auf dem H\"{u}gel 71, 53121, Bonn,
  Germany \and Academia Sinica Institute of Astronomy and Astrophysics, P.O. Box 23-141, Taipei 10617, Taiwan}

\begin{abstract}
  {It remains a key challenge to establish the molecular content of
    different components of low-mass protostars, like their envelopes
    and disks, and how this depends on the evolutionary stage and/or
    environment of the young stars. Observations at submillimeter
    wavelengths provide a direct possibility to study the chemical
    composition of low-mass protostars through transitions probing
    temperatures up to a few hundred K in the gas surrounding these
    sources. This paper presents a large molecular line survey of the
    deeply embedded protostellar binary IRAS~16293-2422 from the
    Submillimeter Array (SMA) -- including images of individual lines
    down to $\approx$~1.5--3\arcsec\ (190--380~AU) resolution. More
    than 500 individual transitions are identified related to 54
    molecular species (including isotopologues) probing temperatures
    up to about 550~K. Strong chemical differences are found between
    the two components in the protostellar system with a separation
    between, in particular, the sulfur- and nitrogen-bearing species
    and oxygen-bearing complex organics. The action of protostellar
    outflow on the ambient envelope material is seen in images of CO
    and SiO and appear to influence a number of other species,
    including (deuterated) water, HDO. The effects of cold gas-phase
    chemistry is directly imaged through maps of CO, N$_2$D$^+$ and
    DCO$^+$, showing enhancements of first DCO$^+$ and subsequently
    N$_2$D$^+$ in the outer envelope where CO freezes-out on dust
    grains.}{}{}{}{}
\end{abstract}

\offprints{Jes K.\,J{\o}rgensen}\mail{jeskj@nbi.dk}
\maketitle

\section{Introduction}\label{introduction}
Developing the understanding of the chemical structure and evolution
of star-forming regions remains an important task. Understanding the
molecular composition of protostars and, in particular, the innermost
regions of the circumstellar envelopes and disks relates to some of
the key scientific questions concerning star and planet formation, for
example, what level of chemical complexity can arise around protostars
\citep[e.g.,][]{vandishoeck98,ceccarellippv,herbst09}. Also, for
studying the physics of the star formation process, it is desired to
know which molecular species are tracing specific components of young
stellar objects, such as their envelope, disks etc. This paper
presents the results of a large Submillimeter Array survey of the
molecular line emission on few hundred AU scales toward the deeply
embedded low-mass protostar IRAS~16293-2422 and discusses some of the
signatures of the physics and chemistry occurring in this deeply
embedded protobinary system.

IRAS~16293-2422 has long been considered one of the ``template''
sources for astrochemistry. Being the deeply embedded (Class~0)
low-mass protostar with the richest line spectrum, it has been the
subject of many targeted (sub)millimeter wave spectroscopic studies
\citep[e.g.,][]{blake94,vandishoeck95,ceccarelli98h2o,ceccarelli00a,cazaux03,caux11}
as well as specialized modeling efforts attempting to establish its
chemical composition -- in particular, variation in its molecular
abundances as function of radius
\citep[e.g.,][]{ceccarelli00a,schoeier02}. The detections of complex
organics toward this source
\citep[e.g.,][]{cazaux03,bottinelli04iras16293,kuan04,bisschop08} have
sparked new interest in the physical processes that can lead to the
evaporation of icy grain mantles on small scales of protostars -- and
thereby also the chemical processes determining their molecular
compositions.

However, IRAS~16293-2422 has also illustrated some of the inherent
difficulties in relating the larger scale line emission picked up by
single-dish telescopes to the source structures on few hundred AU
scales, e.g., revealed by millimeter interferometric
observations. After the identification of IRAS~16293-2422 as a binary
through high resolution centimeter and millimeter wavelength continuum
observations \citep{wootten89,mundy92}, it has been the target of many
studies trying to relate the structure of the two main components to
their line emission and place them in an evolutionary context. The
southeastern of the two components, ``IRAS~16293A'', appears resolved
in continuum observations, breaking into a number of different
components at subarcsecond scales \citep{chandler05,pech10}. The
northwestern component, ``IRAS~16293B'', in contrast appears
unresolved on these scales. In terms of line emission the two sources
also show significant differences: both show detection of complex
organic molecules for example
\citep{bottinelli04iras16293,kuan04,remijan06,bisschop08} -- but the
relative line strengths and widths vary between the two
sources. Whereas it is generally agreed that the IRAS~16293A component
is protostellar in nature -- it has been suggested that the IRAS~16293B
component either represented a more evolved (T Tauri) star
\citep[e.g.,][]{stark04,takakuwa07iras16293} or alternatively a very
young object, possibly before starting accretion/mass loss
\citep{chandler05}.

In this paper we present a large survey of the line emission in the
230~GHz and 345~GHz atmospheric windows of IRAS~16293-2422 from the
Submillimeter Array. The paper is laid out as follows:
\S~\ref{observations} describes the details of the observations and
\S~\ref{analysis} presents an overview of the line emission in global
terms. \S~\ref{discussion} discusses a few of the key aspects that can
be derived from just visual inspection of the molecular line emission
maps and \S~\ref{summary} summarizes the main conclusions of the
paper.

\section{Observations}\label{observations}
IRAS~16293-2422 was observed in a number of spectral settings between
2004 and 2007 using the Submillimeter Array (SMA; \citealt{ho04}). We
here focus on four sets of observations covering different spectral
setups at 220 and 340~GHz. The log of the observations, beam sizes and
noise levels are summarized in Table~\ref{obslog}. Previous papers by
\cite{yeh08} and \cite{bisschop08} presented part of these data from
2005 Feb 18, focusing on the CO outflow emission and selected complex
organic molecules, respectively.

\begin{figure}
\resizebox{\hsize}{!}{\includegraphics{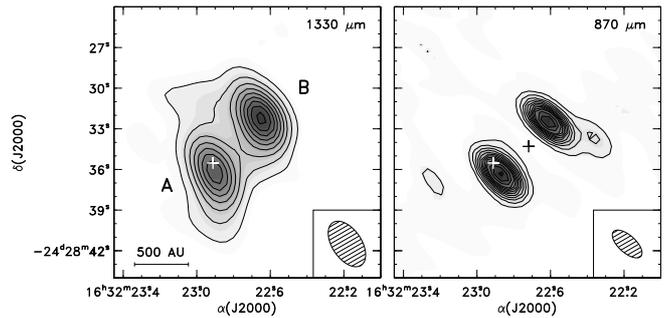}}
\caption{Continuum images of IRAS~16293-2422 at 1.3~mm (left) and
  0.87~mm (right). In the left panel the contours are given in steps
  of 0.1~Jy~beam$^{-1}$ to 1.0~Jy~beam$^{-1}$ and from there in steps
  of 0.2~Jy~beam$^{-1}$ and in the right in steps of
  0.2~Jy~beam$^{-1}$ to 2.0~Jy~beam$^{-1}$ and from there in steps of
  0.4~Jy~beam$^{-1}$ (in the continuum data the noise level is
  determined by the interferometer dynamic range). The white
  plus-signs indicate the pointing centers of the observations, except
  those from 2007 March 22 for which the pointing center is indicated
  by the black plus-sign in the right panel.}\label{contfig}
\end{figure}

The data were taken with the SMA in its compact or compact-North
configuration resulting in average beam sizes of $\approx$~2--4$''$
(250--500~AU at a typical distance to Ophiuchus of 125~pc). For three
of the four datasets a pointing center at $\alpha$=16\fh32\fm22.91\fs,
$\delta$=-24\fd28\arcmin35\farcs5 [J2000] was used. For the last
dataset (2007 March 22) the pointing center was
$\alpha$=16\fh32\fm22.72\fs, $\delta$=-24\fd28\arcmin34\farcs3
[J2000]. The positions for the two main continuum peaks seen in these
images (Fig.~\ref{contfig}) are at $\alpha$=16\fh32\fm22.87\fs,
$\delta$=-24\fd28\arcmin36\farcs4 [J2000] (IRAS16293A or ``A'') and
$\alpha$=16\fh32\fm22.62\fs, $\delta$=-24\fd28\arcmin32\farcs4 [J2000]
(IRAS16293B or ``B'') with an agreement between the fitted positions
of about 0.2\arcsec. The field of view of the SMA observations cover a
region of 30--50$''$ (3750--6250~AU; diameter) at 345--230 GHz.

Two of the four datasets (the 220/230~GHz and 337/347~GHz datasets
from 2005 February 18 and 2007 March 22, respectively) were taken as
part of dual receiver observations with the high frequency receiver
tuned to lines at 690~GHz. For these datasets the excellent weather
conditions required for the 690~GHz observations also result in
significantly improved RMS noise levels in the lower frequency data
discussed here. The correlator was configured with uniform spectral
coverage over the (at the time) $\approx$~2~GHz bandwidth in each
sideband of the SMA receivers. Each 2~GHz sideband was covered by 24
chunks of the correlator with a width of 128~MHz each -- and each
chunk split into 256~channels (128~channels for the dual receiver
observations).

The data were calibrated using the standard recipes -- including
calibration of the complex gains by observations of the nearby quasar
J1743-038 and flux and passband calibration by observations of planets
(Uranus, in particular). The initial data reduction was performed
using the MIR package \citep{qimir} and continuum subtracted line maps
were created using the Miriad package \citep{sault95} with which
further analysis was also done.
\begin{table*}[!htb]
\caption{Log of observations. \label{obslog}}
\begin{tabular}{llllll} \hline\hline
Frequency range & Observed date    & Beam size\tablefootmark{a} & {
  Spectral resolution} & {Sensitivity} & {Projected baselines}\tablefootmark{b} \\
     $[$GHz$]$                  &                  &          &  [km~s$^{-1}$] &  [Jy~beam$^{-1}$~chan$^{-1}$] & [k$\lambda$] \\ \hline
215.6--217.6 / 225.6--227.6 & 2004-07-23 & 5.5\arcsec$\times$3.2\arcsec &  0.56           & 0.24   & \phantom{1}6--91      \\
219.4--221.4 / 229.4--231.4 & 2005-02-18 & 4.0\arcsec$\times$2.4\arcsec &  1.1            & 0.062  & \phantom{1}7--54      \\
336.8--338.7 / 346.7--348.7 & 2007-03-22 & 2.5\arcsec$\times$1.6\arcsec &  0.72           & 0.11   & 10--141    \\
341.2--343.1 / 351.1--353.1 & 2005-08-14 & 3.5\arcsec$\times$1.9\arcsec &  0.36           & 0.60   & 11--82     \\ \hline
\end{tabular}
\tablefoot{\tablefoottext{a}{With natural weighting.}
  \tablefoottext{b}{Range of projected interferometer baselines observed for given
    dataset measured in k$\lambda$.}}
\end{table*}

\section{Analysis}\label{analysis}
In this paper we focus on the morphology of the line emission from the
IRAS~16293-2422 data and refer to other papers in the literature for a
more in-depth discussion of continuum emission toward the sources. We
make the data publicly available for anyone interested in a more
in-depth analysis of specific molecules: the SMA raw data are
available through CfA Radio Telescope Data
Center\footnote{http://cfa-www.harvard.edu/rtdc} and the spectra
toward the two continuum peaks in FITS format through a dedicated
website\footnote{http://www.nbi.dk/$\sim$jeskj/sma-iras16293.html}
(the full reduced datacubes are available on a collaborative basis).

\begin{figure*}
\resizebox{0.90\hsize}{!}{\includegraphics{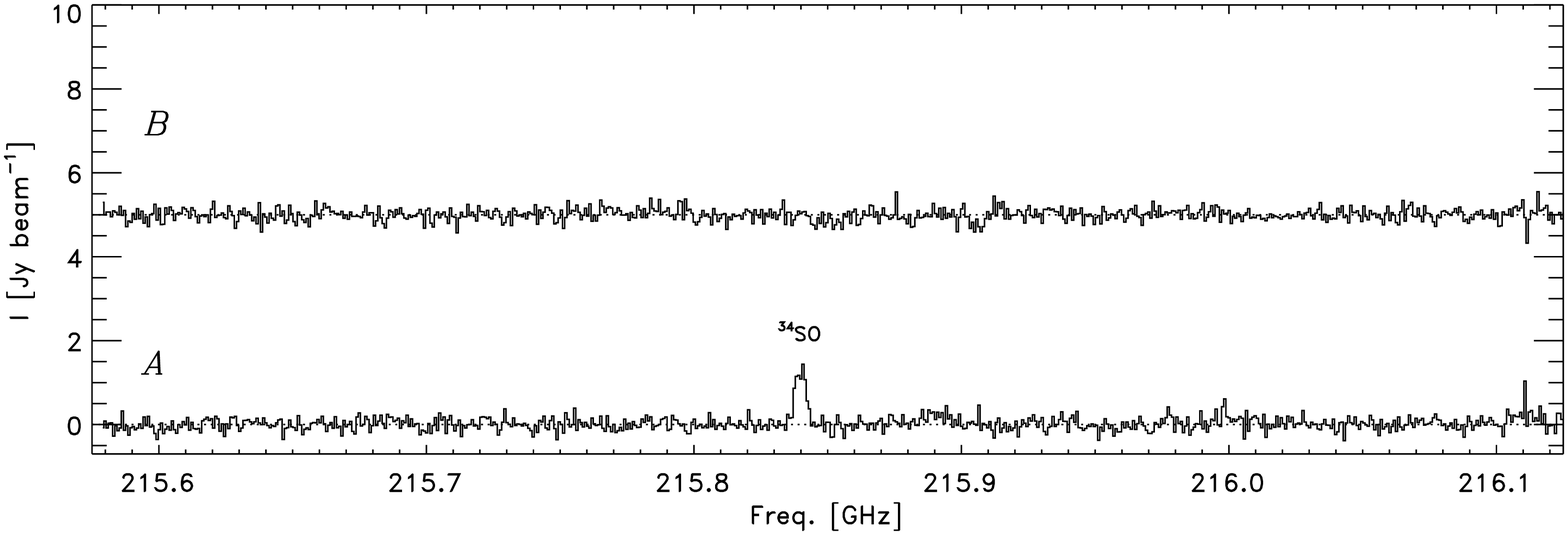}}
\resizebox{0.90\hsize}{!}{\includegraphics{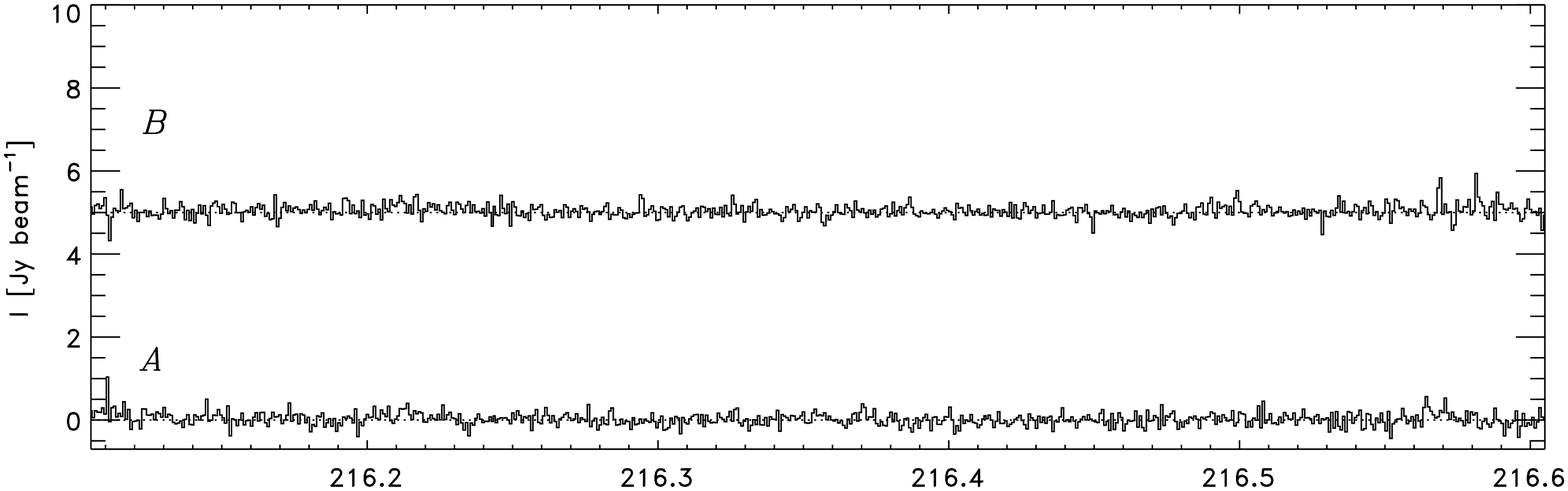}}
\resizebox{0.90\hsize}{!}{\includegraphics{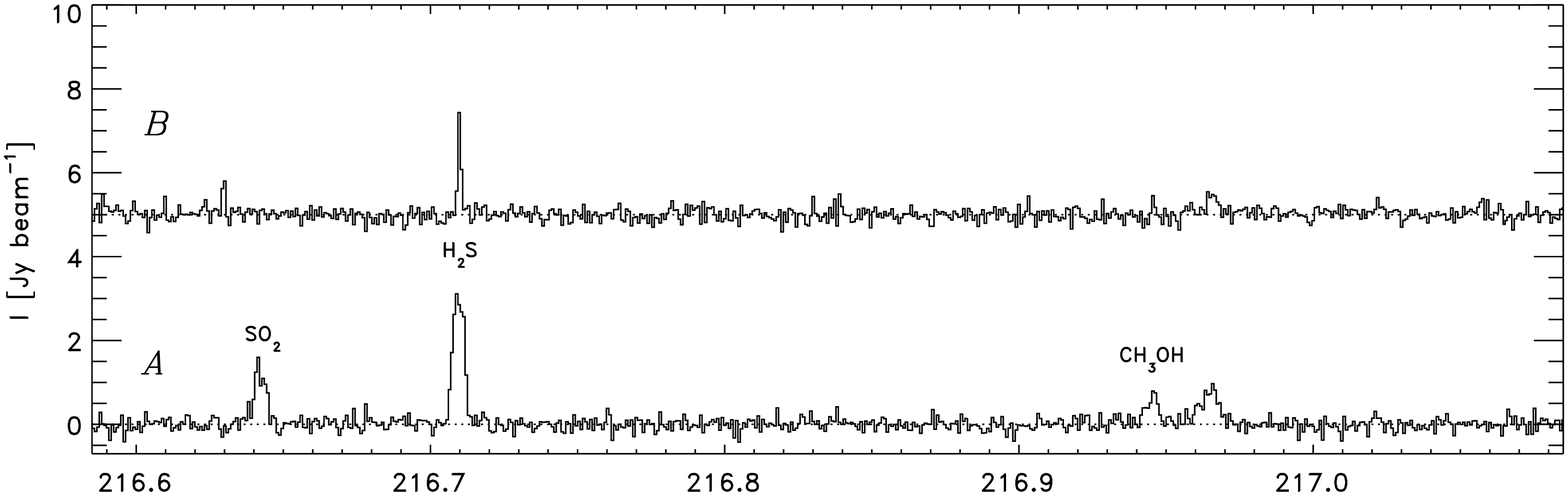}}
\resizebox{0.90\hsize}{!}{\includegraphics{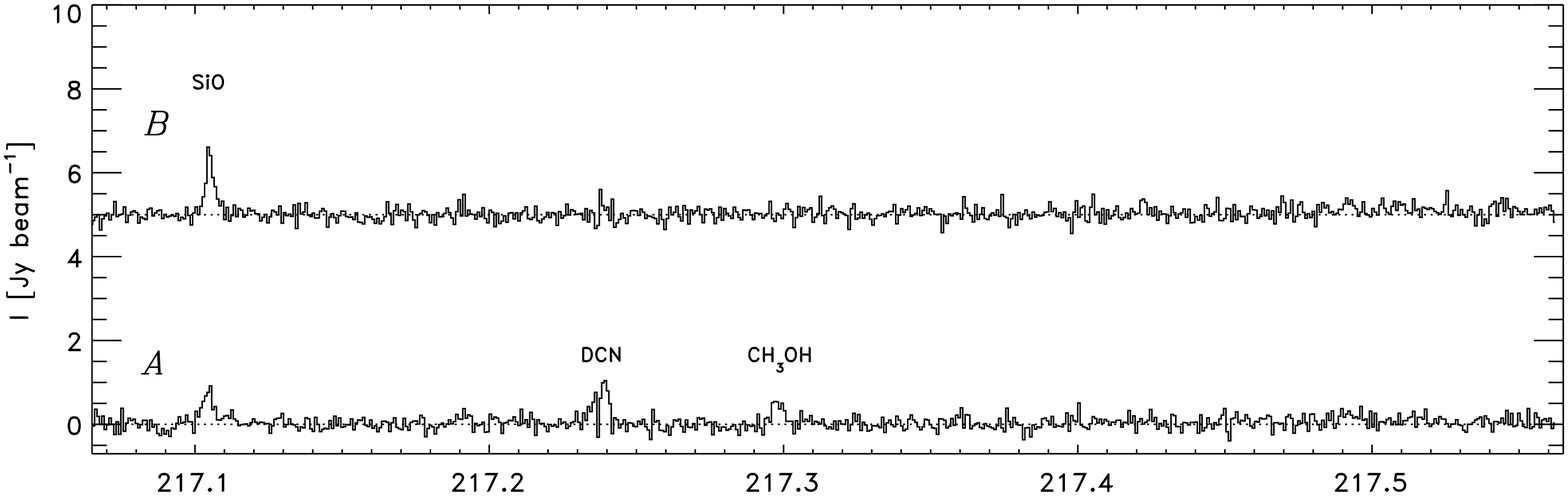}}
\caption{Composite spectrum in the range 215.5--217.5 GHz for the
  central beam (5.5\arcsec$\times$3.2\arcsec; line 1 of
  Table~\ref{obslog}) toward the source IRAS~16293A (at 0 on the
  Y-axis) and IRAS~16293B (offset in the Y-axis
  direction). Transitions of some of the prominent species have been
  identified.} \label{spectrum_first}
\end{figure*}

Fig.~\ref{spectrum_first}--\ref{spectrum_last} show the composite
spectra toward the continuum positions of IRAS~16293A and
IRAS~16293B. Key differences for the line emission in the two sources
also noted in previous papers are clearly illustrated: typically the
IRAS~16293A component show broader and stronger lines, for example,
clearly illustrated in the methanol CH$_3$OH $J_k = 7_k-6_k$ branch at
338.4~GHz. On the other hand, IRAS~16293B shows the presence of some
sets of lines not seen in IRAS~16293A, e.g., in the frequency range
from 346.9--347.3~GHz harboring a number of transitions of
acetaldehyde, CH$_3$CHO. 

A few lines show very complex profiles likely due to combinations of
optical depth effects and spatial resolving out due to the
interferometer's lack of short-spacings: with the shortest projected
baselines of about 8~m length (Table~\ref{obslog}) the SMA
observations for example recover less than 50\% of any emission with a
Gaussian distribution with a FWHM of 10--15\arcsec\ (i.e., molecules
with a similar surface brightness distribution as the envelope traced
by single-dish continuum observations; \citealt{schoeier02}) and an
even smaller fraction for molecules more homogeneously distributed
(e.g., less strongly weighted by temperature than the dust continuum
emission). This is clearly seen for the CO isotopologues -- $^{12}$CO
at 230.538~GHz and $^{13}$CO at 220.398~GHz, as examples -- but the
``absorption'' features seen toward IRAS~16293B for, e.g., N$_2$D$^+$
at 231.321~GHz and CN at 226.875~GHz, also reflect this. In
particular, the absence of CN emission in the maps is a clear example
of the effects of the interferometer resolving out more extended
emission: in pointed JCMT single-dish observations of CN at
226.875~GHz \cite{vandishoeck95} found lines with intensities of
0.6--0.8~\kms\ (10--15~Jy~beam$^{-1}$~\kms), contrasting the absence
of CN emission in the SMA maps presented here. As pointed out by
\citeauthor{vandishoeck95}, the narrow widths of the CN lines suggests
a picture where this species is probing only the outer envelope and
ambient core. This would indeed be on spatial scales resolved out by
the SMA observations.

The interferometer's spatial filtering makes it difficult to use the
interferometric data for quantitative estimates of, e.g., exact column
densities or other physical parameters without a careful treatment of
the amount of resolved-out emission, for example by combining the
interferometric data with short-spacing maps from single-dish
telescopes or more detailed models of the source structure (see, e.g.,
\citealt{hotcorepaper} and \citealt{takakuwa07iras16293}). Still, the
interferometric maps reveal the most prominent structures in the
surface brightness distributions from the molecular lines -- for
example where the largest column densities of the different species
occur -- and therefore allow for a qualitative interpretation of the
relations and differences between the imaged molecular species.

\begin{figure*}
\resizebox{0.90\hsize}{!}{\includegraphics{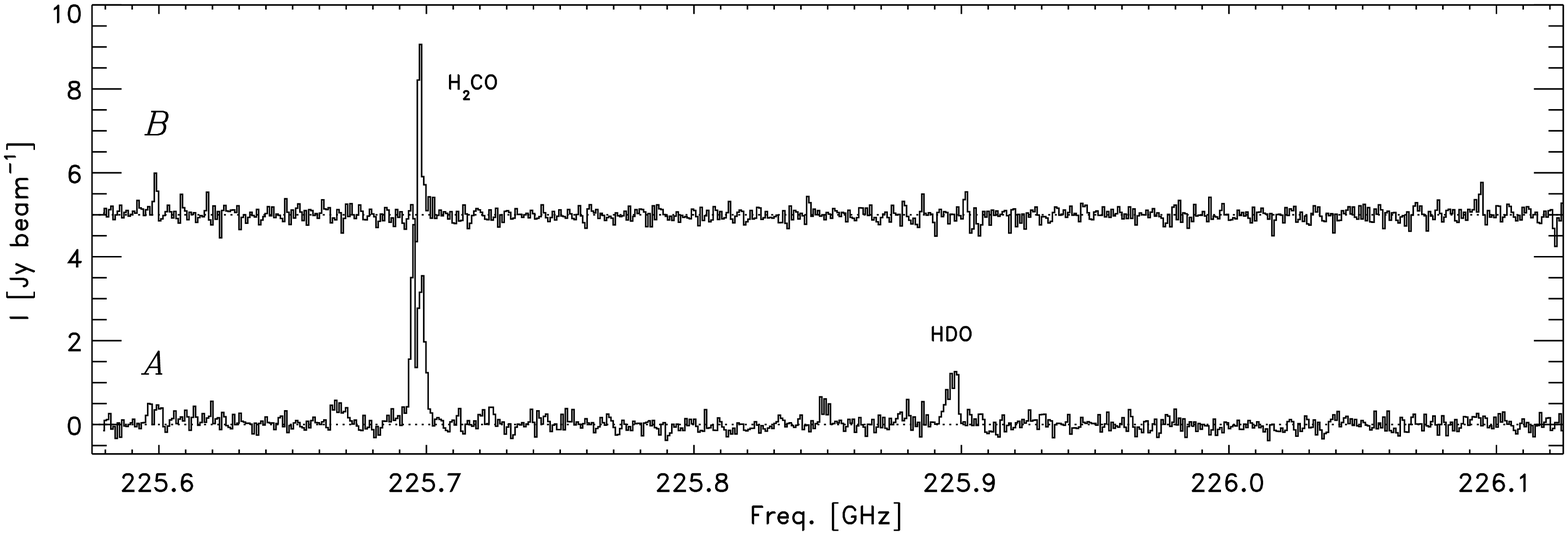}}
\resizebox{0.90\hsize}{!}{\includegraphics{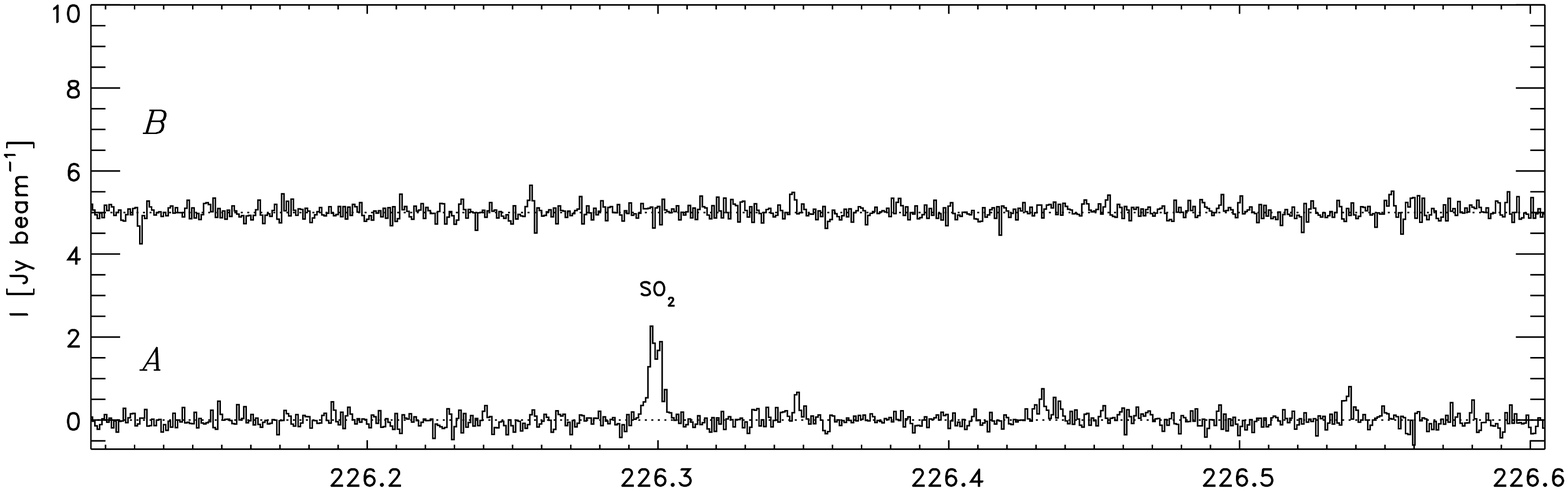}}
\resizebox{0.90\hsize}{!}{\includegraphics{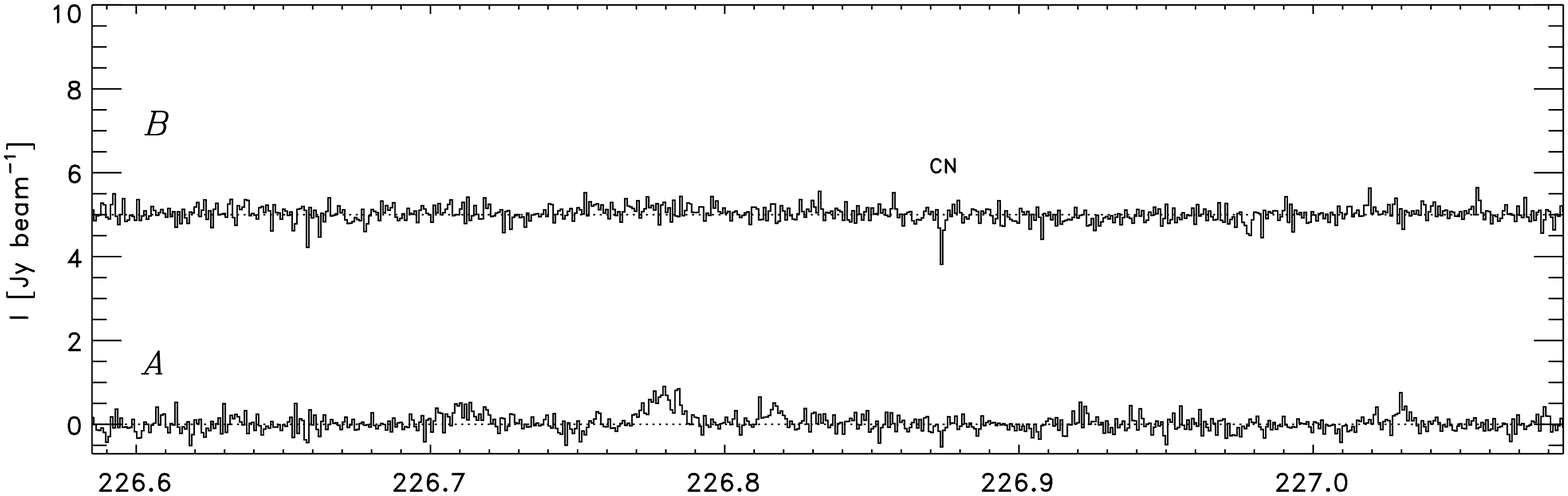}}
\resizebox{0.90\hsize}{!}{\includegraphics{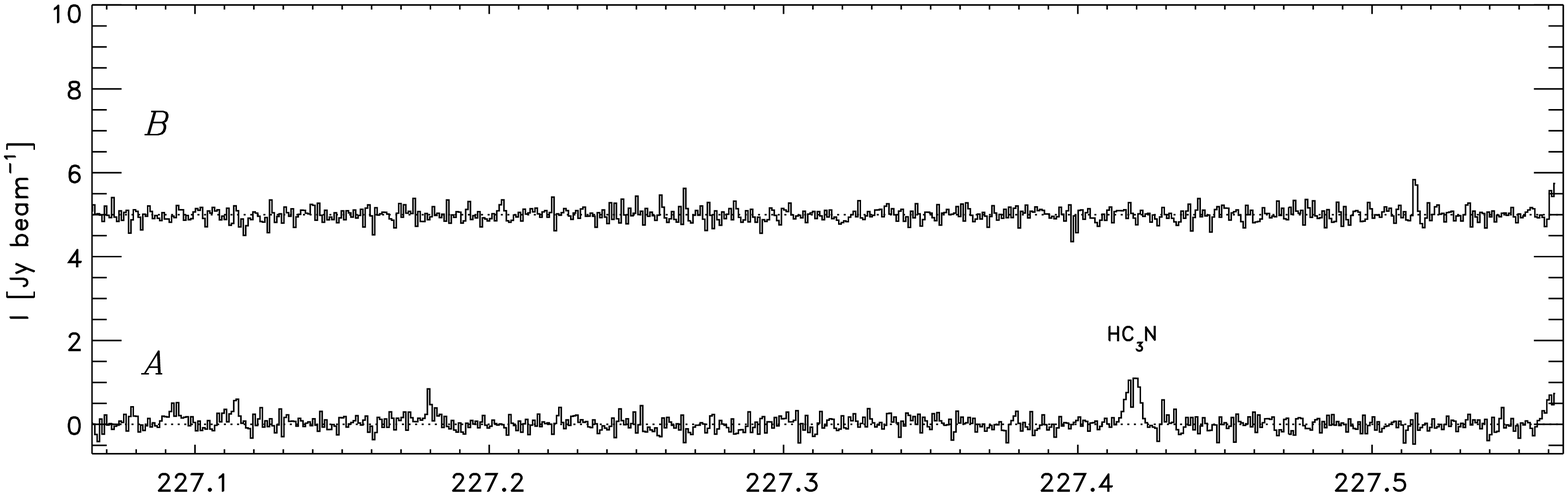}}
\caption{As in Fig.~\ref{spectrum_first} for the range 225.5--227.5
  GHz (beam size 5.5\arcsec$\times$3.2\arcsec; line 1 of Table~\ref{obslog}).}
\end{figure*}
\begin{figure*}
\resizebox{0.90\hsize}{!}{\includegraphics{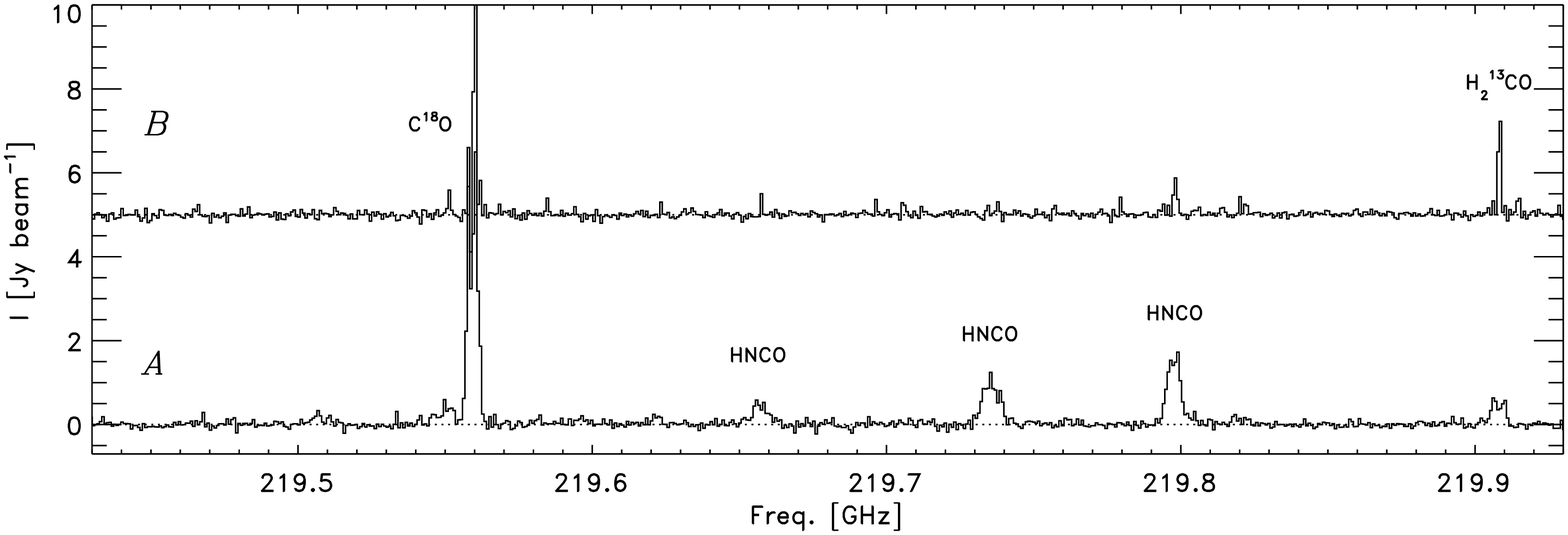}}
\resizebox{0.90\hsize}{!}{\includegraphics{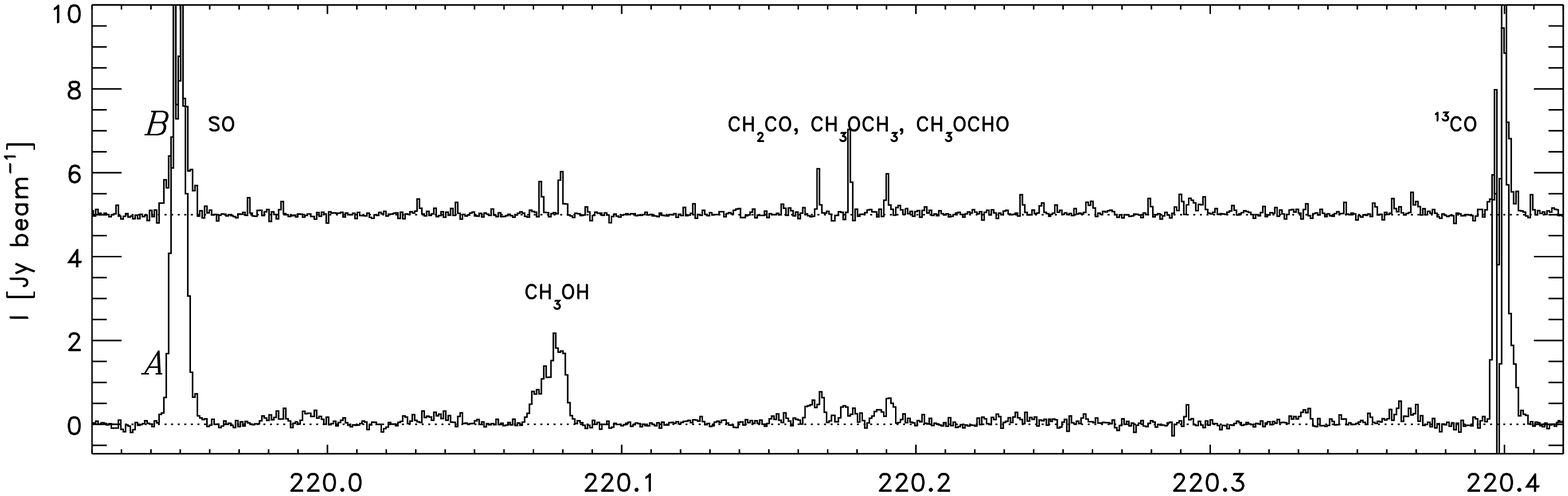}}
\resizebox{0.90\hsize}{!}{\includegraphics{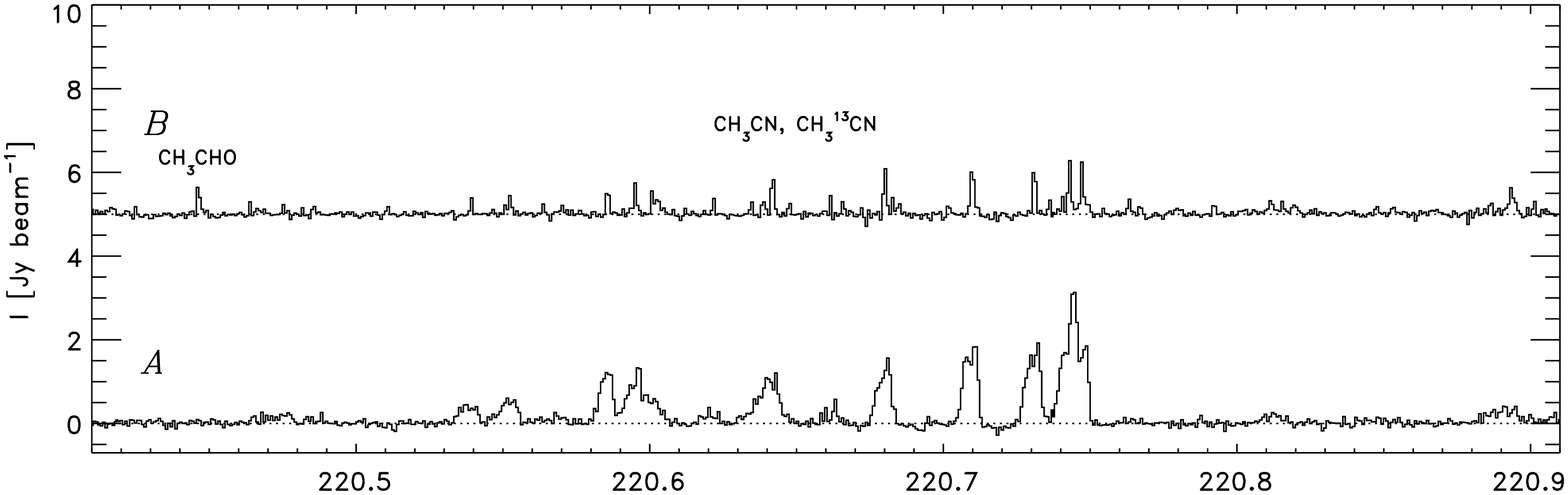}}
\resizebox{0.90\hsize}{!}{\includegraphics{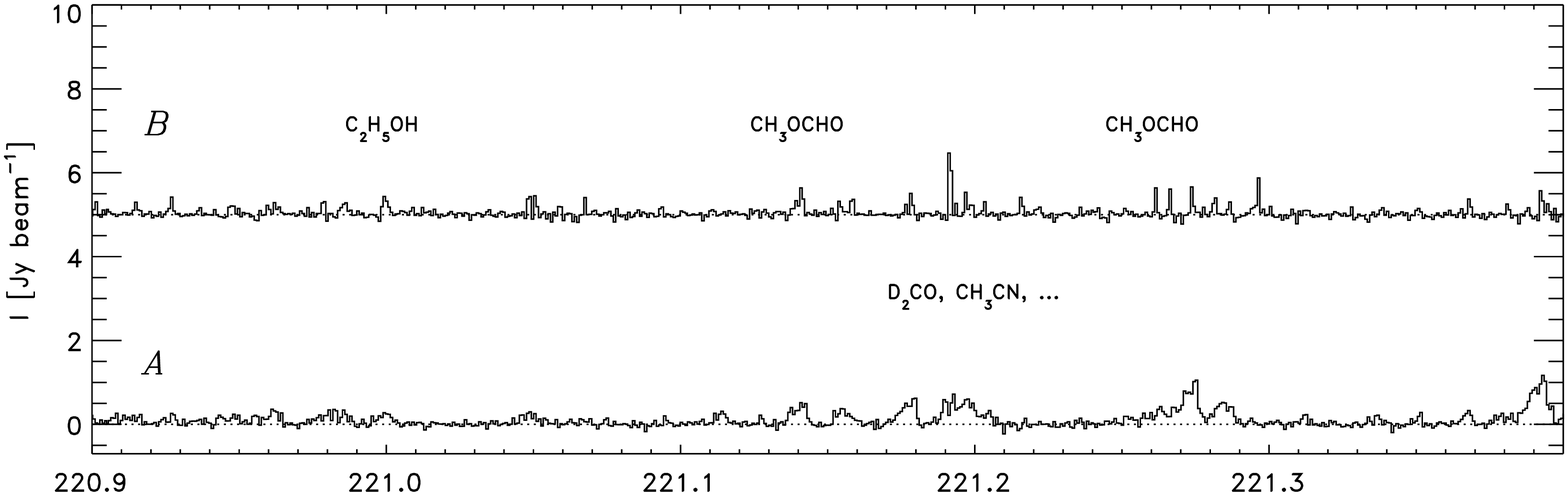}}
\caption{As in Fig.~\ref{spectrum_first} for the range  219.4--221.4
  GHz (beam size 4.0\arcsec$\times$2.4\arcsec; line 2 of Table~\ref{obslog}).}
\end{figure*}
\begin{figure*}
\resizebox{0.90\hsize}{!}{\includegraphics{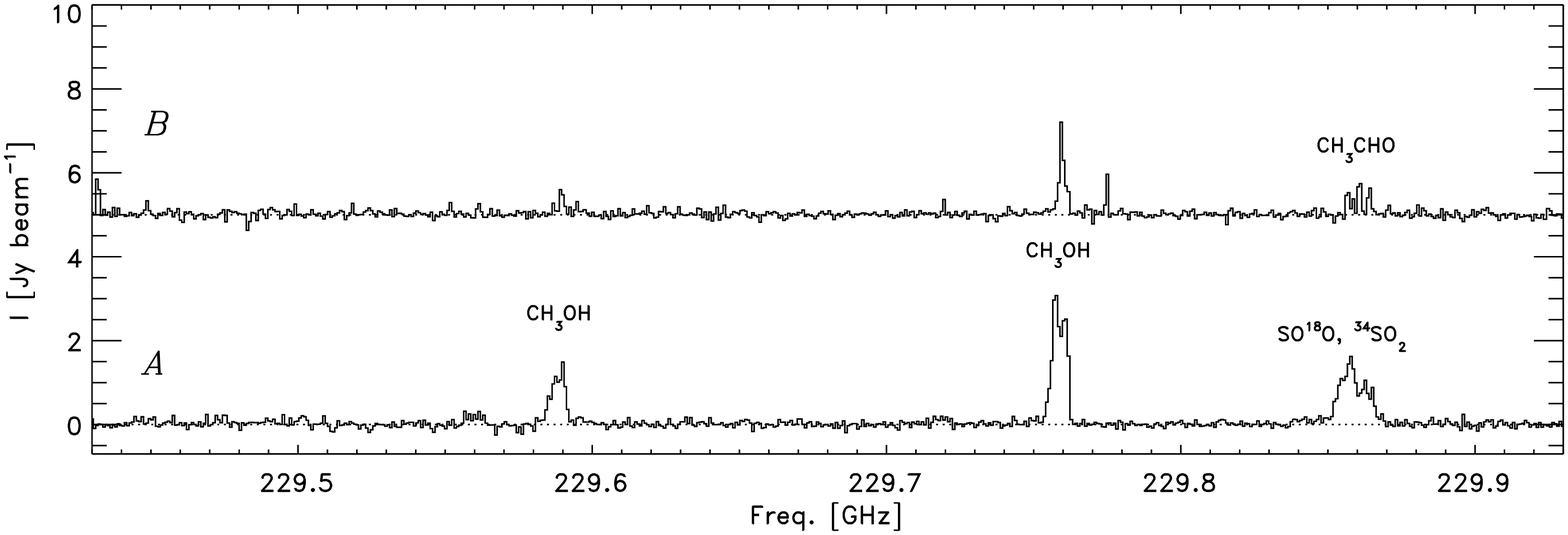}}
\resizebox{0.90\hsize}{!}{\includegraphics{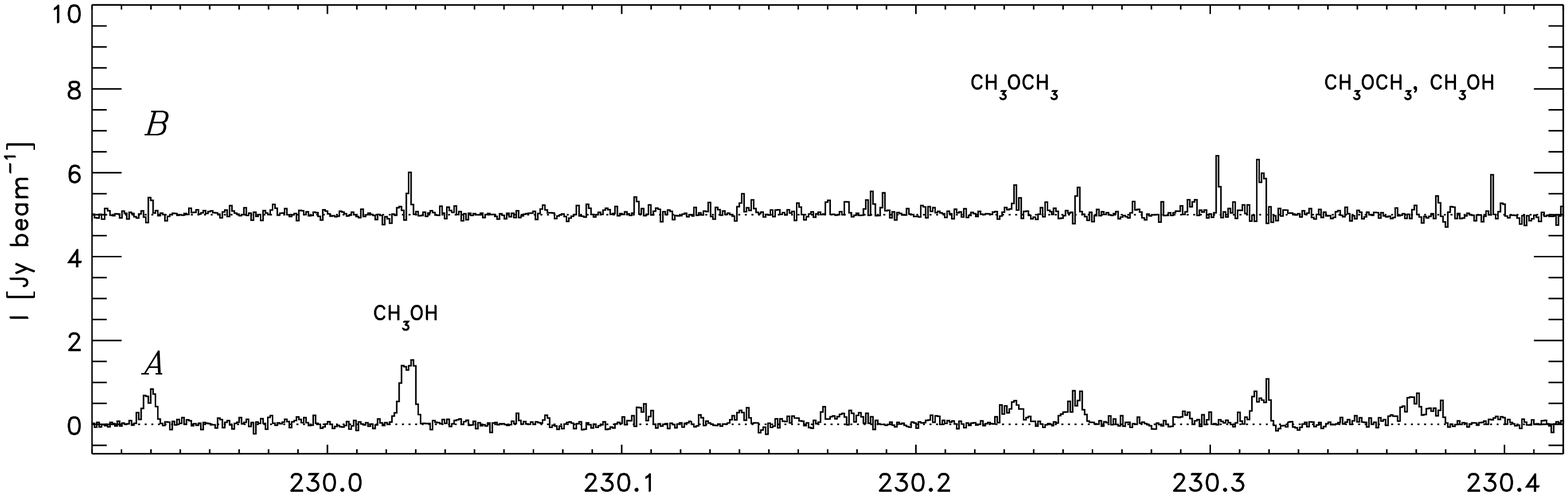}}
\resizebox{0.90\hsize}{!}{\includegraphics{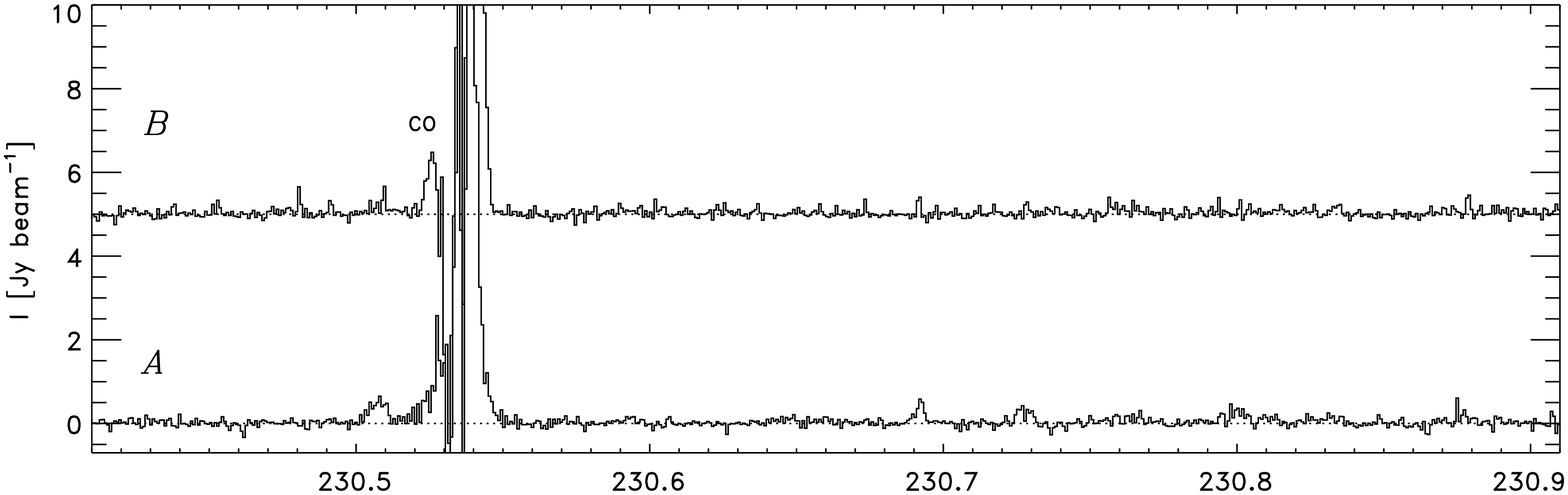}}
\resizebox{0.90\hsize}{!}{\includegraphics{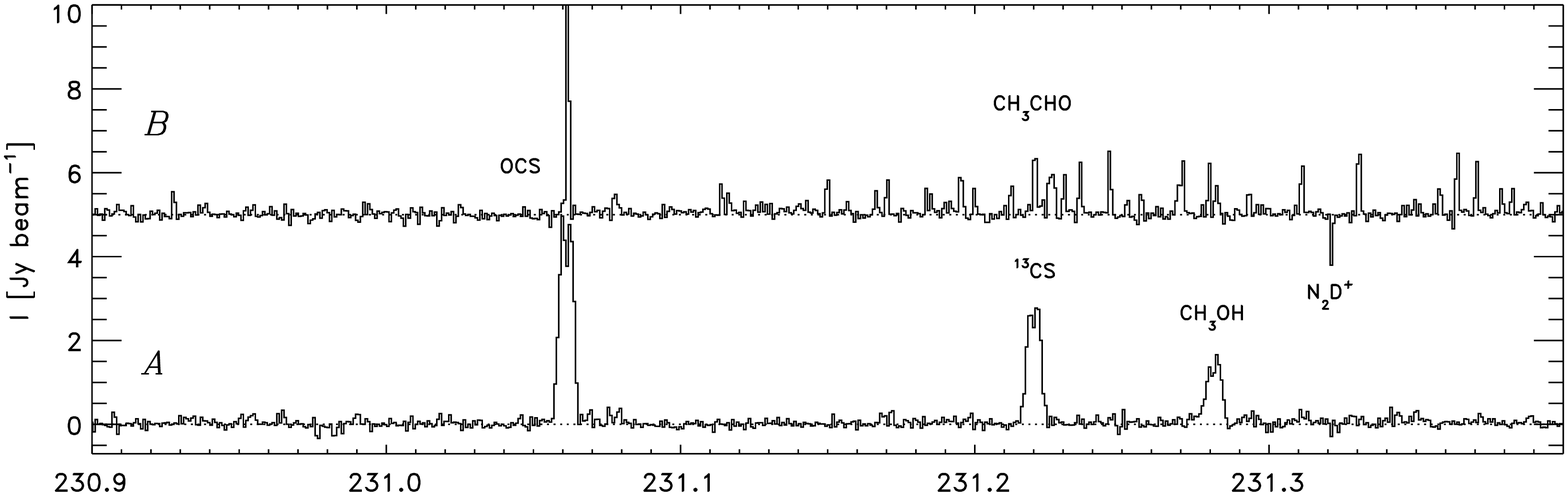}}
\caption{As in Fig.~\ref{spectrum_first} for the range  229.4--231.4
  GHz (beam size 4.0\arcsec$\times$2.4\arcsec; line 2 of Table~\ref{obslog}).}
\end{figure*}
\begin{figure*}
\resizebox{0.90\hsize}{!}{\includegraphics{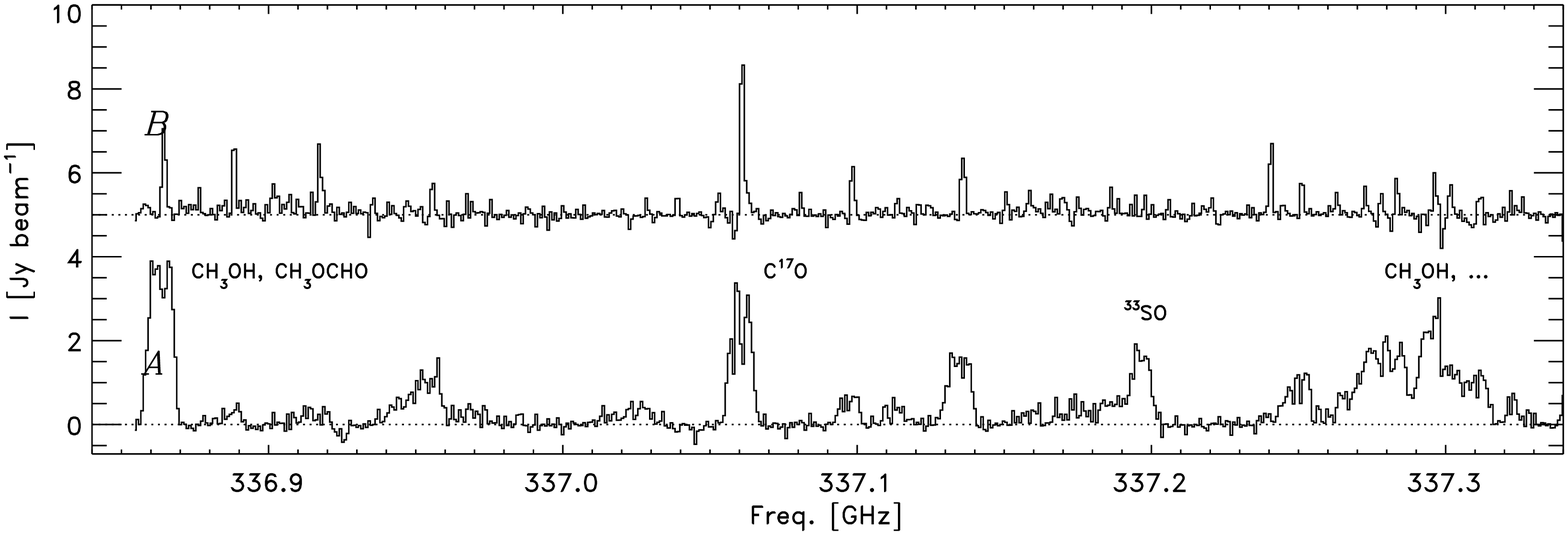}}
\resizebox{0.90\hsize}{!}{\includegraphics{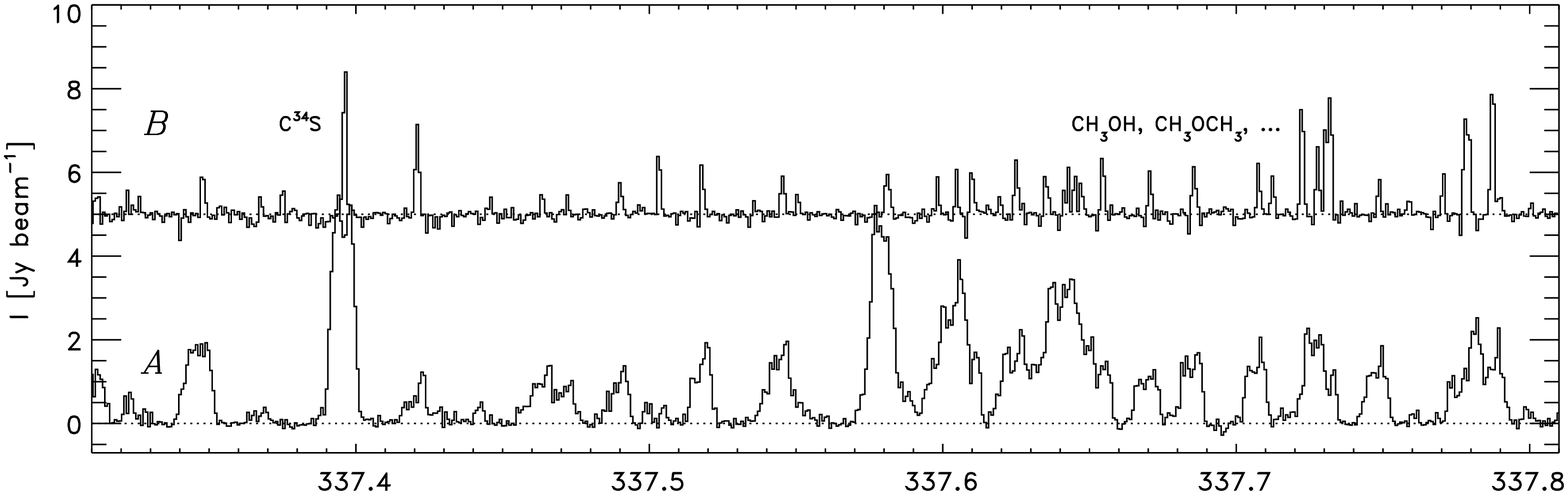}}
\resizebox{0.90\hsize}{!}{\includegraphics{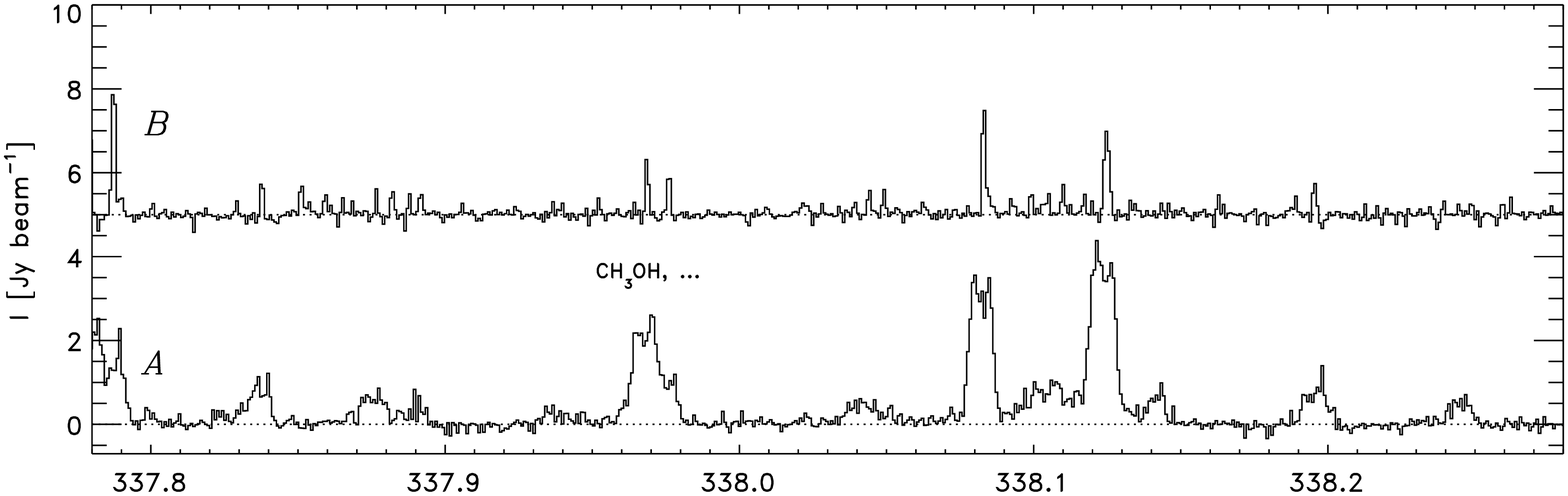}}
\resizebox{0.90\hsize}{!}{\includegraphics{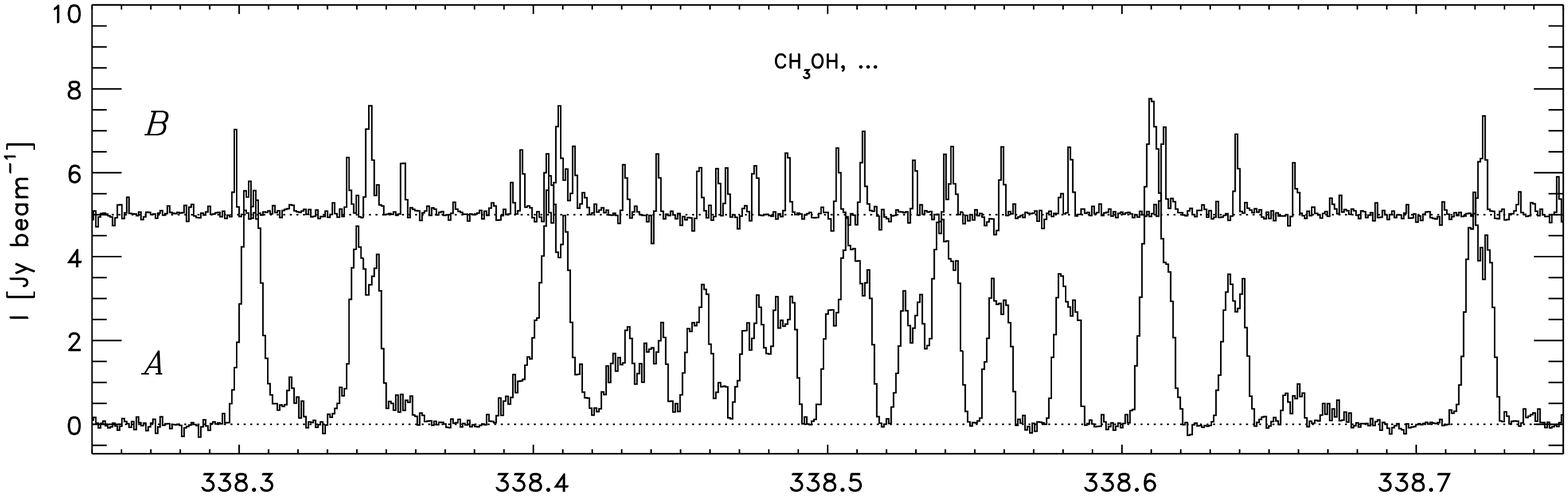}}
\caption{As in Fig.~\ref{spectrum_first} for the range 336.85--338.85
  GHz (beam size 2.5\arcsec$\times$1.6\arcsec; line 3 of Table~\ref{obslog}).}
\end{figure*}
\begin{figure*}
\resizebox{0.90\hsize}{!}{\includegraphics{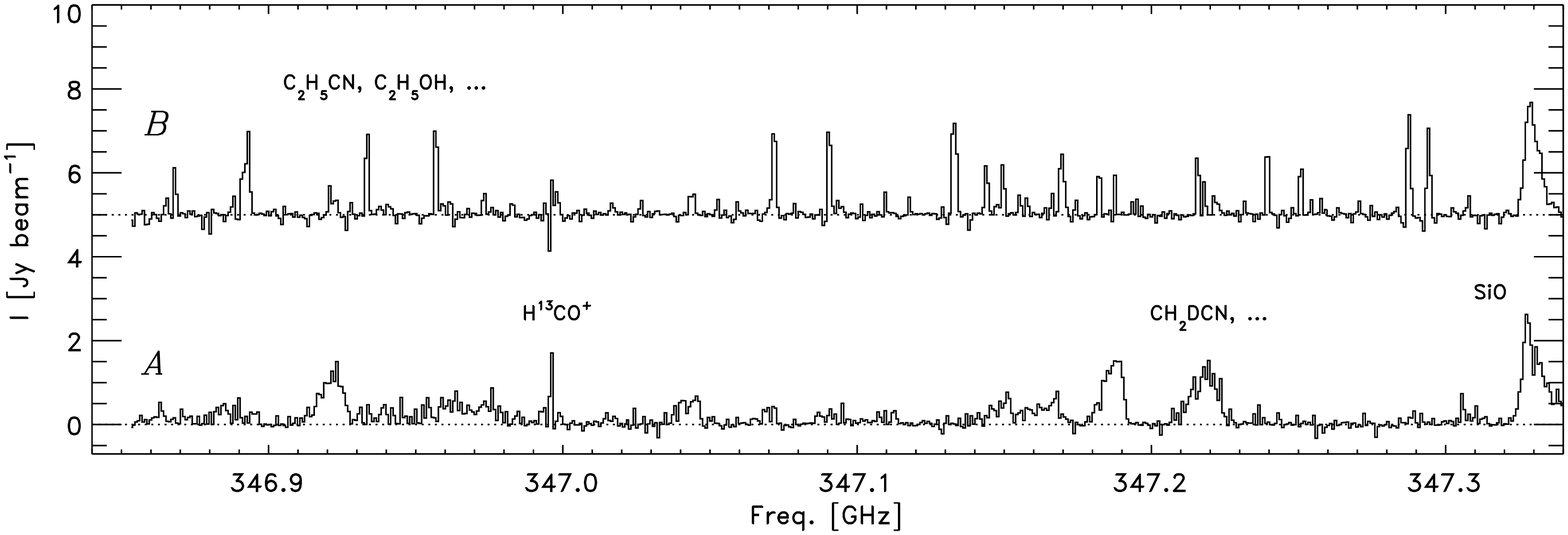}}
\resizebox{0.90\hsize}{!}{\includegraphics{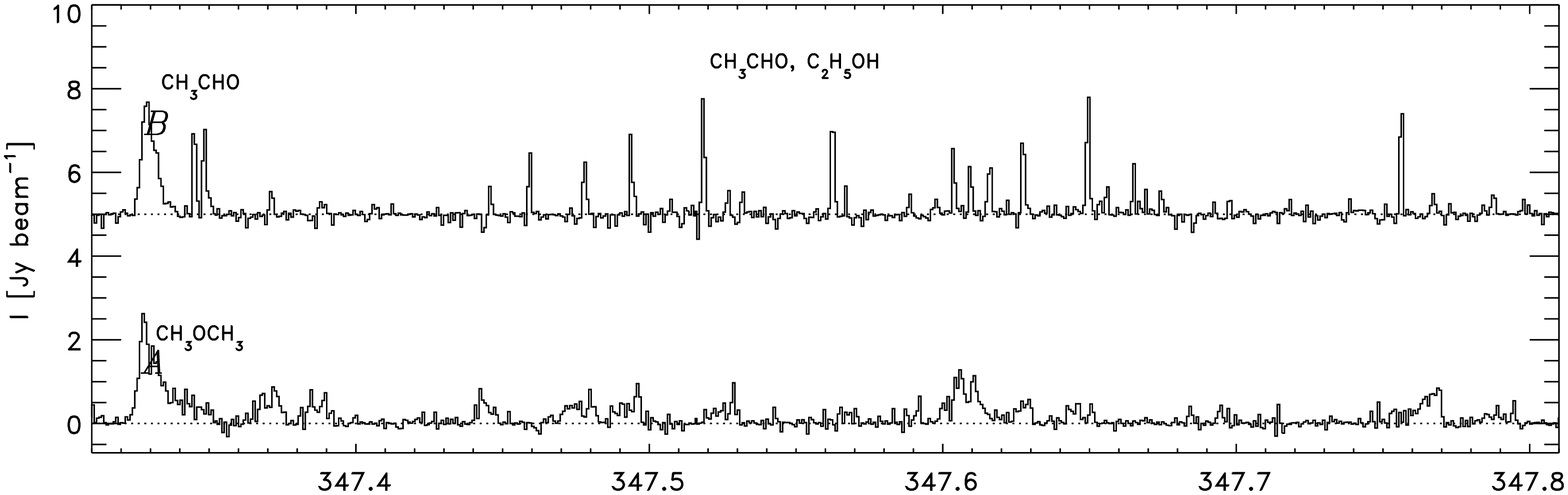}}
\resizebox{0.90\hsize}{!}{\includegraphics{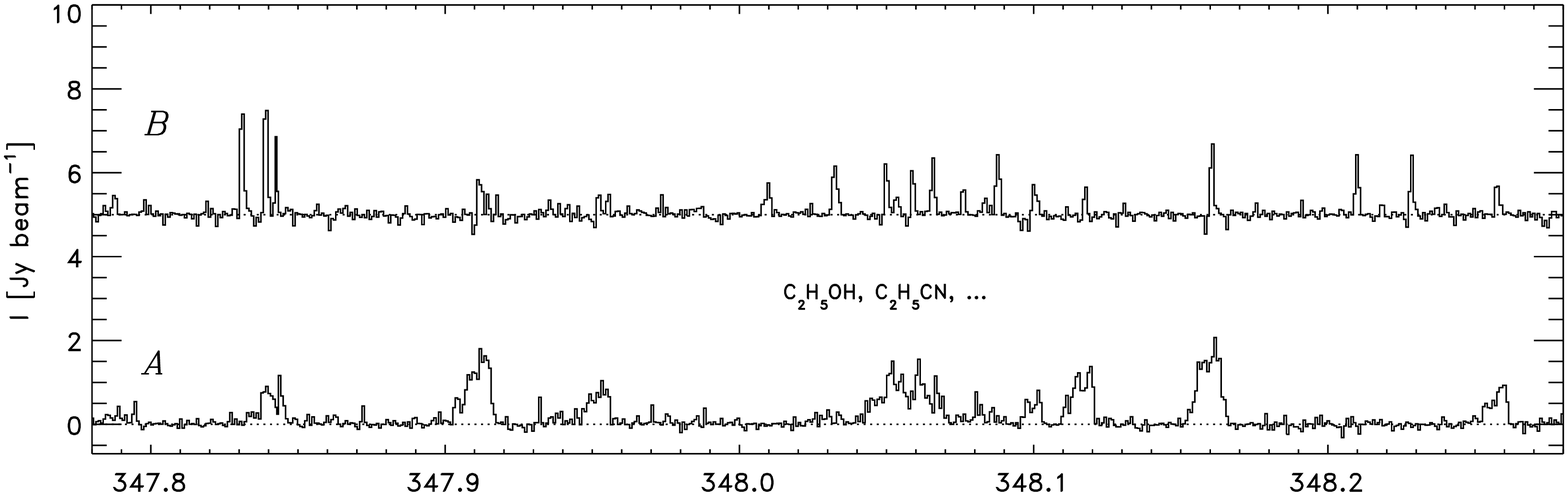}}
\resizebox{0.90\hsize}{!}{\includegraphics{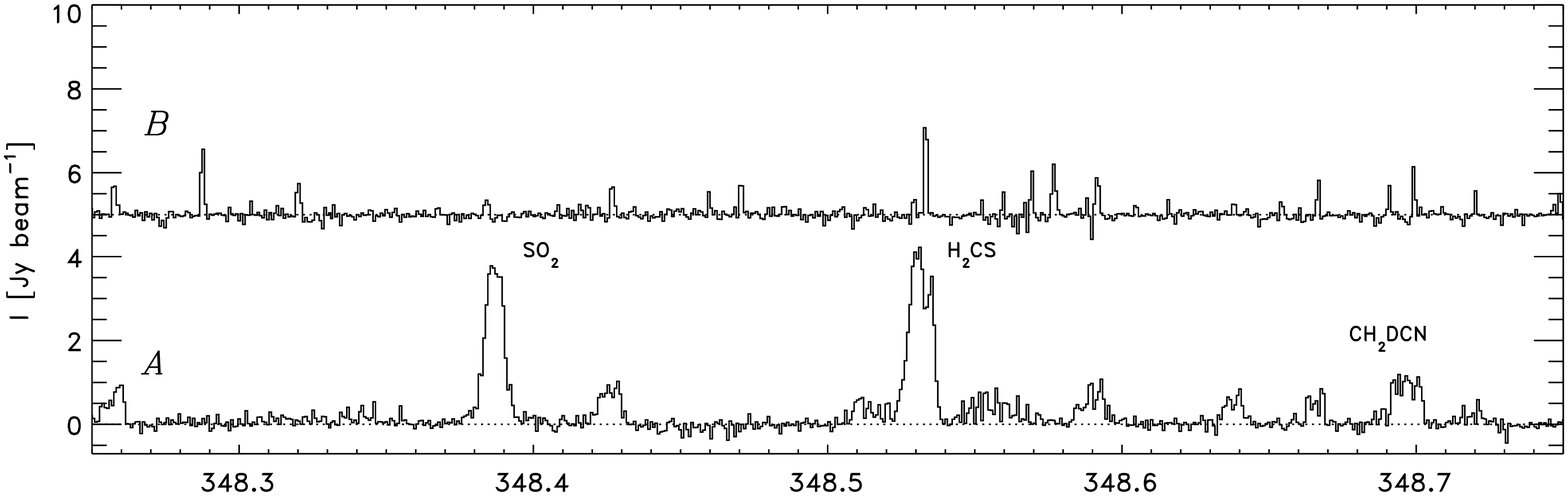}}
\caption{As in Fig.~\ref{spectrum_first} for the range 346.85--348.85
  GHz (beam size 2.5\arcsec$\times$1.6\arcsec; line 3 of Table~\ref{obslog}).}
\end{figure*}
\begin{figure*}
\resizebox{0.90\hsize}{!}{\includegraphics{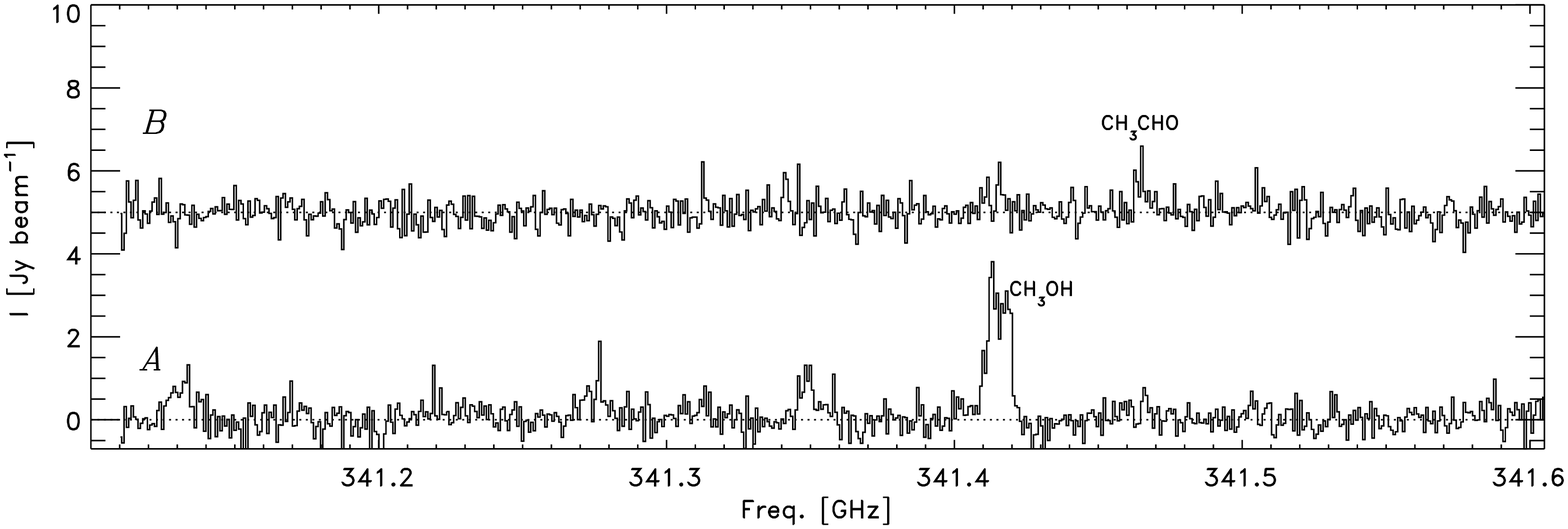}}
\resizebox{0.90\hsize}{!}{\includegraphics{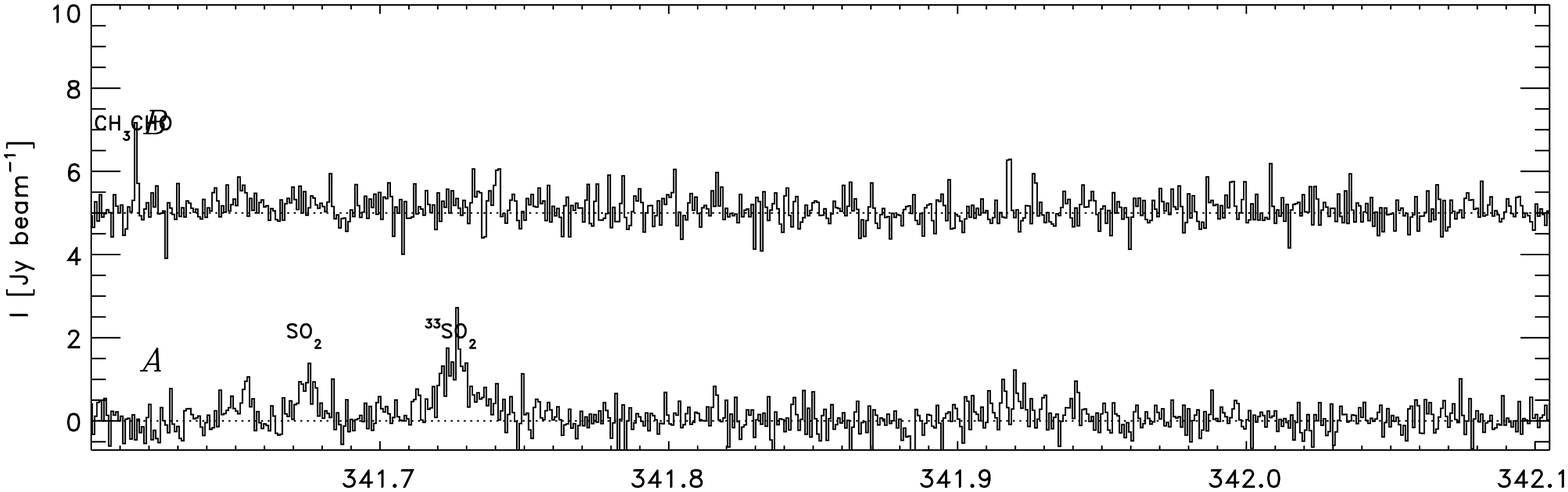}}
\resizebox{0.90\hsize}{!}{\includegraphics{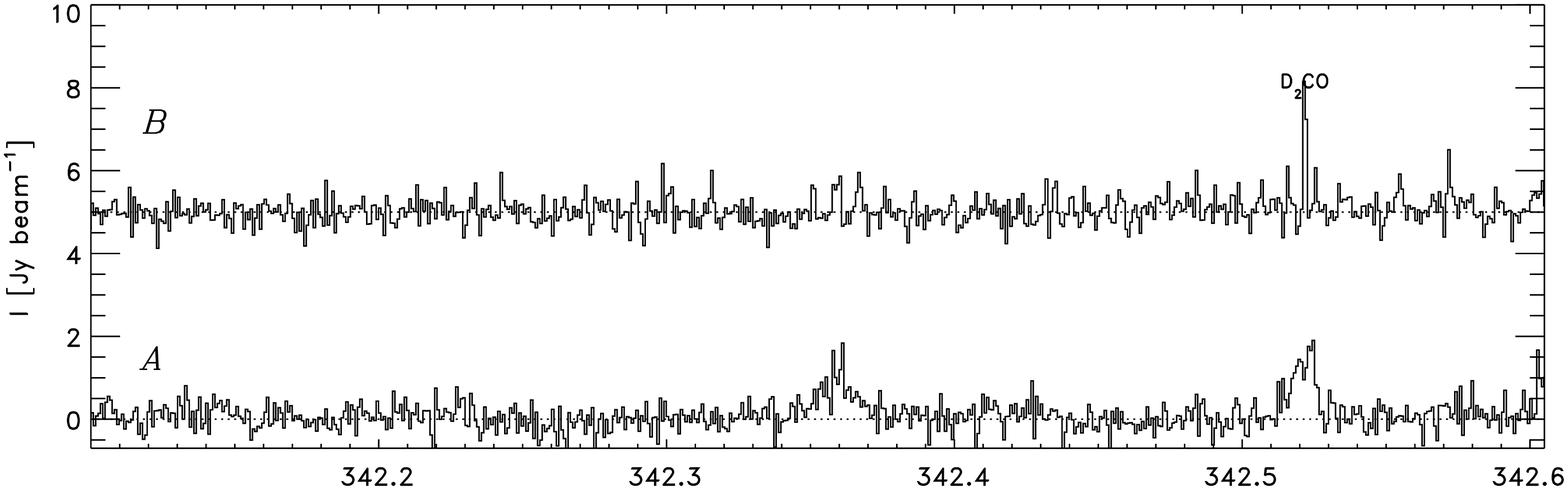}}
\resizebox{0.90\hsize}{!}{\includegraphics{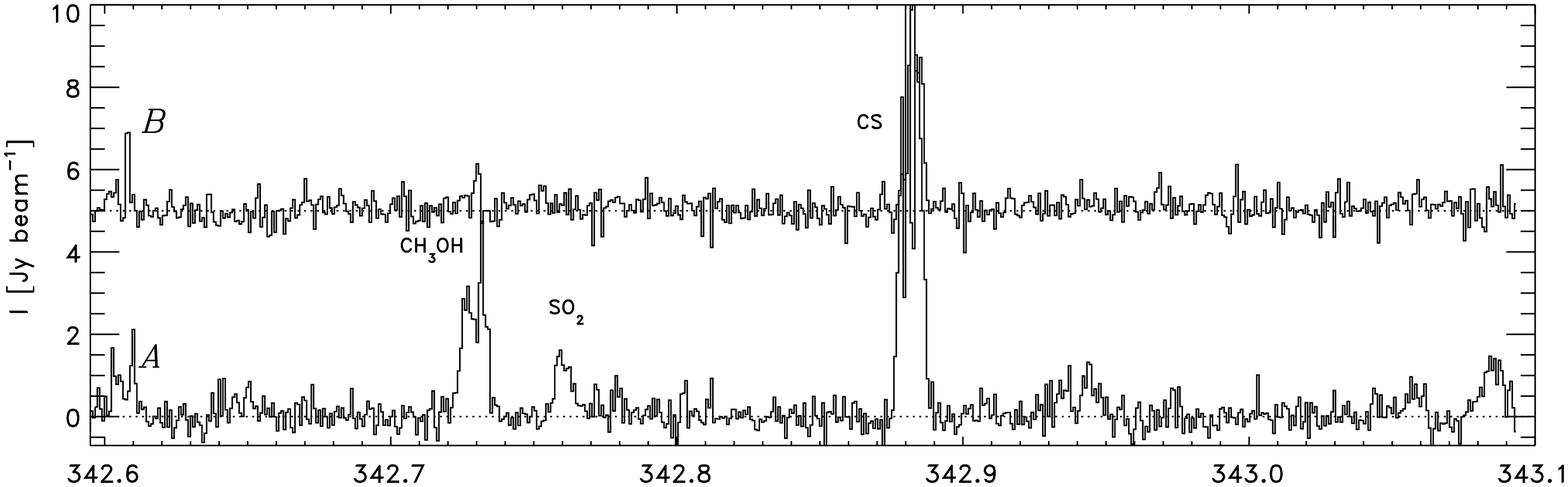}}
\caption{As in Fig.~\ref{spectrum_first} for the range 341.1--343.3
  GHz (beam size 3.5\arcsec$\times$1.9\arcsec; line 4 of Table~\ref{obslog}).}
\end{figure*}
\begin{figure*}
\resizebox{0.90\hsize}{!}{\includegraphics{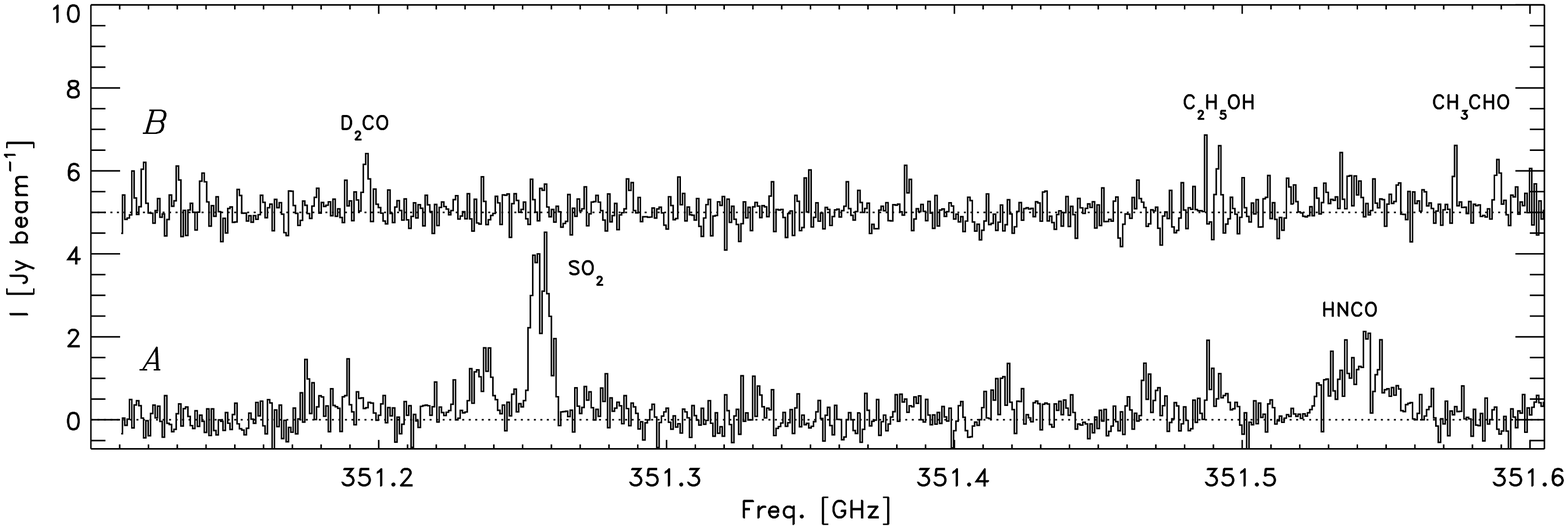}}
\resizebox{0.90\hsize}{!}{\includegraphics{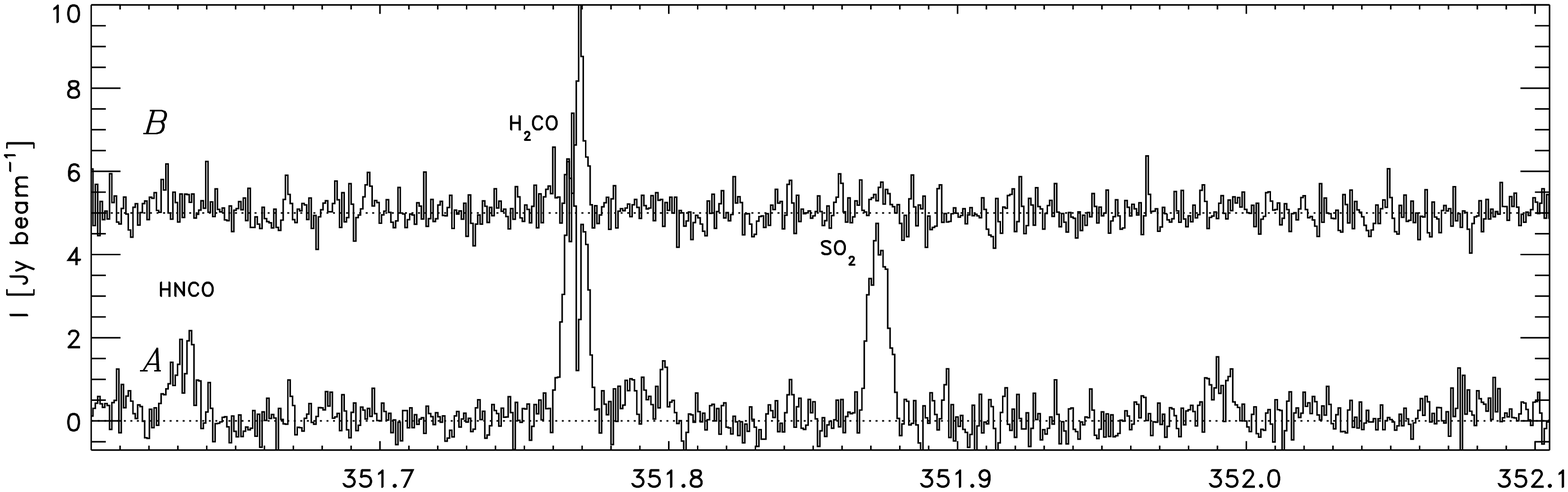}}
\resizebox{0.90\hsize}{!}{\includegraphics{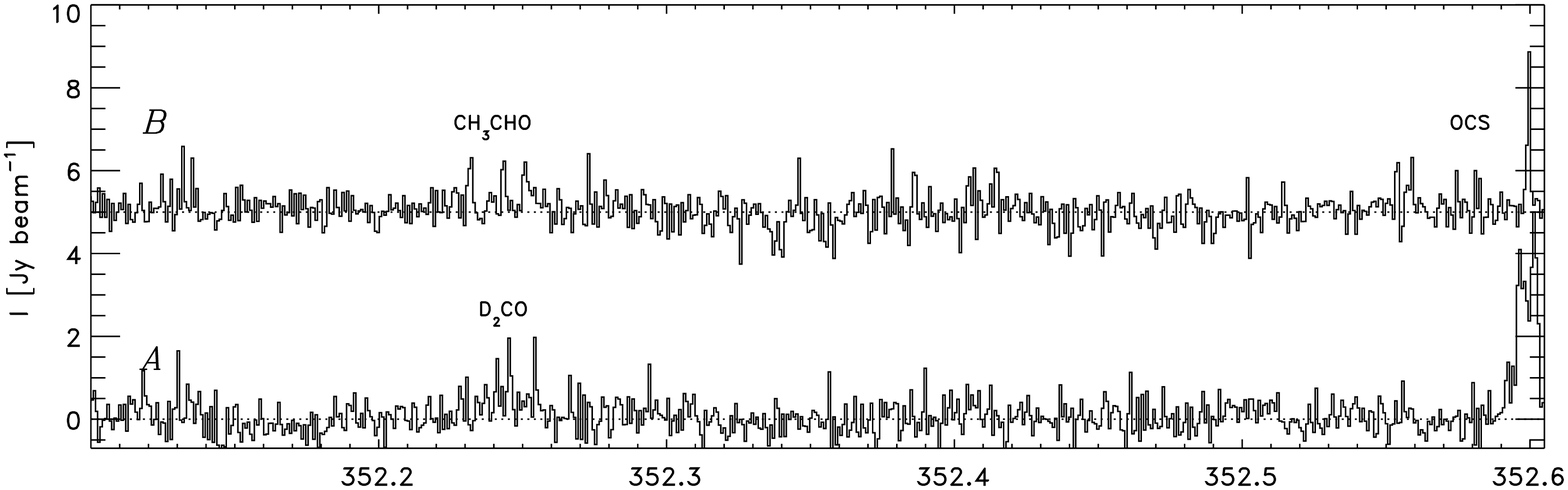}}
\resizebox{0.90\hsize}{!}{\includegraphics{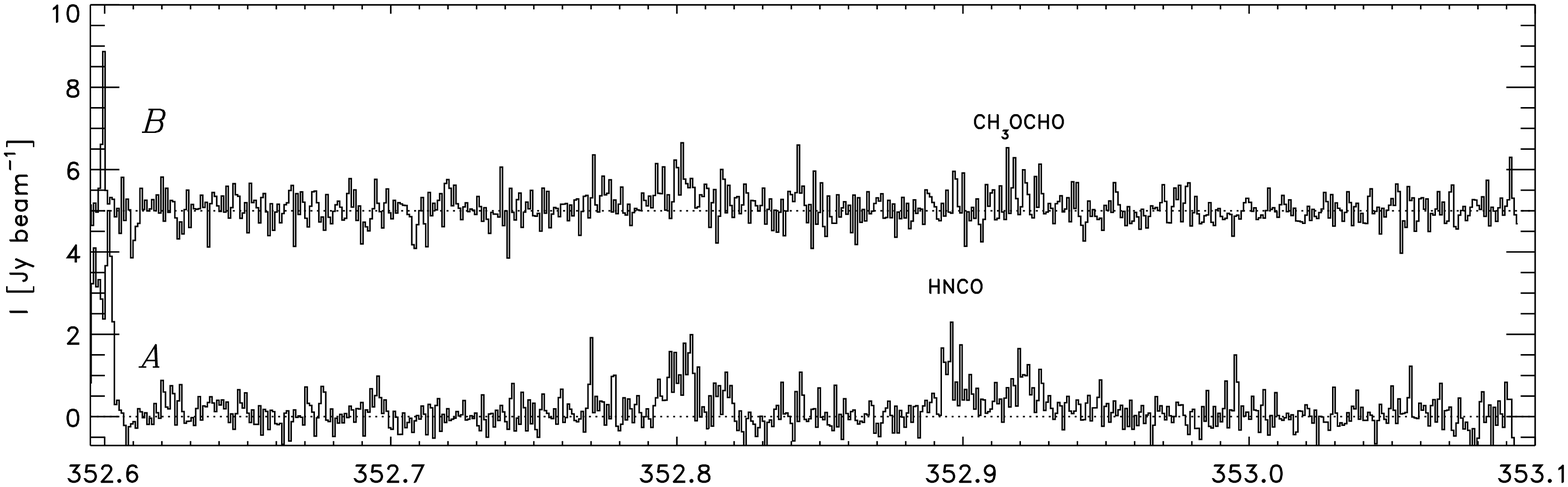}}
\caption{As in Fig.~\ref{spectrum_first} for the range 351.1--353.1
  GHz (beam size 3.5\arcsec$\times$1.9\arcsec; line 4 of Table~\ref{obslog}). }\label{spectrum_last}
\end{figure*}

Lines present in the spectra toward each of the two continuum peak
positions were identified using the CDMS \citep{cdms1,cdms2} and JPL
\citep{jpl} spectroscopic catalogs: the catalog entries were
downloaded for the molecular species expected toward the two
protostars, and compared to the observed spectra. Line identifications
were made by eye taking into account the excitation energy levels and
intrinsic line strengths when evaluating the detection of a given
transition. This process was done iteratively: that is, for detected
species lower excited or stronger lines unassigned in the first round
were searched for, e.g., looking at possible line blends with stronger
lines for other species. Unassigned lines were checked against both
catalogs and if a sufficient number of lines of a given molecule was
detected, it was included in the search list.

In this manner a total of 515 transitions (996 including tabulated
hyperfine components) from 54 molecular species (including
isotopologues) were identified. About 10\% of the features in the
spectra remain unassigned. Table~\ref{detectmols} summarizes the
detected molecules, the number of lines and the energy ranges for the
detected species -- while Table~\ref{detectlines} compiles all the
identified lines.
\begin{table*}
\caption{Detected molecules. \label{detectmols}}
\begin{tabular}{lcc@{\hspace{0.2in}}|@{\hspace{0.2in}}lcc@{\hspace{0.2in}}|@{\hspace{0.2in}}lcc} \hline\hline
{Molecules} & {$n_{\rm lines}$} & {$E_u$} {[K]} & {Molecules} & {$n_{\rm lines}$} &
{$E_u$} {[K]} & {Molecules} & {$n_{\rm lines}$} & {$E_u$} {[K]} \\ \hline
\multicolumn{9}{c}{CO, HCO$^+$ and N-bearing species}\\\hline
CO                        &   1 &    16.6 &
$^{13}$CO                 &   1 &    15.9 &
C$^{18}$O                 &   1 &    15.8 \\

C$^{17}$O                 &   1 &   32.4 &
H$^{13}$CO$^{+}$          &   1 &   41.6 &
HC$_{3}$N                 &   1 &   141.9 \\

H$_2$CN                   &   4 &   21.1--175.4 &
CH$_{3}$CN                &  20 &   68.9--1291.5 &
CH$_{3}$$^{13}$CN         &   8 &   68.8--646.6 \\

C$_{2}$H$_{5}$CN          &  34 &   79.1--681.3 &
HNCO                      &  12 &   58.0--835.7 &
HNC$^{18}$O               &   1 &  195.5 \\\hline

\multicolumn{9}{c}{O-bearing species}\\\hline
H$_{2}$CO                 &   3 &   33.4--174.0 &
H$_{2}$$^{13}$CO          &   1 &   32.9 &
H$_{2}$C$^{18}$O          &   5 &   97.4--239.6 \\

CH$_{3}$OH                &  62 &   39.8--802.2 &
$^{13}$CH$_{3}$OH         &   5 &   87.1--302.5 &
C$_{2}$H$_{5}$OH          &  37 &   74.4--1289.1 \\

HCOOH                     &   1 &  332.7 &
CH$_{2}$CO                &   1 &   76.5 &
CH$_{3}$CHO               & 101 &   22.9--864.6 \\

CH$_{3}$OCH$_{3}$         &  26 &   48.0--952.8 &
CH$_{3}$OCHO-A            &  35 &   81.4--604.3 &
CH$_{3}$OCHO-E            &  43 &   56.9--629.8 \\

(CH$_3$)$_2$CO            &  28 &   29.9--700.8 &
& & &
& &\\\hline

\multicolumn{9}{c}{S-bearing species}\\\hline
CS                        &   1 &   65.8 &
C$^{34}$S                 &   1 &   64.8 &
$^{13}$CS                 &   1 &   33.3 \\

H$_{2}$S                  &   1 &   84.0 &
H$_{2}$CS                 &   2 &   102.4--105.2 &
H$_{2}$C$^{34}$S          &   1 &   141.8 \\

HCS$^{+}$                 &   1 &   73.7 &
OCS                       &   2 &   110.9--253.9 &
O$^{13}$CS                &   2 &   110.5--253.1 \\

SO                        &   1 &   35.0 &
$^{33}$SO                 &   5 &   25.4--80.5 &
$^{34}$SO                 &   2 &   25.3--34.4 \\

SO$_{2}$                  &  10 &   35.9--678.5 &
$^{33}$SO$_{2}$           &   5 &   52.4--118.5 &
$^{34}$SO$_{2}$           &   5 &   18.7--247.8 \\

SO$^{18}$O                &   3 &   49.0--522.1 &
SO$^{17}$O                &   2 &   102.8--103.3 &
& & \\\hline

\multicolumn{9}{c}{SiO and deuterated species}\\\hline
SiO                       &   2 &   31.3--75.0 &
N$_{2}$D$^{+}$            &   1 &    22.2 &
DCO$^{+}$                 &   1 &    20.7 \\

DCN                       &   1 &    20.9 &
HDCS                      &   1 &   51.5 &
D$_{2}$CS                 &   3 &   105.8--207.5 \\

DCS$^{+}$                 &   1 &   36.3 &
D$_{2}$CO                 &   6 &   32.0--285.5 &
D$_{2}$$^{13}$CO          &   1 &   57.2 \\

CH$_{2}$DCN               &   9 &    24.1--309.7 &
HCOOD                     &   2 &   34.3--386.6 &
HDO                       &   1 &  167.6 \\
\hline
\end{tabular}
\end{table*}

\begin{figure}
\resizebox{\hsize}{!}{\includegraphics{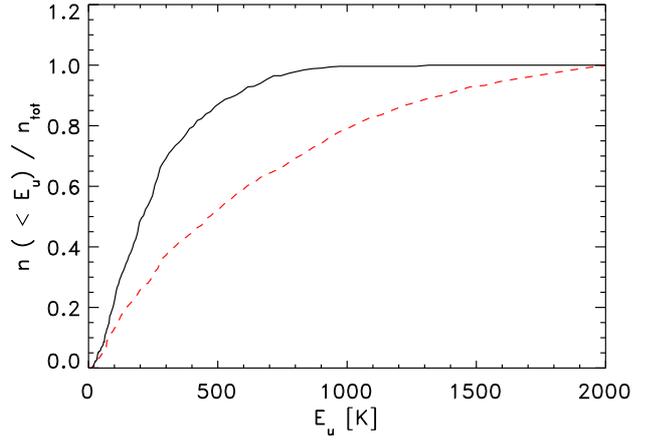}}
\caption{Cumulative distribution of the upper energy levels for the
  detected transitions (solid line) and all transitions in the search
  frequency range for the detected molecules (dashed
  line).}\label{energy_range}
\end{figure}

Fig.~\ref{energy_range} shows the cumulative distribution of the upper
energy levels, $E_u$, for the detected molecules -- compared to the
full catalog entries for the same species over the measured frequency
ranges. The figure clearly illustrates the upper limit for the
temperatures of the detected species -- with approximately 90\% of
  the detected molecular lines having $E_u$ lower than 550~K, compared
  to 50\% of the catalog entries having $E_u$ lower than this
  temperature.

\begin{figure}
\resizebox{\hsize}{!}{\includegraphics{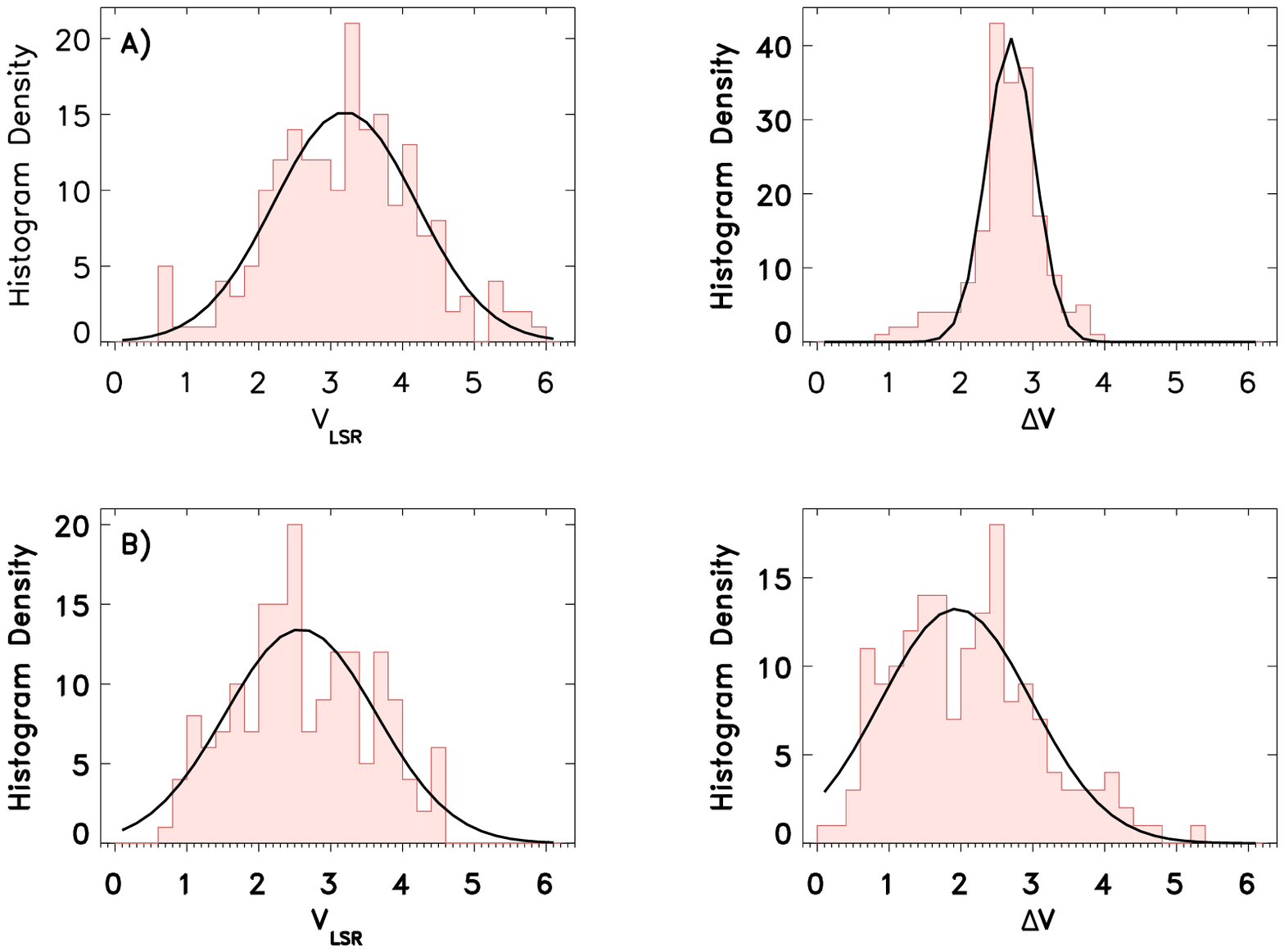}}
\caption{Distribution of centroid velocities and widths (dispersions)
  in the left and right panels, respectively for emission lines
  detected toward the A source (upper panels) and B source (lower
  panels). Gaussian fits to the distributions are over-plotted
  ($V_{\rm LSR}$ of 3.2 and 2.7~\kms\ and $\Delta V$ of 2.6 and 1.9
  for A and B respectively). The $V_{\rm LSR}$ and $\Delta V$
  distributions for the two sources are significantly different
  according to a standard T test for distributions with unequal
  variances. }\label{vel_dist}
\end{figure}

For each detected line, the first and second moments (centroid
velocity and dispersion) were estimated toward the positions of the
two components by numerical integrations of the spectra over velocity
intervals of $\pm$5~\kms\ around the expected systemic velocity of
each line at $\approx$~3~\kms.  Fig.~\ref{vel_dist} summarizes the
distribution of these velocities. As expected from the above comments
and previous findings in literature, the lines toward the IRAS~16293A
continuum peak are on average wider and slightly more red-shifted than
toward the IRAS~16293B peak (the peaks of the distributions in
Fig.~\ref{vel_dist} shifted toward higher velocities). However, as
illustrated by the figure, the distributions of both centroid
velocities and dispersions are broader toward IRAS~16293B than
IRAS~16293A. This suggests that the emission toward IRAS~16293B can be
divided into a set of lines localized for this source with another set
of lines probing the more general environment of both sources in
common. It is possible that uncertainties in the determined line rest
frequencies contribute to some of the scatter in this plot. However,
since a large number of the lines are detected in both sources, it is
unlikely that these uncertainties introduce the systematic shift and
broadening of the lines between the two sources.

\begin{figure*}
\begin{minipage}[!h]{0.15\linewidth}\phantom{xxx}\end{minipage}
\hspace{0.5cm}
\begin{minipage}[!h]{0.85\linewidth}
\resizebox{0.245\hsize}{!}{\includegraphics{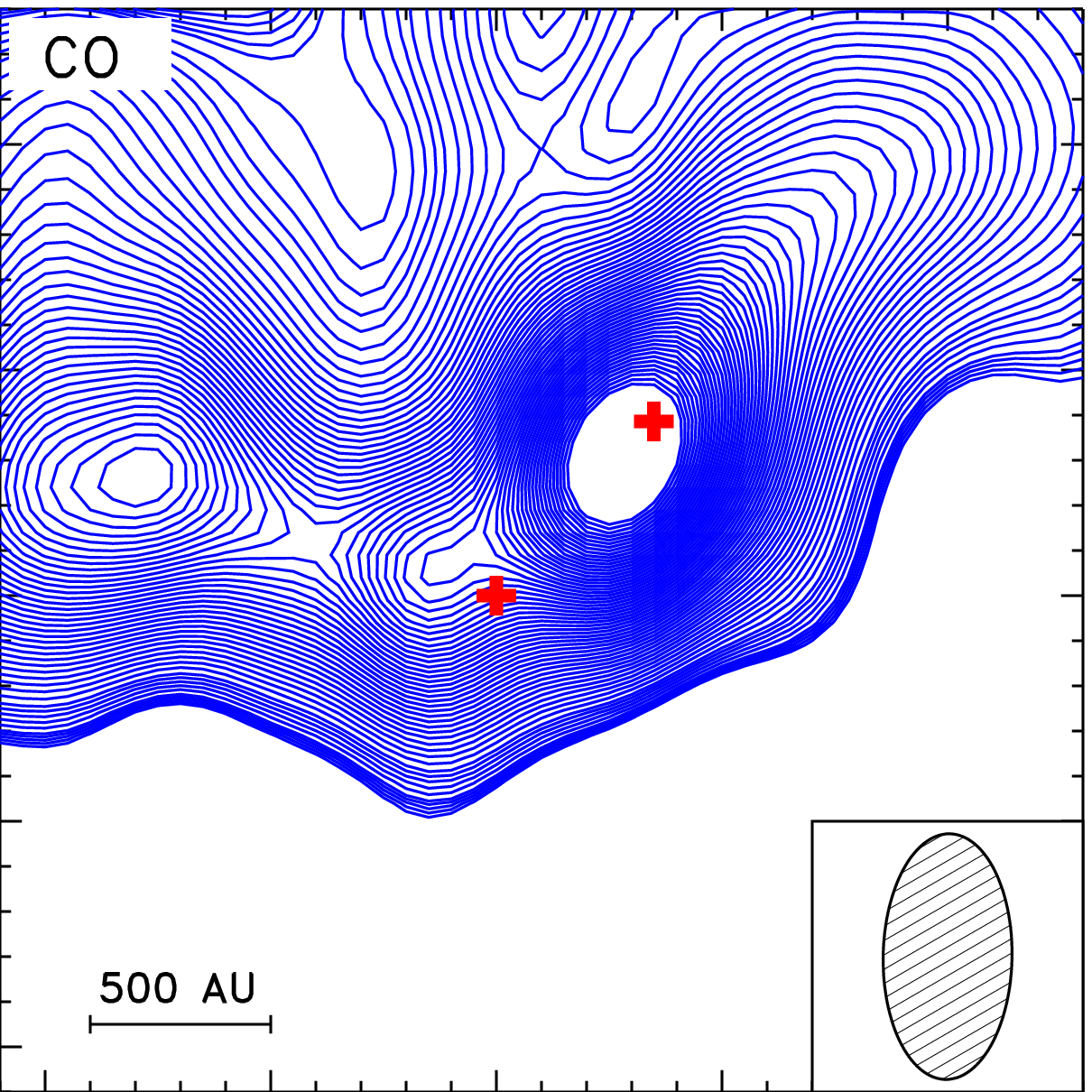}}
\resizebox{0.245\hsize}{!}{\includegraphics{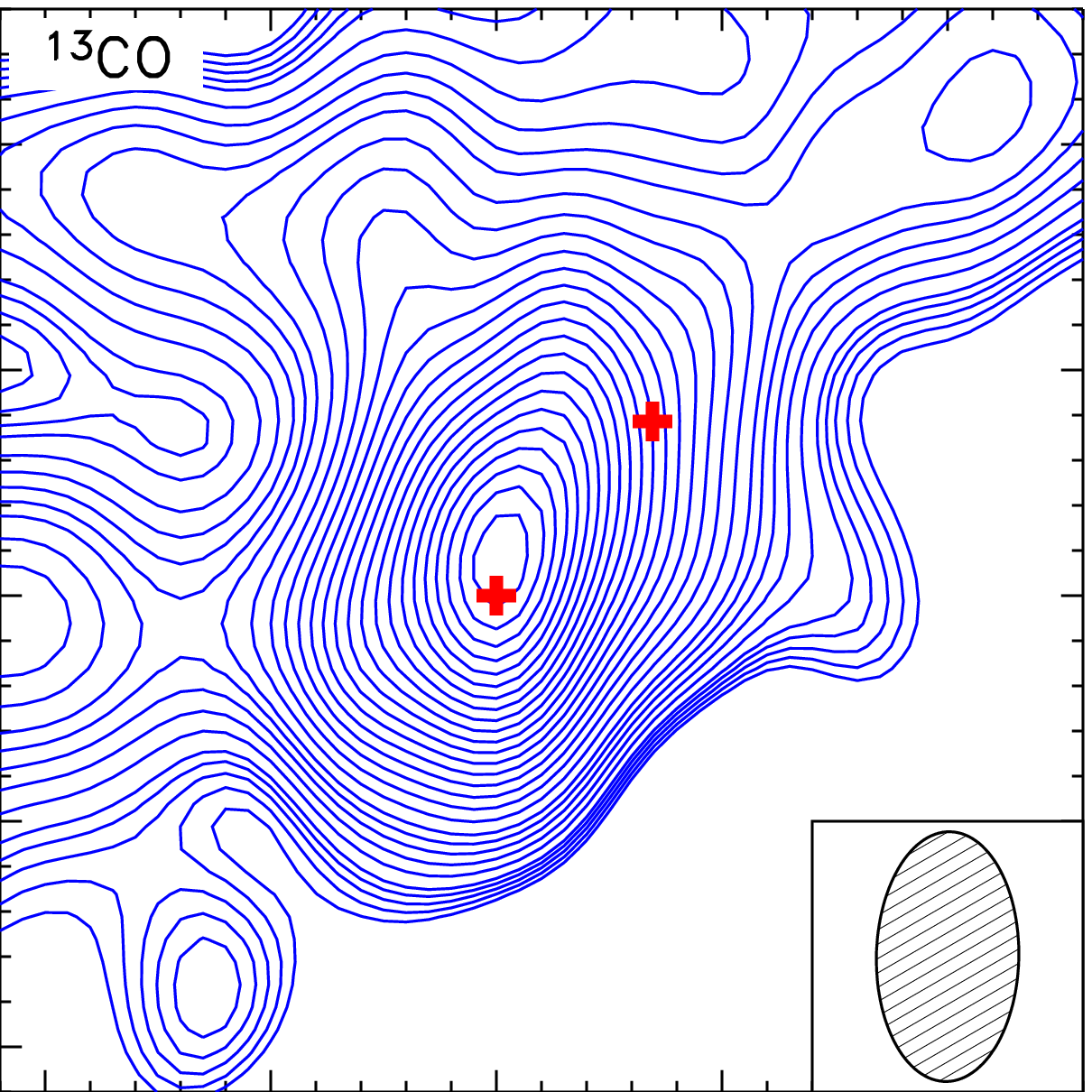}}
\resizebox{0.245\hsize}{!}{\includegraphics{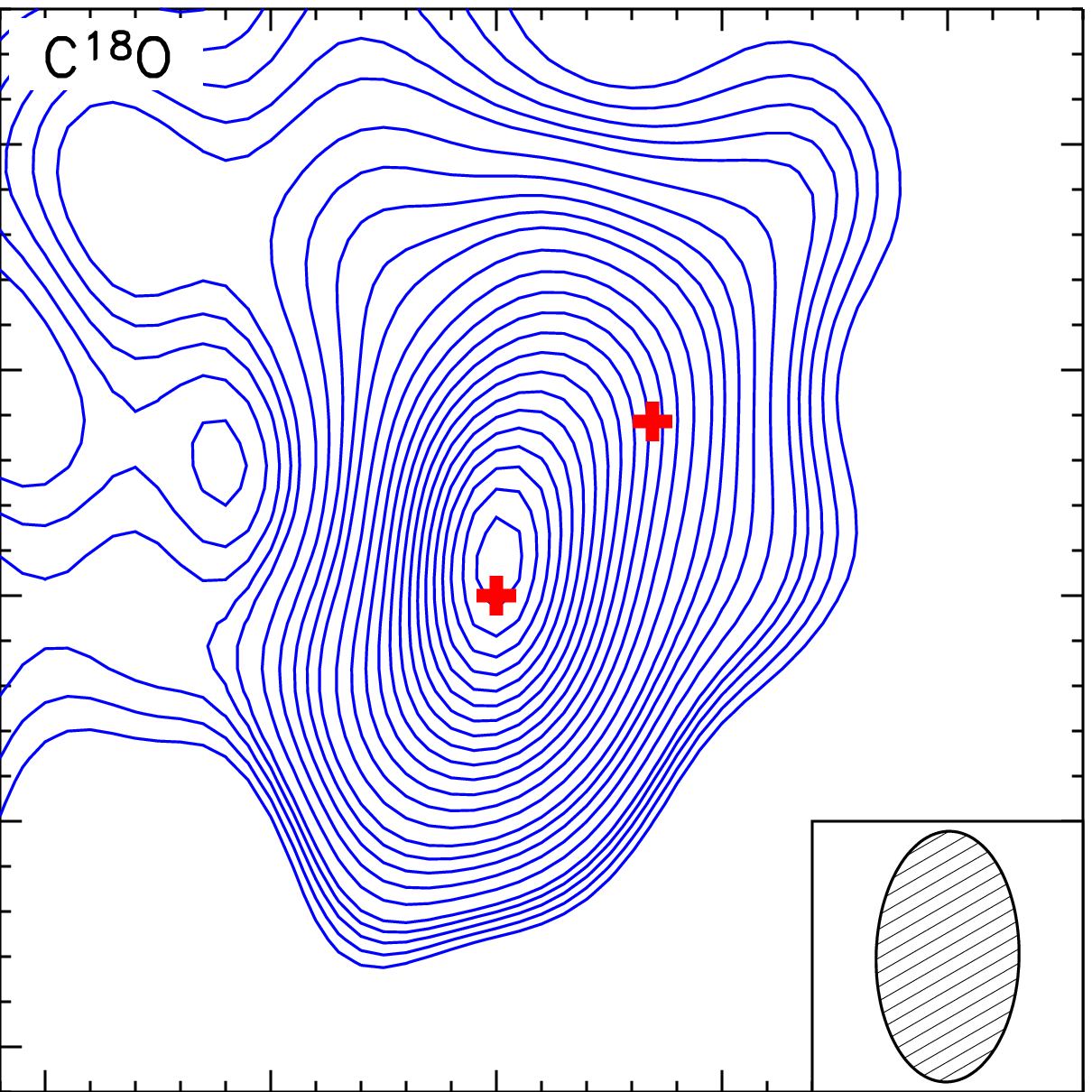}}
\resizebox{0.245\hsize}{!}{\includegraphics{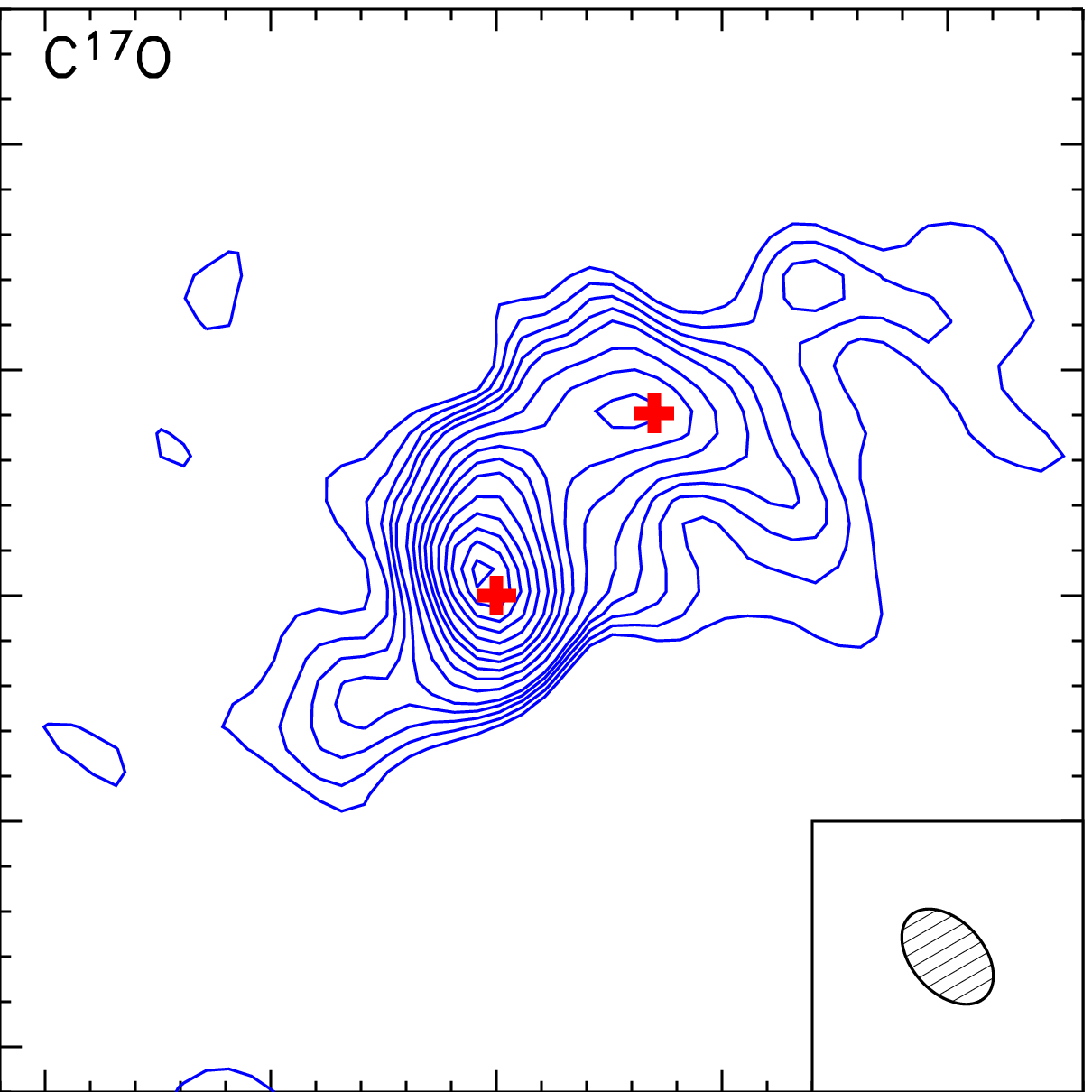}}
\resizebox{0.245\hsize}{!}{\includegraphics{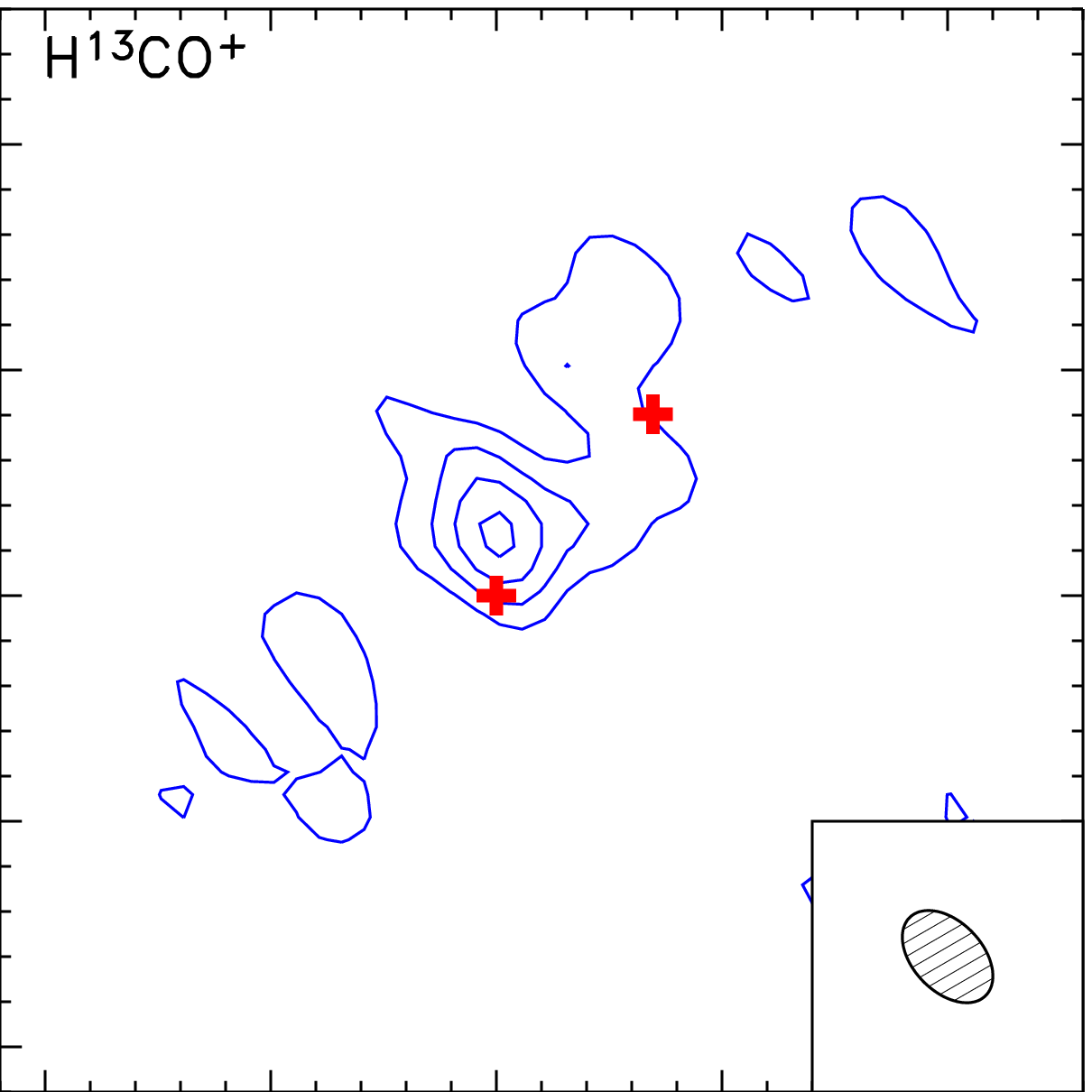}}
\resizebox{0.245\hsize}{!}{\includegraphics{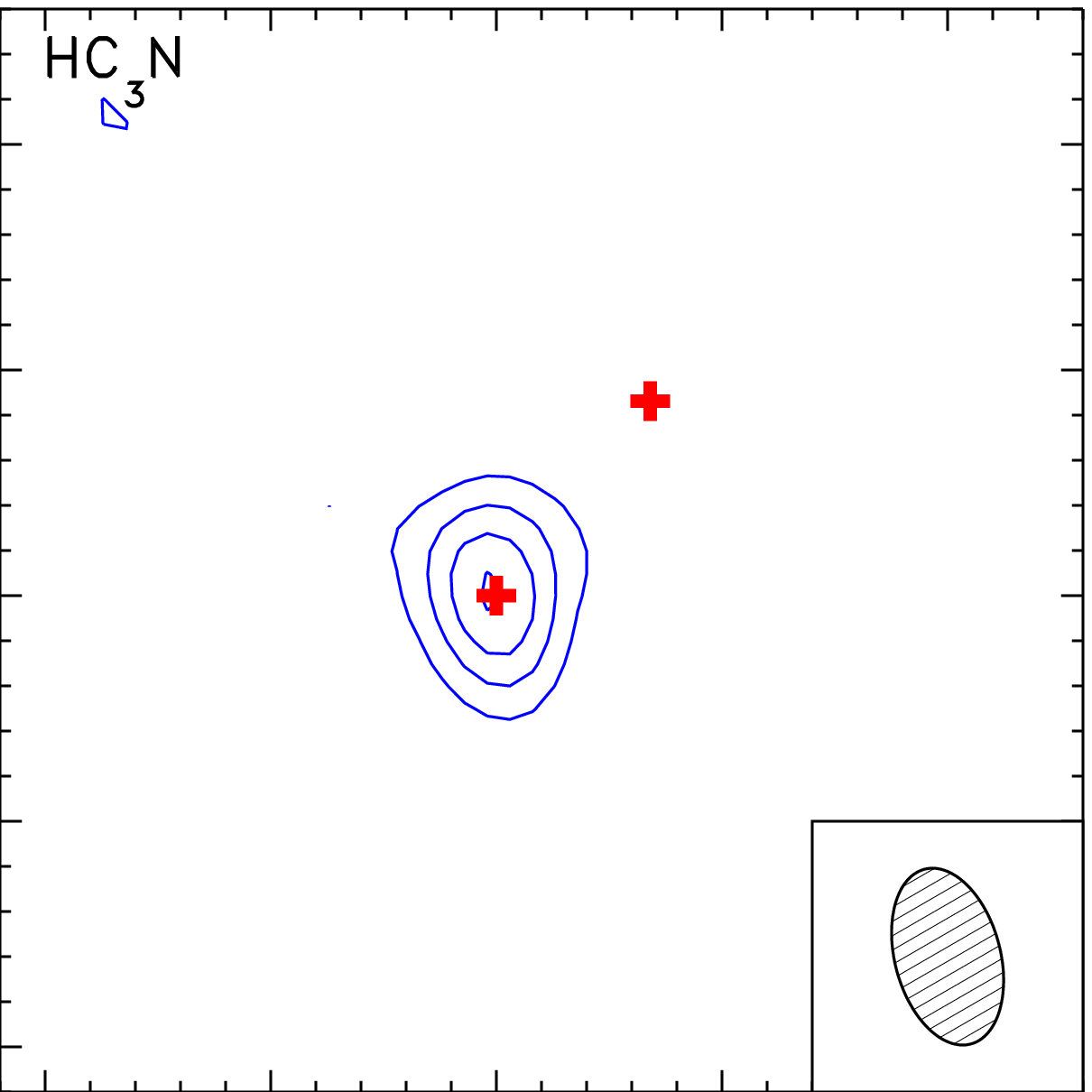}}
\resizebox{0.245\hsize}{!}{\includegraphics{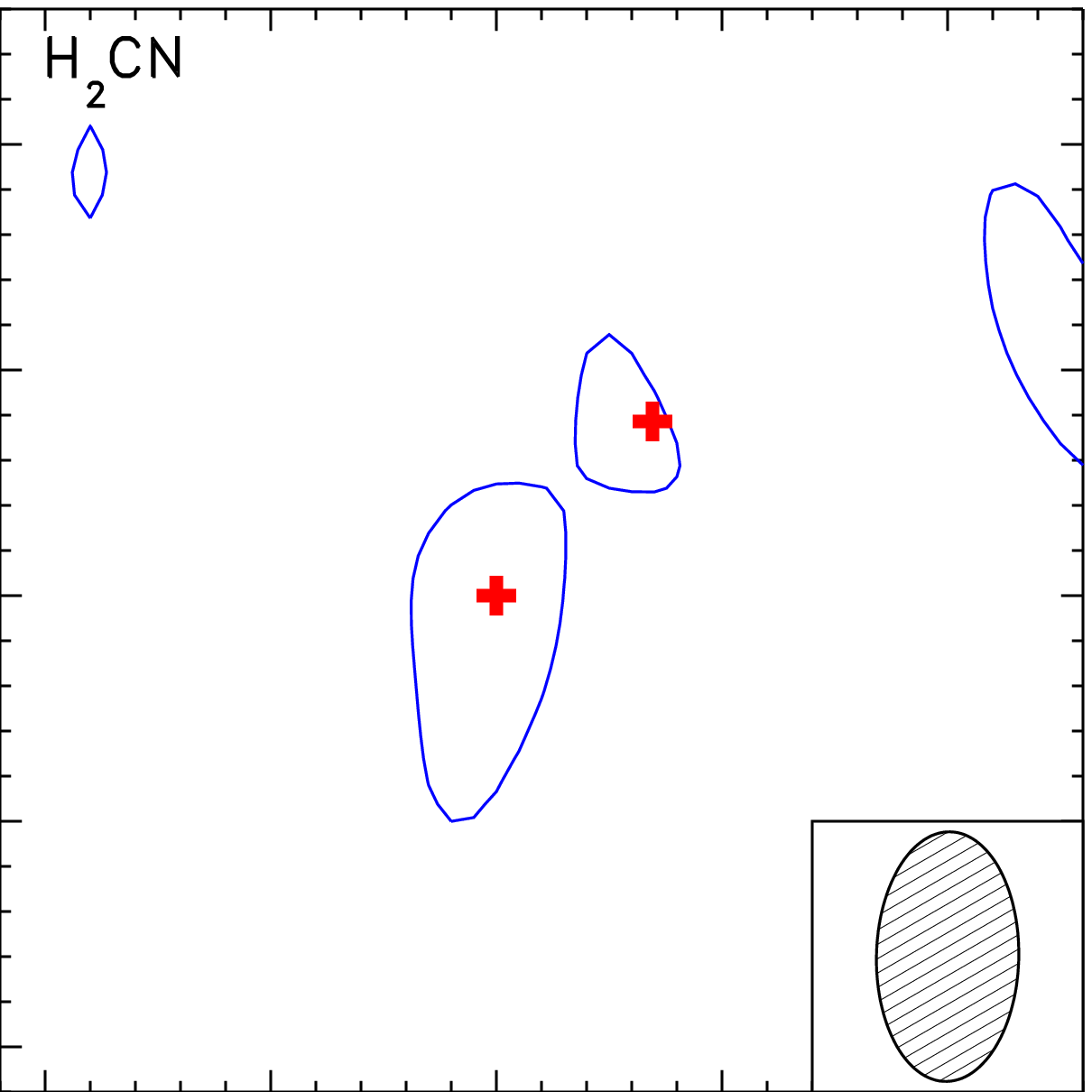}}
\resizebox{0.245\hsize}{!}{\includegraphics{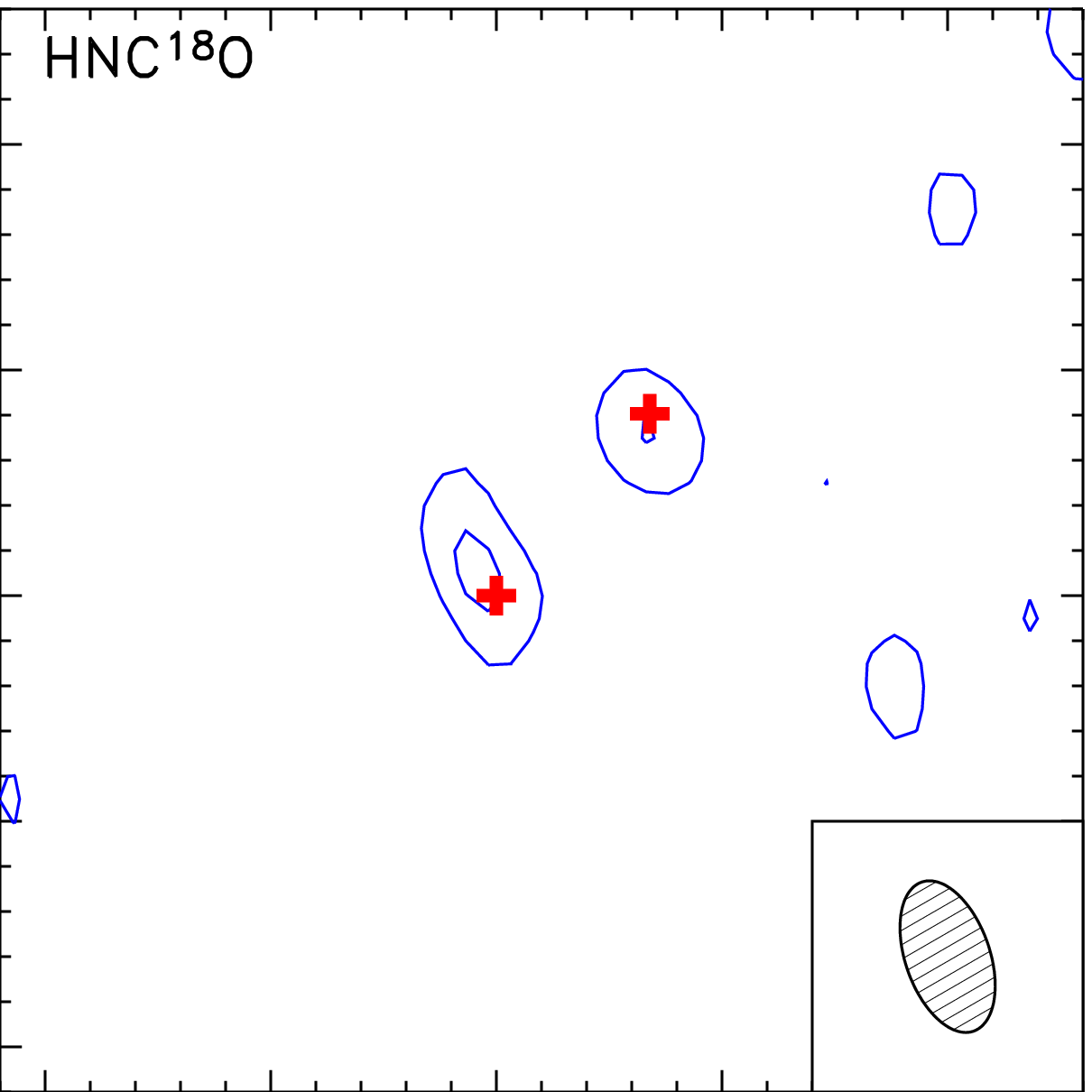}}
\resizebox{0.245\hsize}{!}{\includegraphics{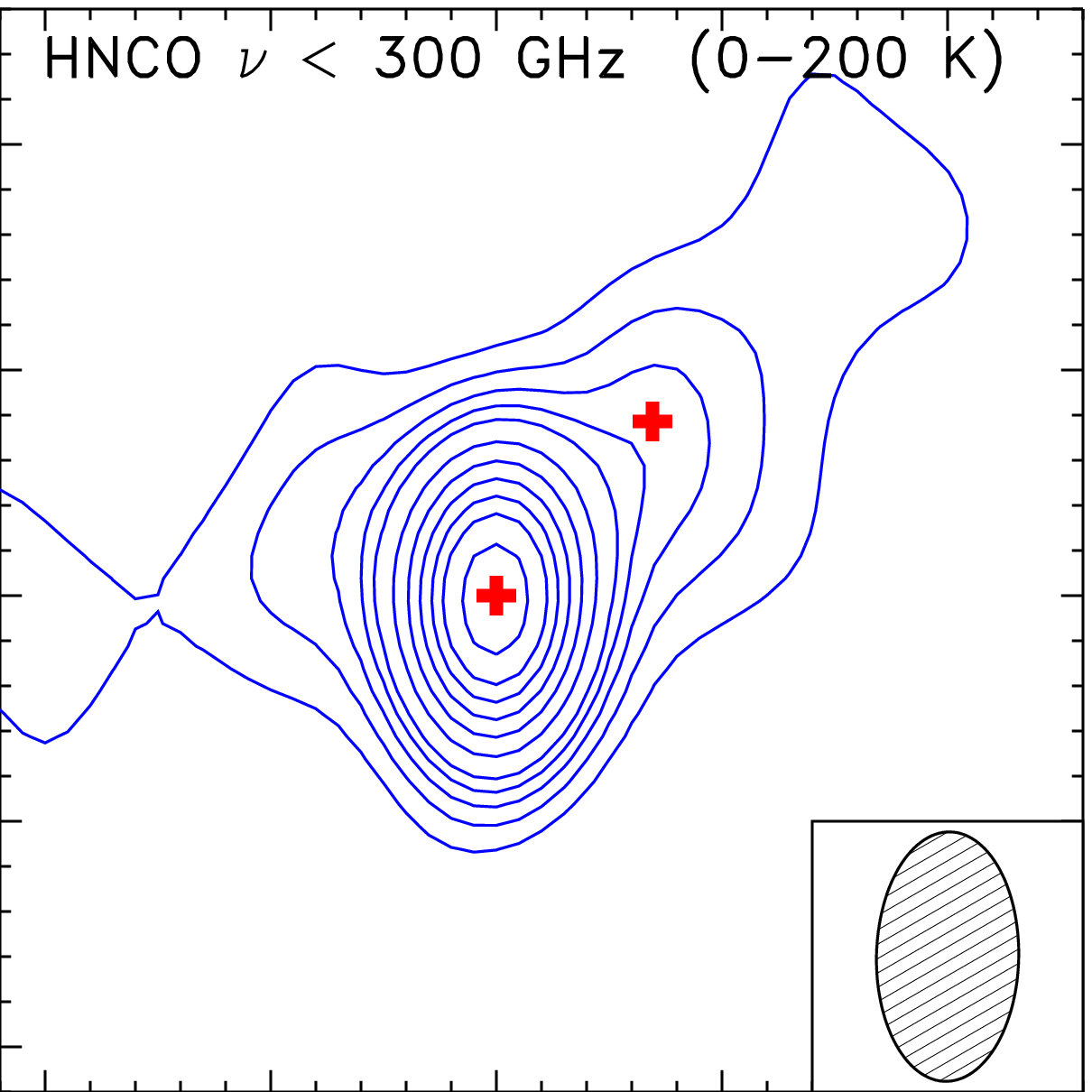}}
\resizebox{0.245\hsize}{!}{\includegraphics{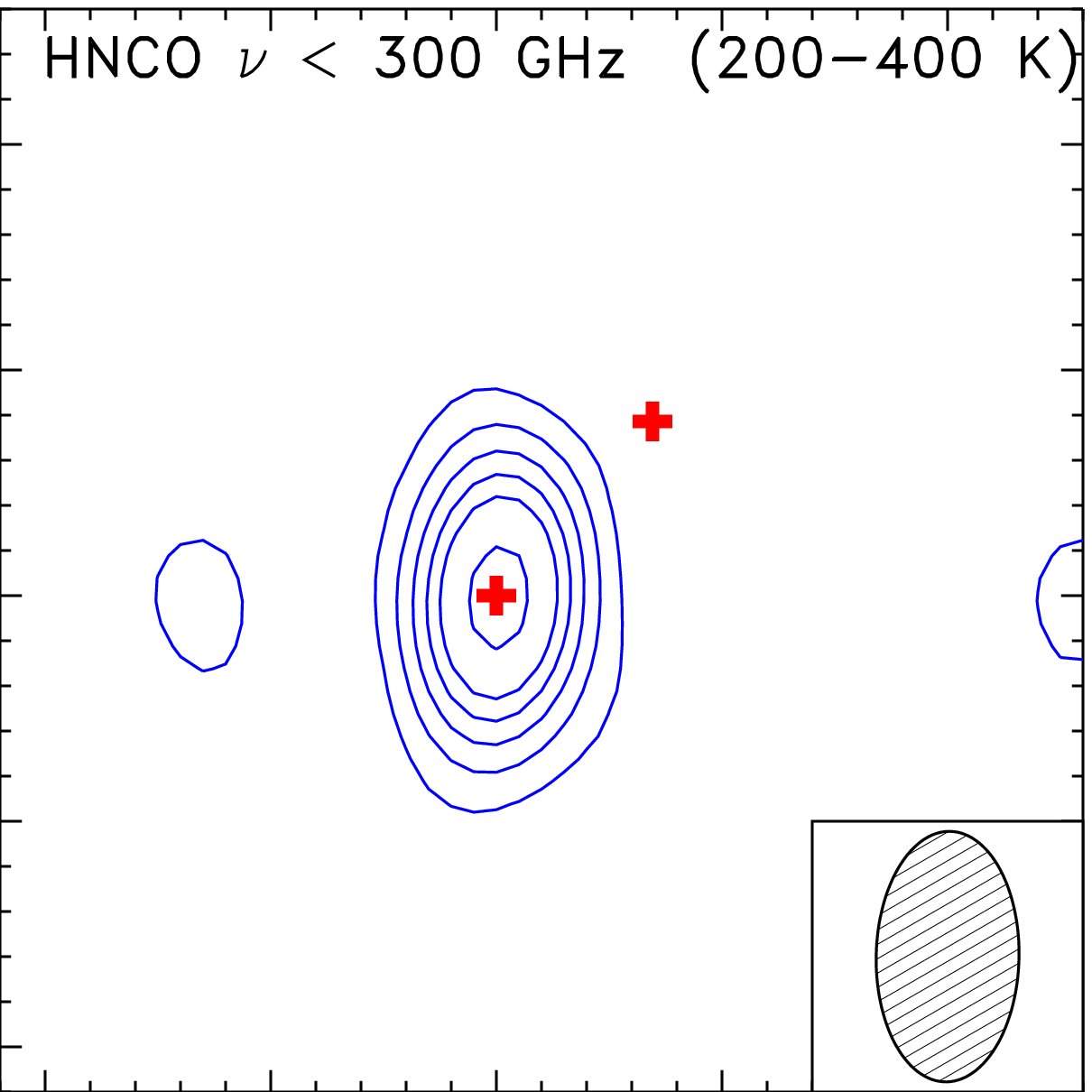}}
\resizebox{0.245\hsize}{!}{\includegraphics{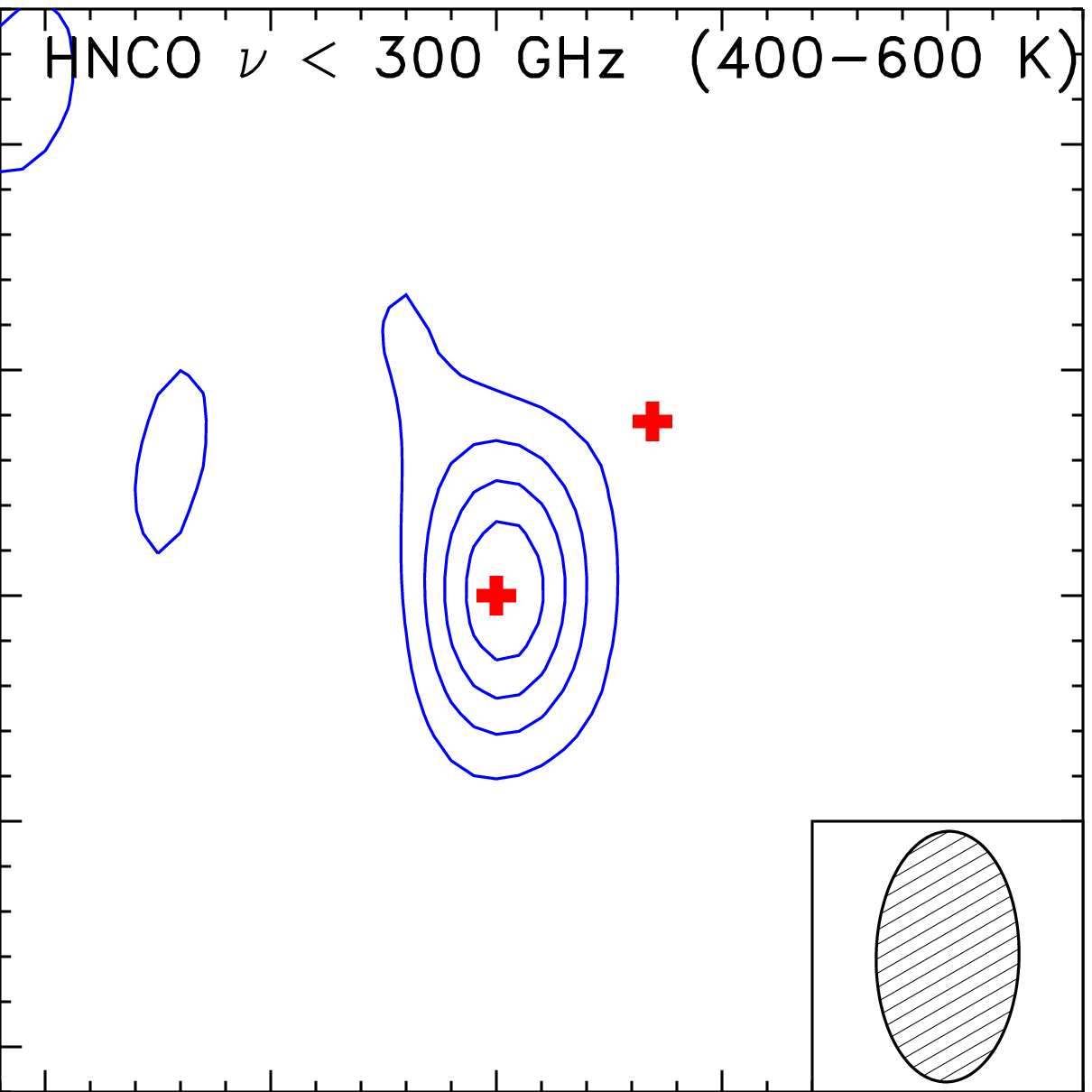}}
\resizebox{0.245\hsize}{!}{\includegraphics{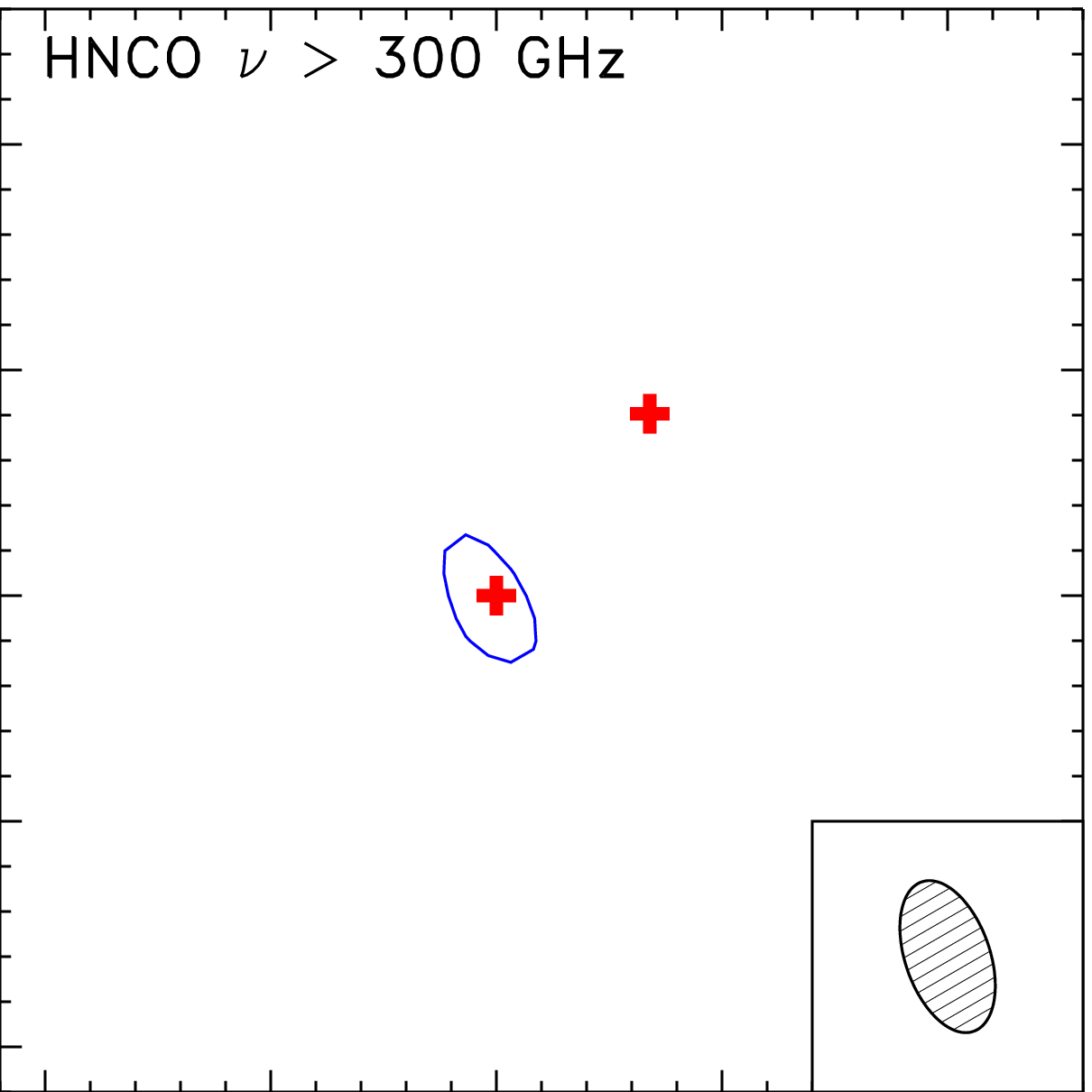}}
\resizebox{0.245\hsize}{!}{\includegraphics{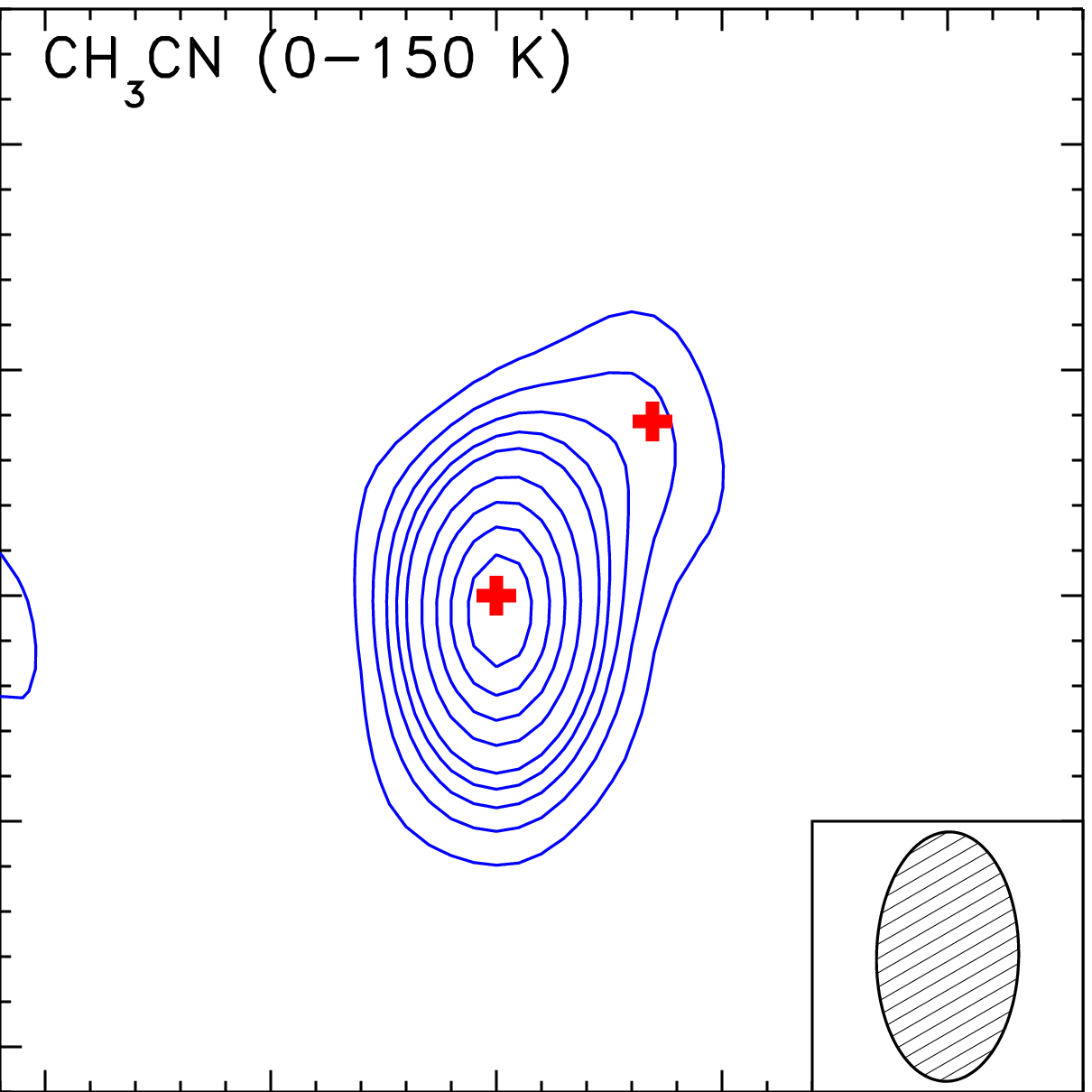}}
\resizebox{0.245\hsize}{!}{\includegraphics{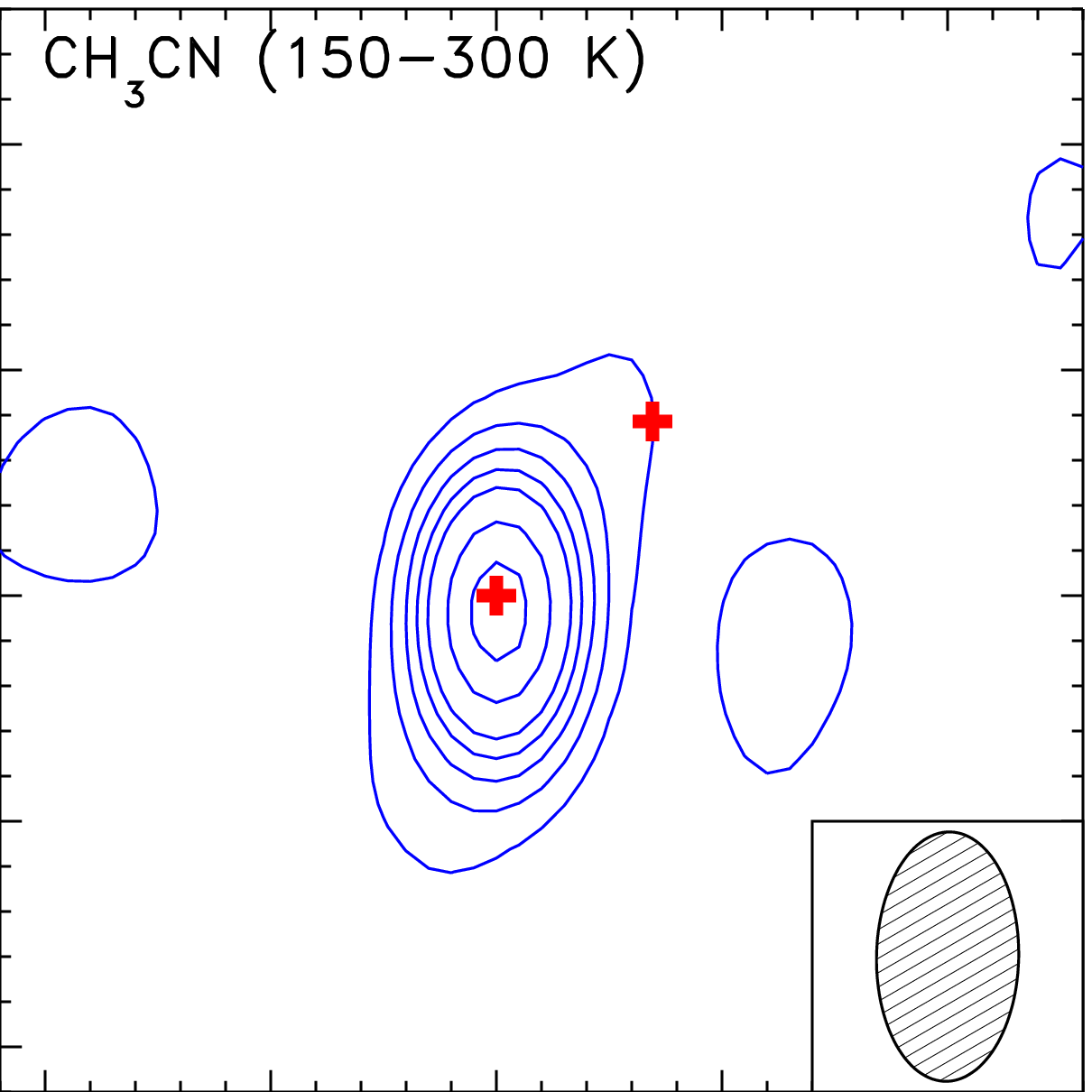}}
\resizebox{0.245\hsize}{!}{\includegraphics{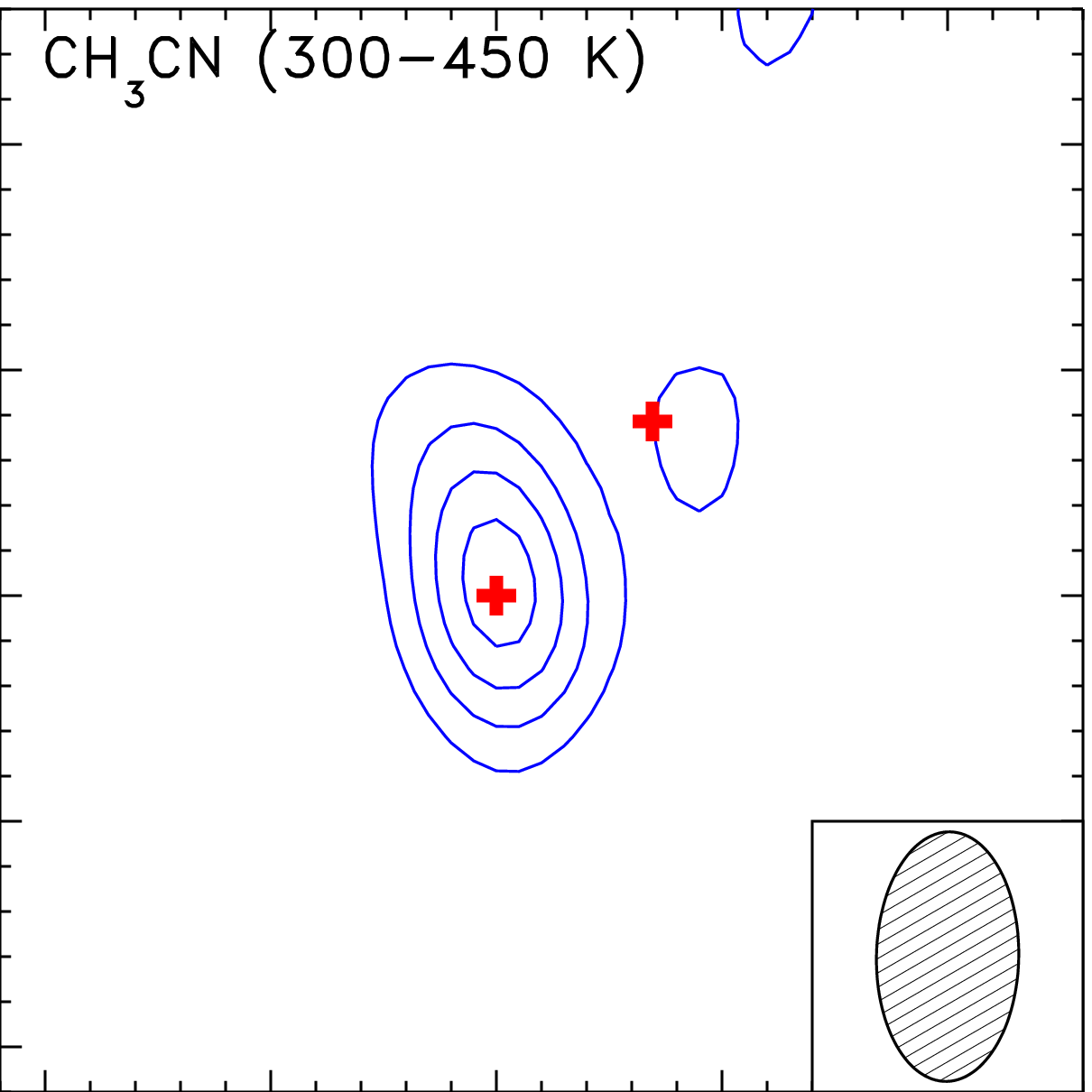}}
\resizebox{0.245\hsize}{!}{\includegraphics{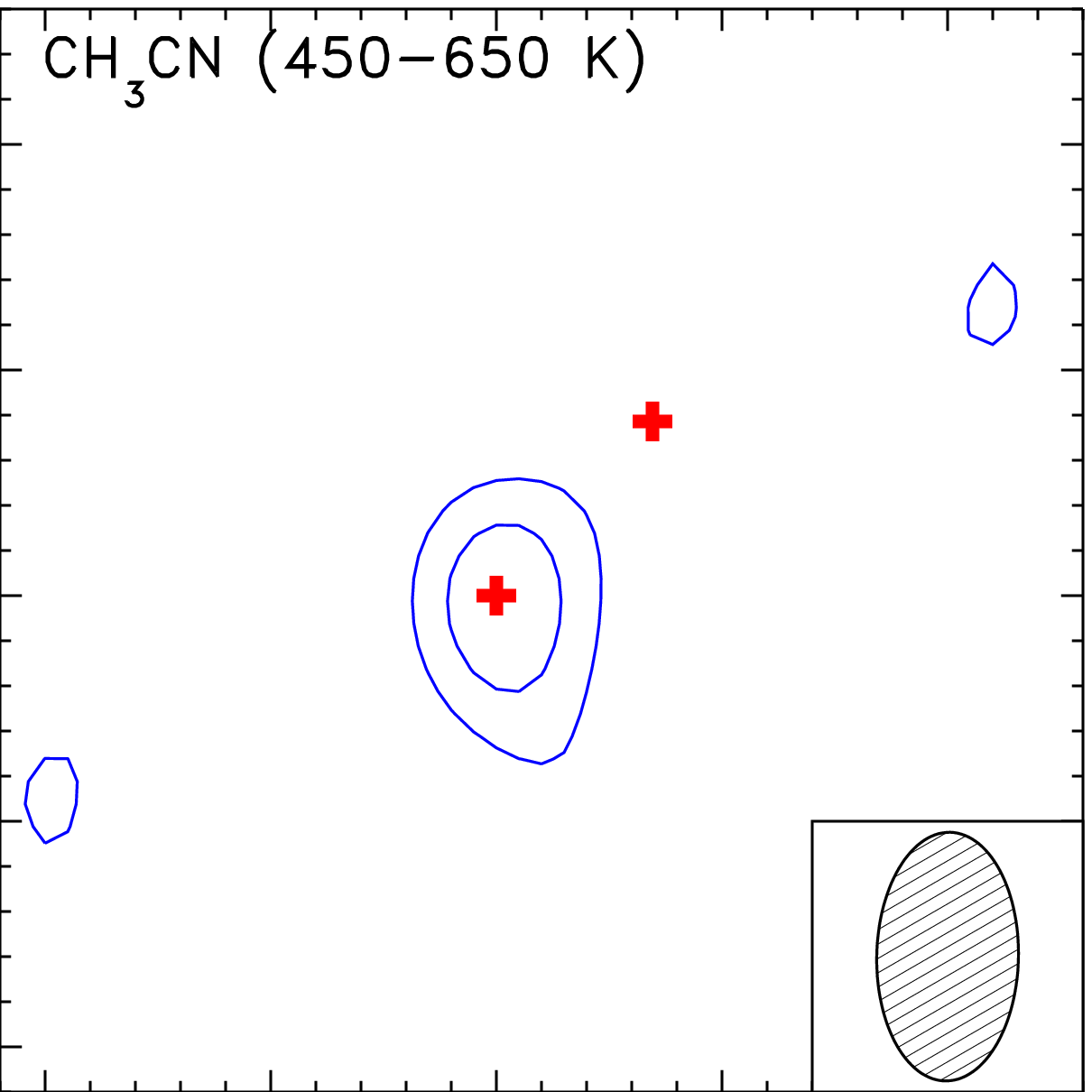}}
\resizebox{0.245\hsize}{!}{\includegraphics{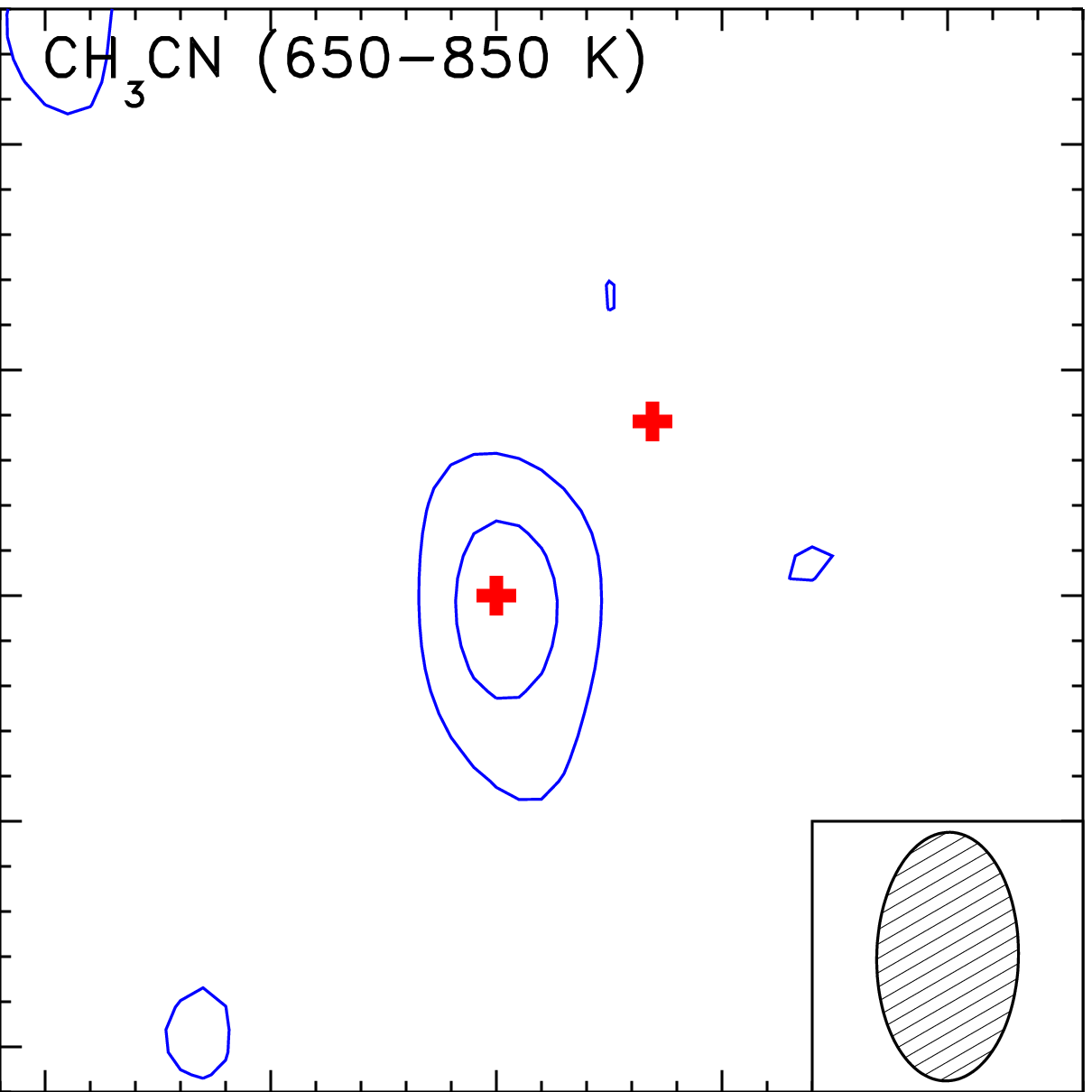}}
\resizebox{0.245\hsize}{!}{\includegraphics{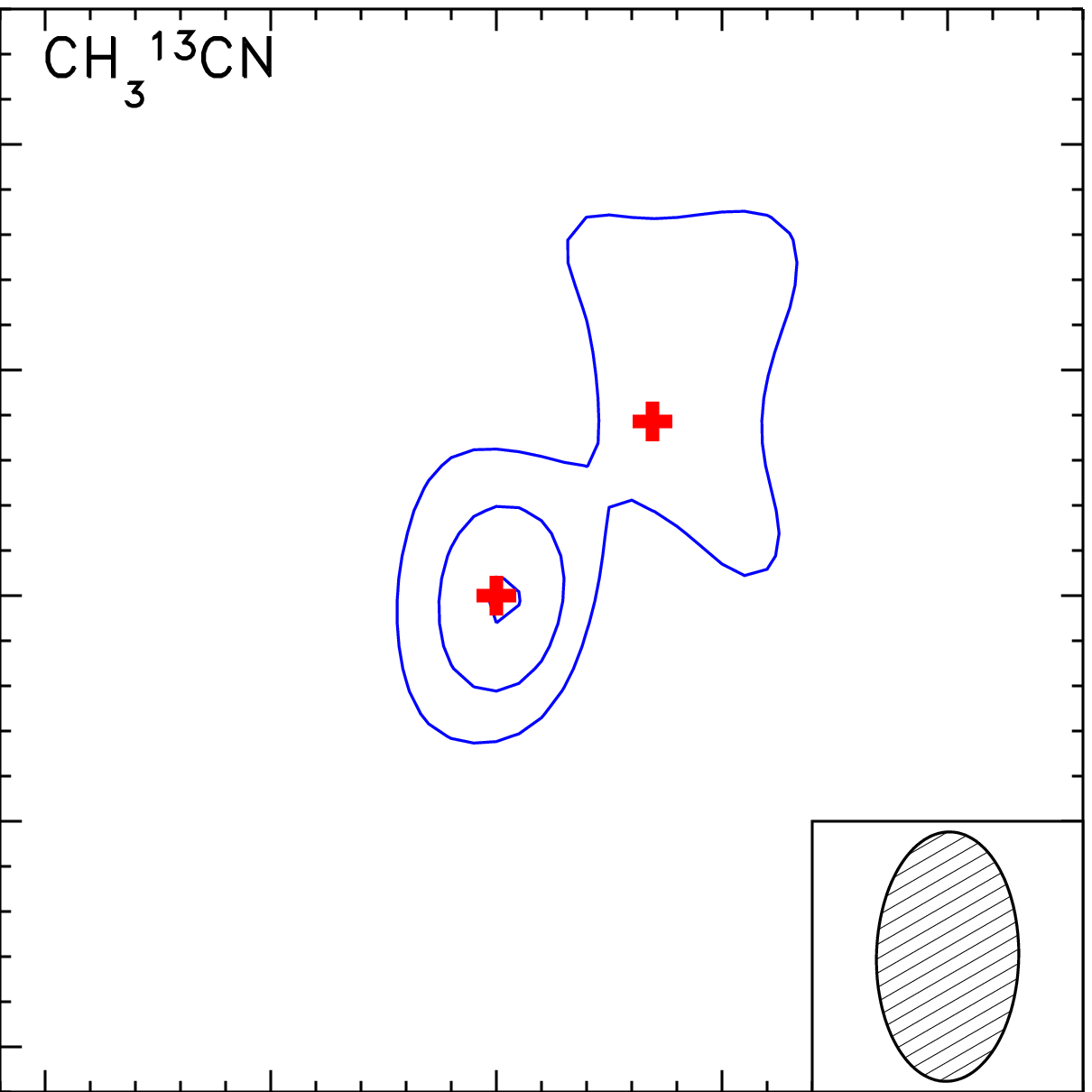}}
\resizebox{0.245\hsize}{!}{\includegraphics{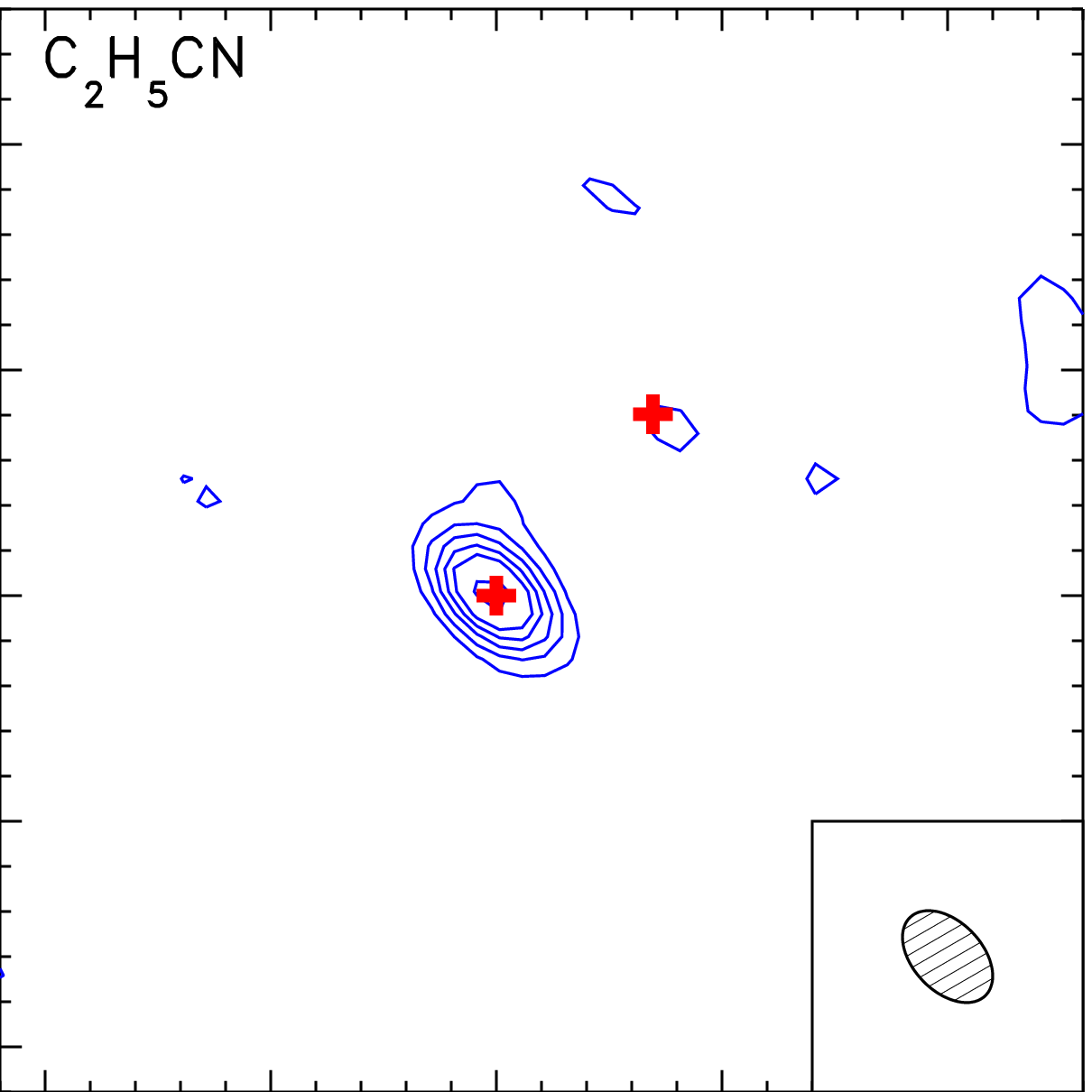}}

\phantom{xxx}
\caption{Emission for CO, HCO$^+$ and the N-bearing organic
  molecules. For each molecule, except where noted otherwise, the
  emission has been integrated over all identified lines in one
  selected frequency band - isolated by at least 10~MHz from other
  species (see Table~\ref{detectlines} for specific lines). The
  integration is performed over the width of the ``A'' component
  (i.e., from 0 to 6~\kms) and contours are shown in steps of
  3~$\sigma$ to 15$\sigma$ and in steps of 6$\sigma$ thereafter, where
  $\sigma$ is the RMS noise level for the integrated line intensity;
  the RMS per channel is given in Table~\ref{obslog}. A scale-bar is
  shown in the upper left panel. The beam size at the frequency
    of the selected transitions for the given molecule is shown in the
    lower right corner of each panel.}\label{image_first}
\end{minipage}
\end{figure*}

For each molecular species the emission from detected, isolated lines
were integrated over $V_{\rm sys} \pm \Delta V$ where $V_{\rm sys}$ is
the average centroid velocity for the IRAS~16293A component and
$\Delta V$ the average line width. Compilations of the maps from the
inner 24\arcsec$\times$24\arcsec\ are shown in
Fig.~\ref{image_first}--\ref{image_last}.  In these plots, maps from
different transitions for the same molecule have been added together
weighted by their noise. For molecules with transitions in multiple
frequency bands, data from one band only were used to keep the spatial
resolution similar (see Table~\ref{detectlines}). For molecules
with a large number of detected lines with high S/N spanning a range
of energy levels, the integrations were subdivided into integrations
over lines in different energy ranges. \clearpage

\begin{figure*}
\begin{minipage}[!h]{0.15\linewidth}\phantom{xxx}\end{minipage}
\hspace{0.5cm}
\begin{minipage}[!h]{0.85\linewidth}
\resizebox{0.245\hsize}{!}{\includegraphics{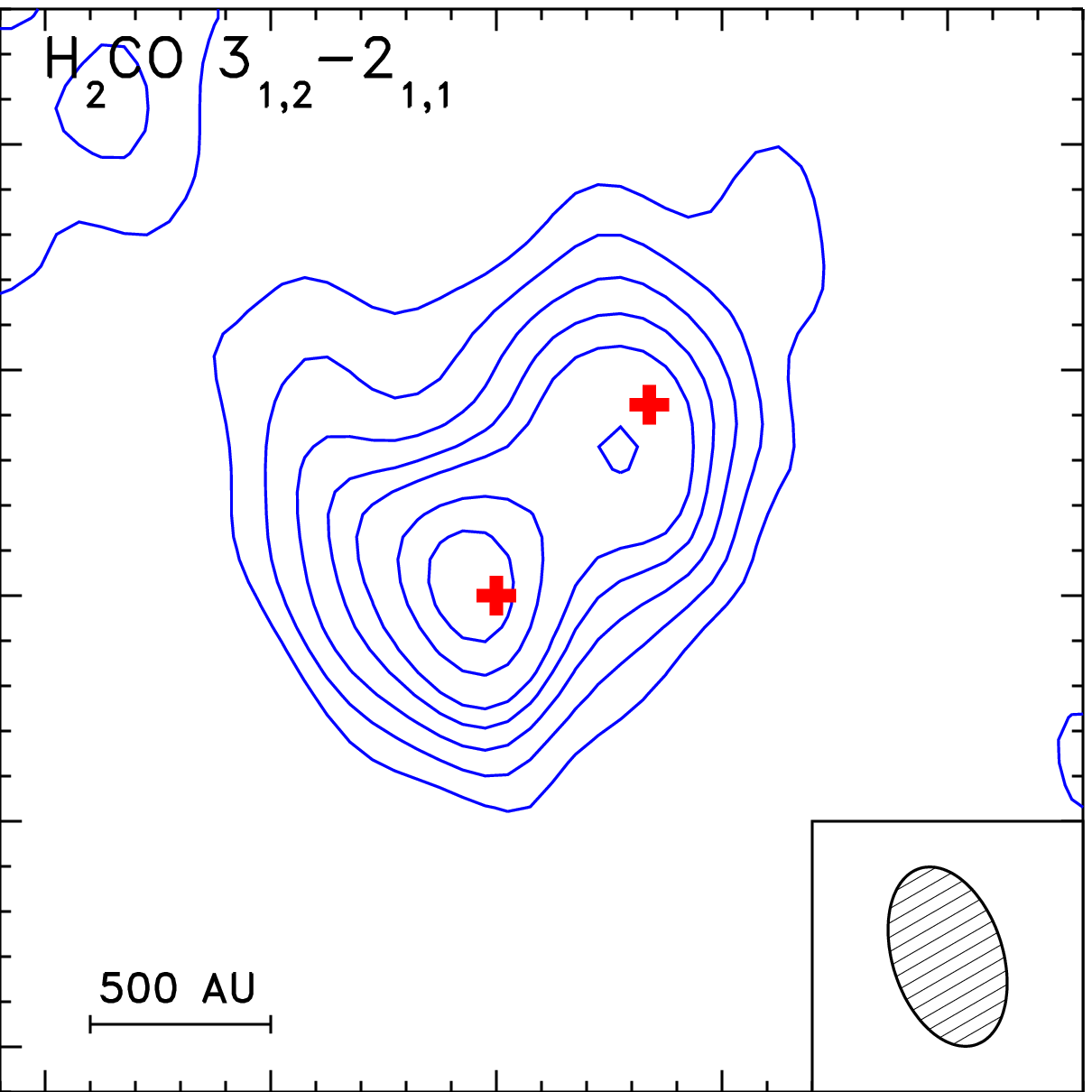}}
\resizebox{0.245\hsize}{!}{\includegraphics{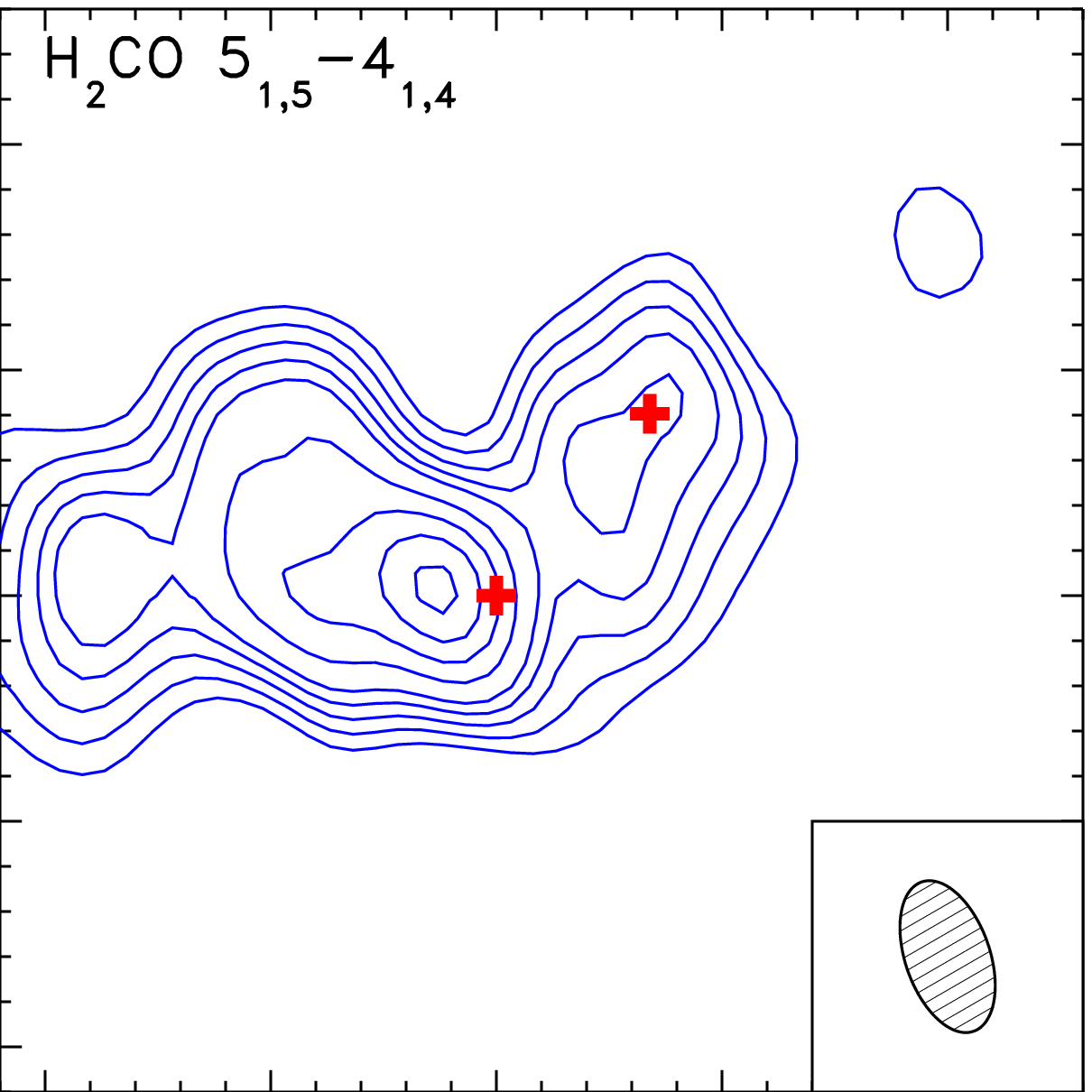}}
\resizebox{0.245\hsize}{!}{\includegraphics{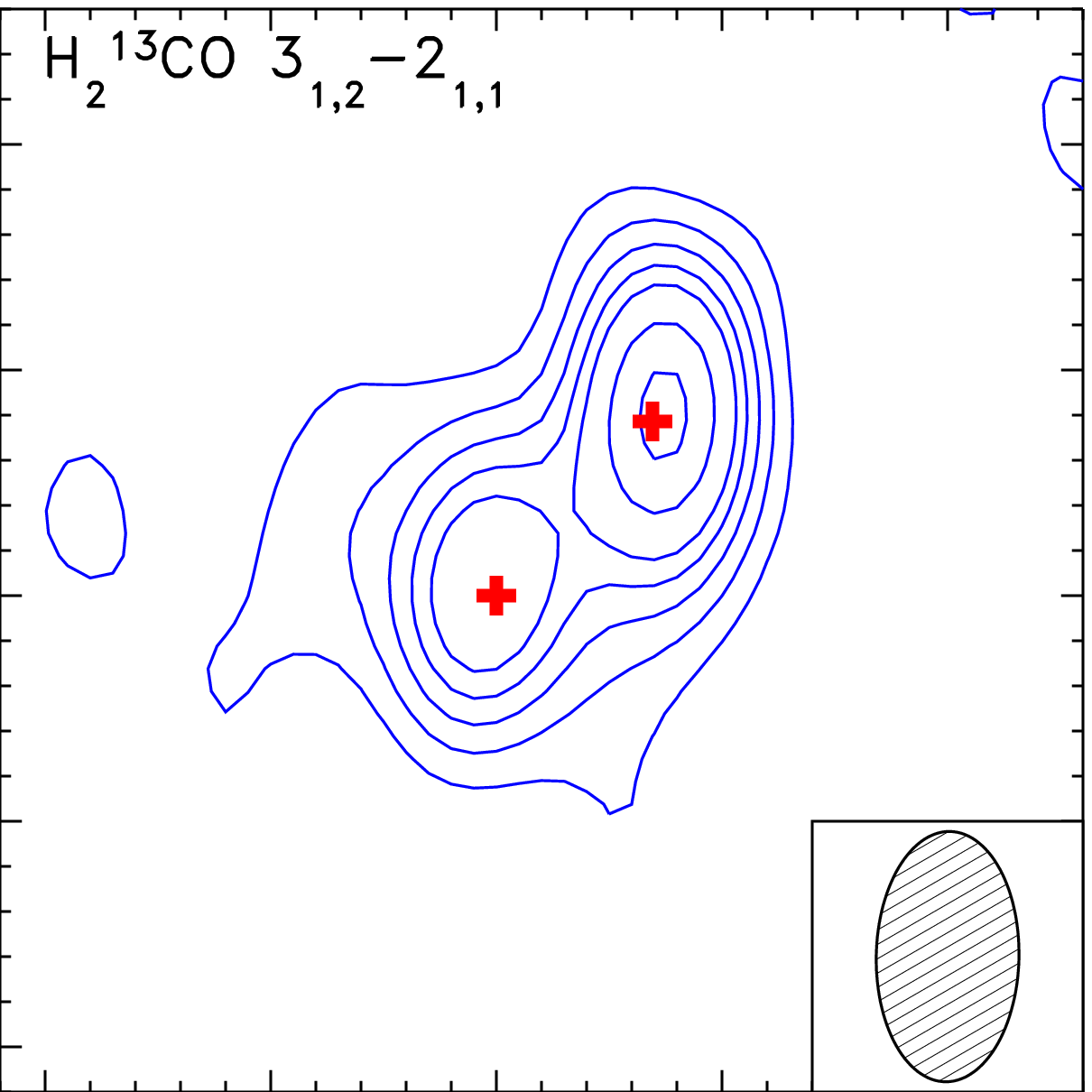}}
\resizebox{0.245\hsize}{!}{\includegraphics{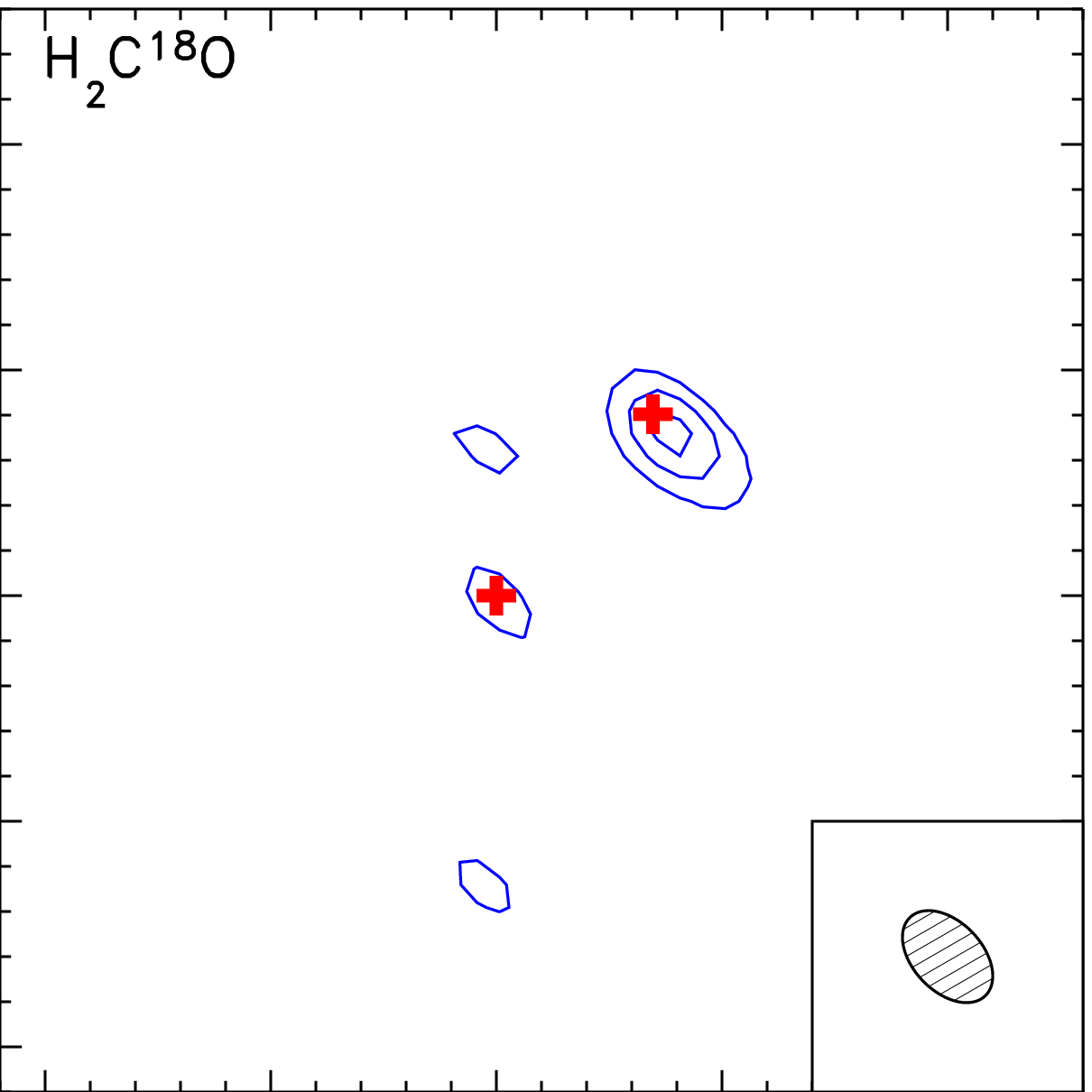}}
\resizebox{0.245\hsize}{!}{\includegraphics{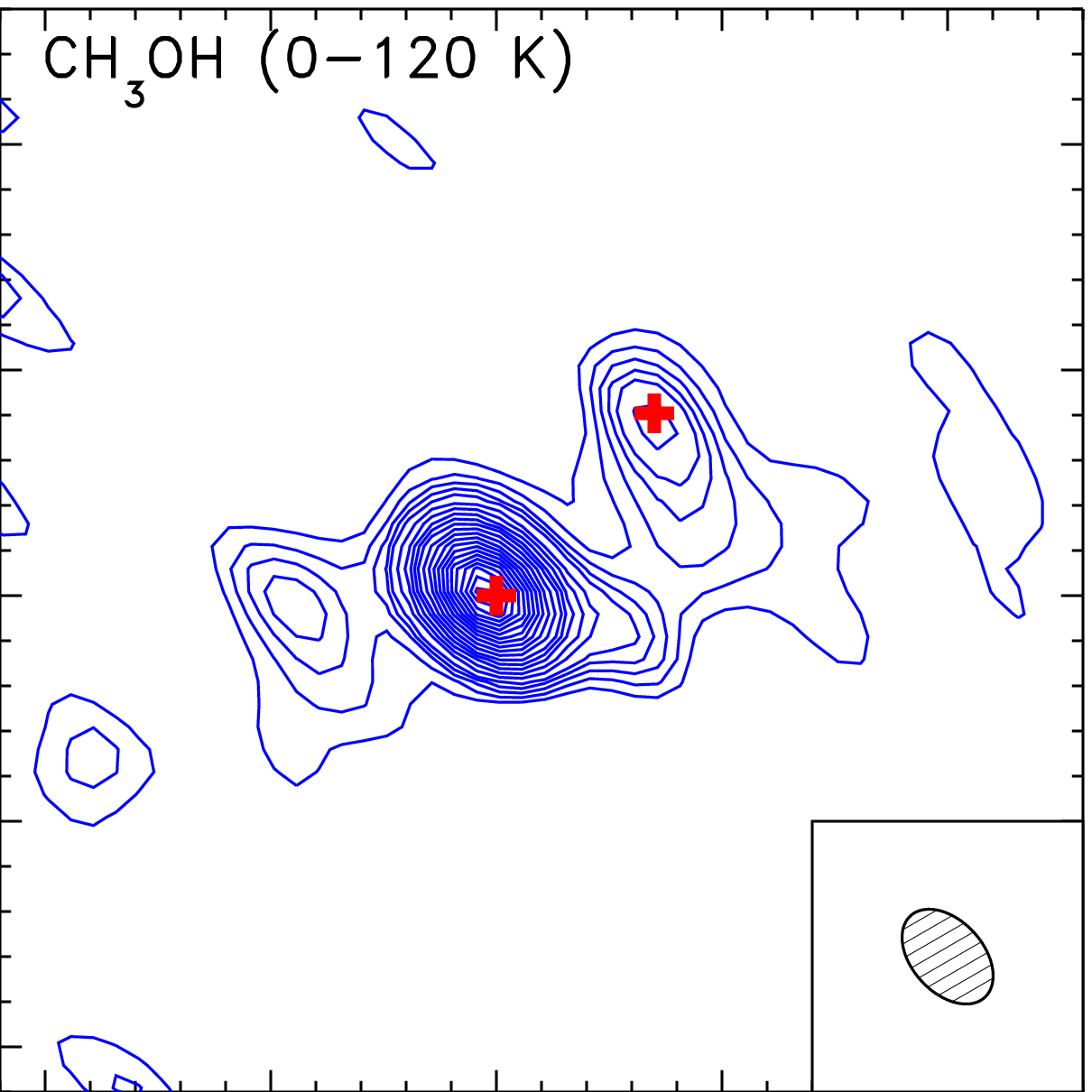}}
\resizebox{0.245\hsize}{!}{\includegraphics{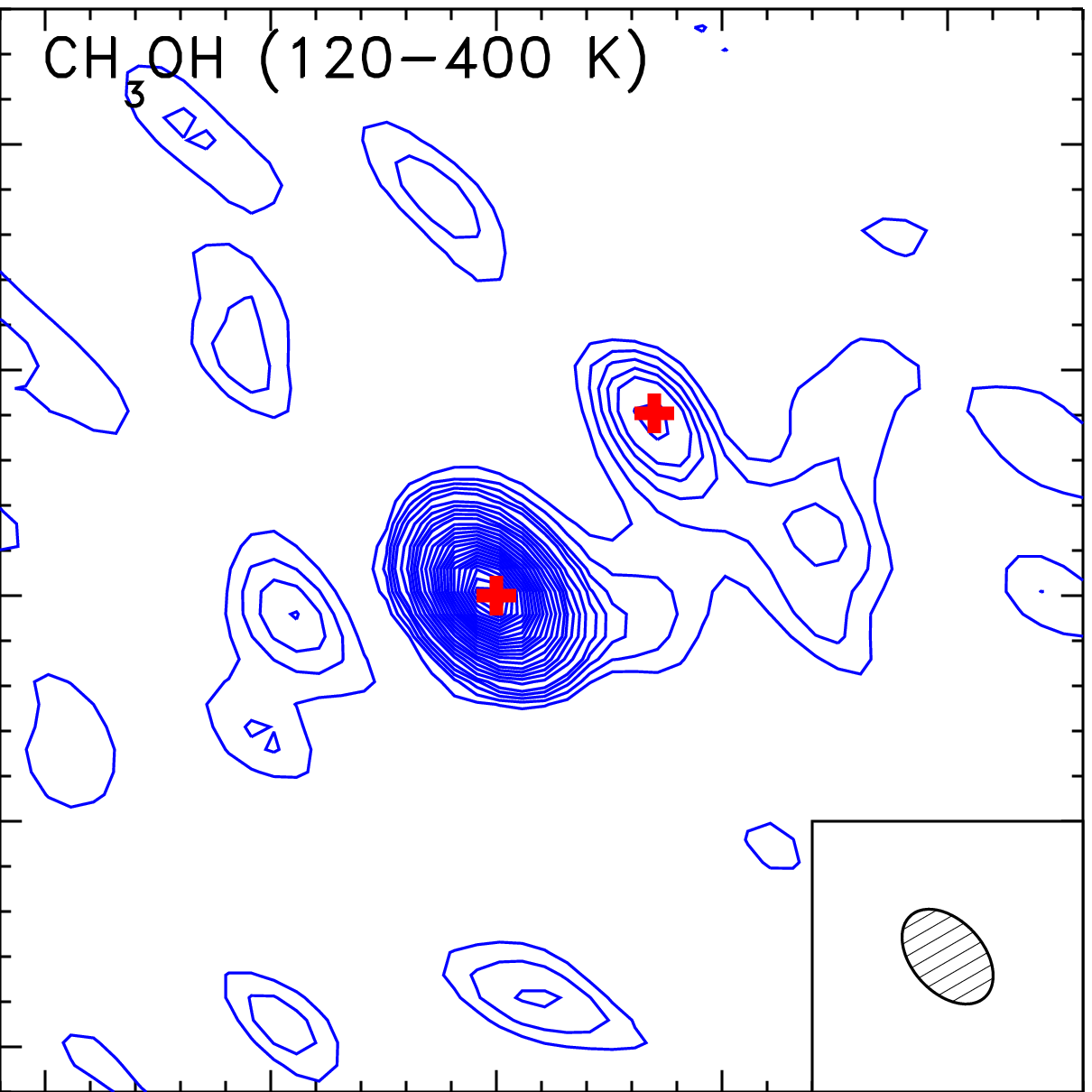}}
\resizebox{0.245\hsize}{!}{\includegraphics{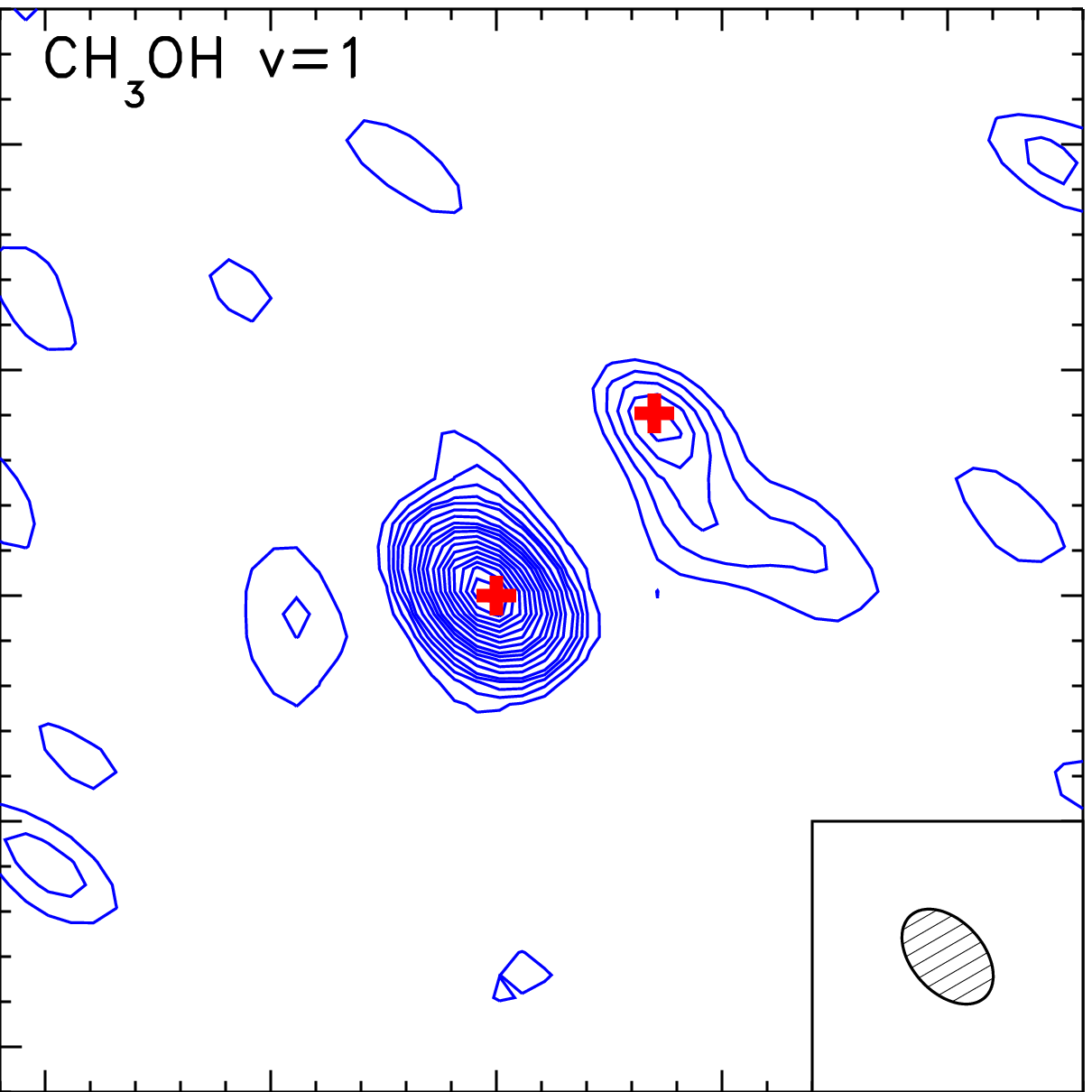}}
\resizebox{0.245\hsize}{!}{\includegraphics{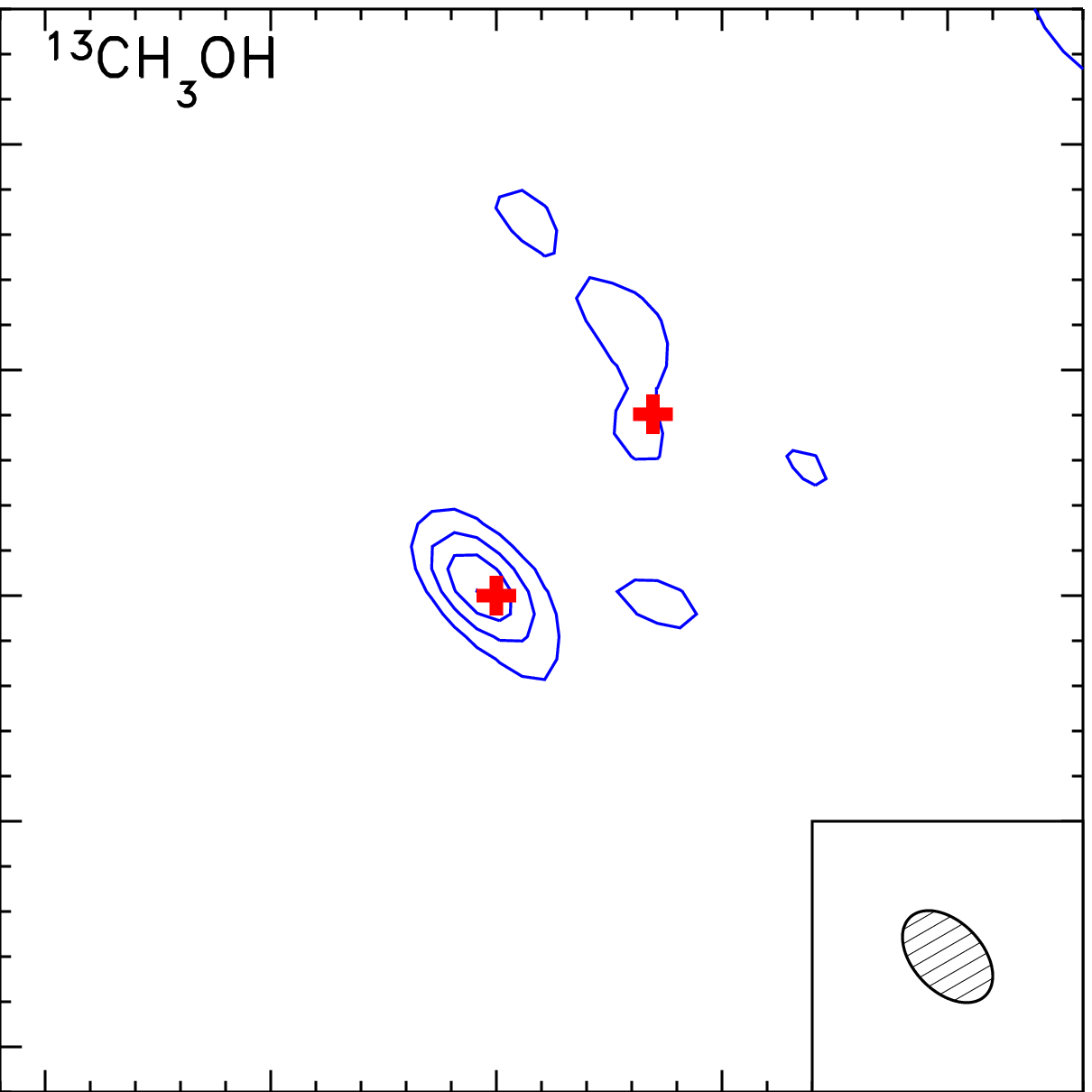}}
\resizebox{0.245\hsize}{!}{\includegraphics{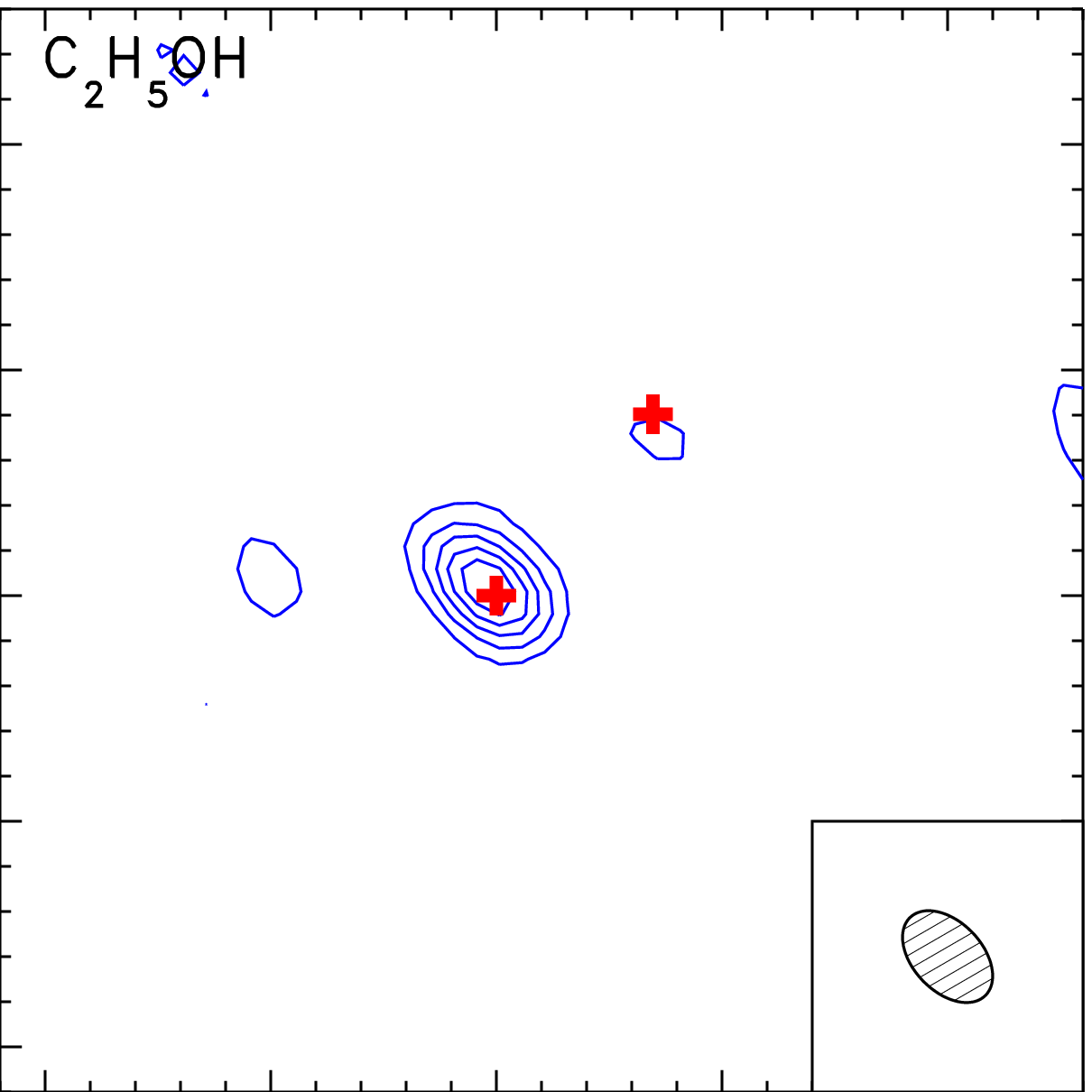}}
\resizebox{0.245\hsize}{!}{\includegraphics{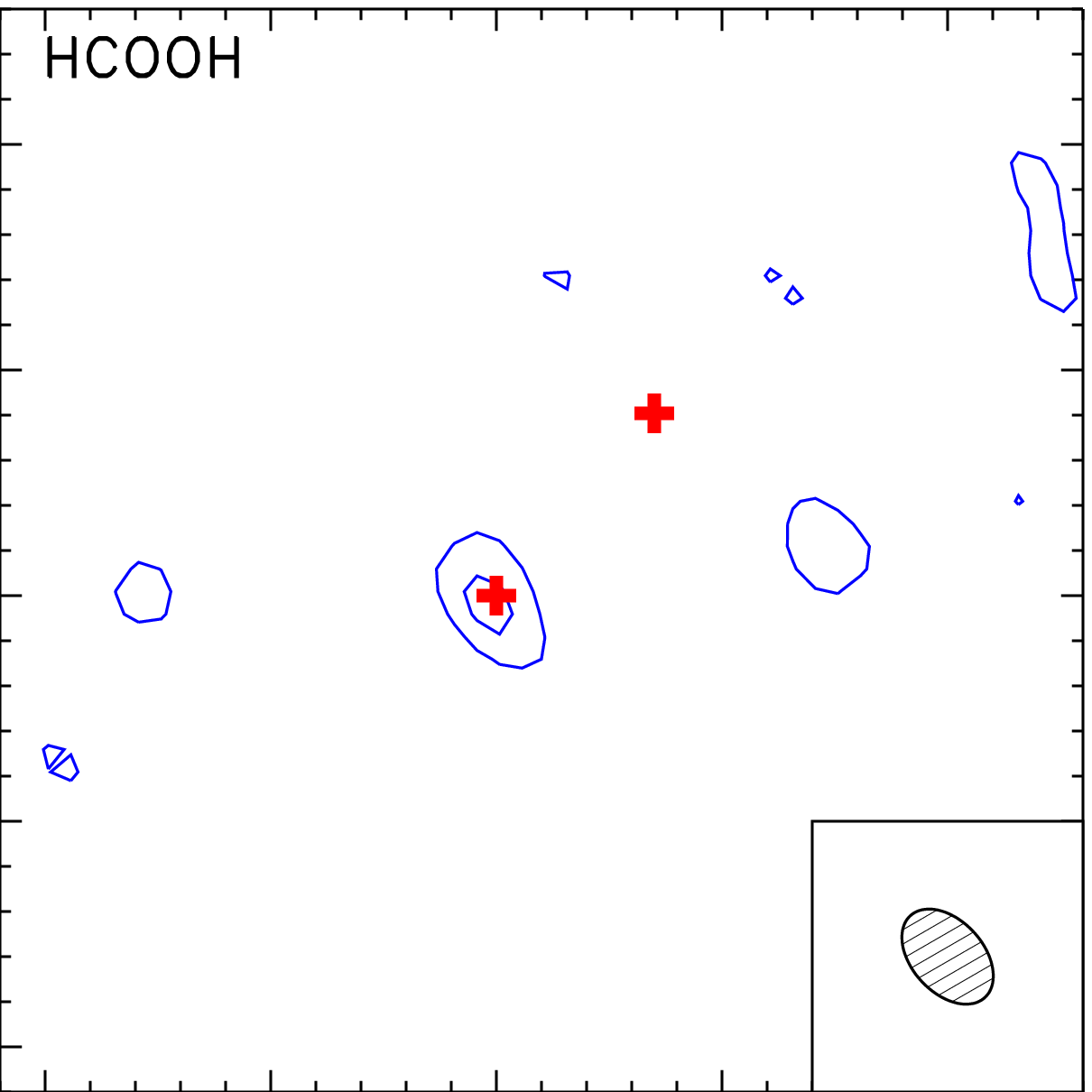}}
\resizebox{0.245\hsize}{!}{\includegraphics{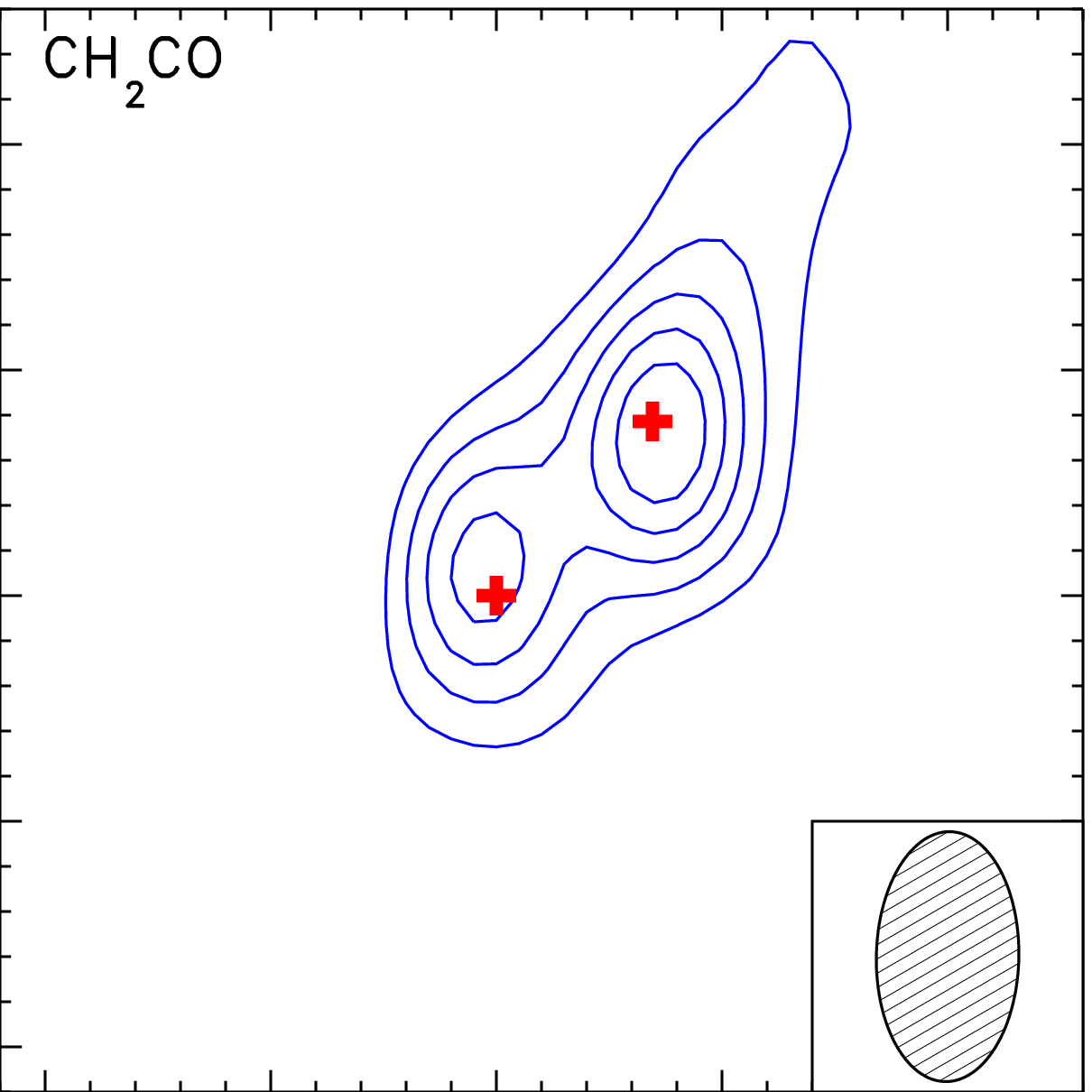}}
\resizebox{0.245\hsize}{!}{\includegraphics{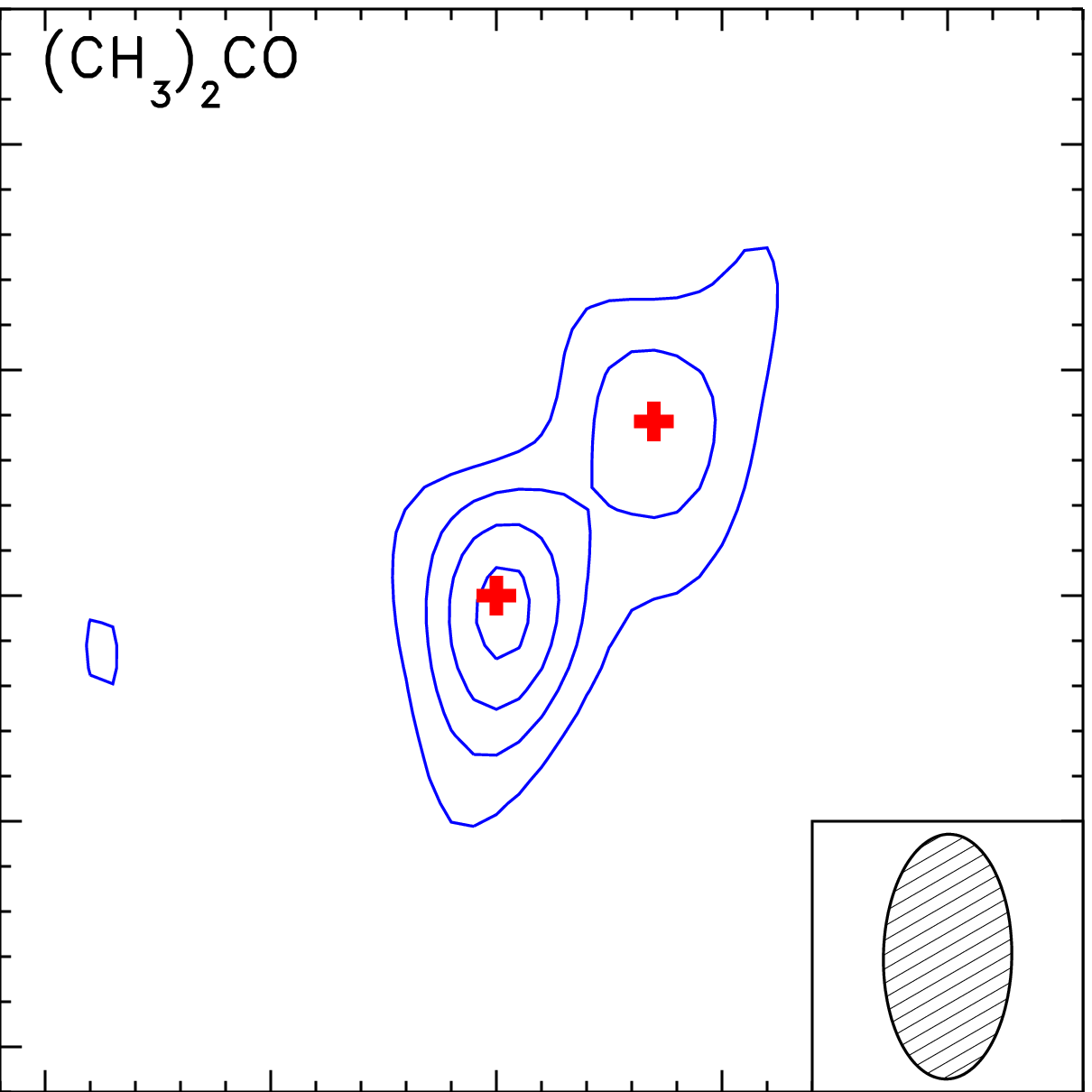}}
\resizebox{0.245\hsize}{!}{\includegraphics{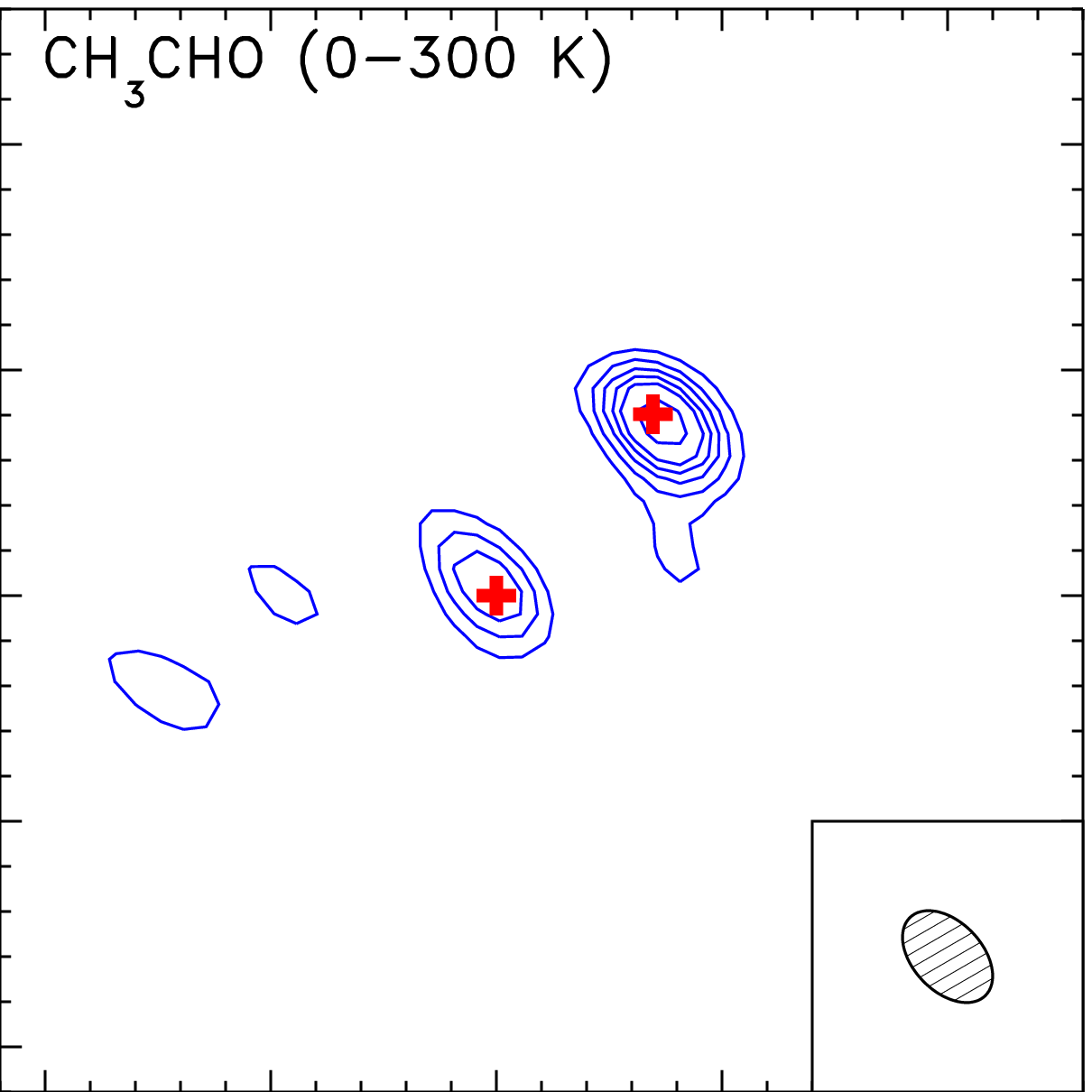}}
\resizebox{0.245\hsize}{!}{\includegraphics{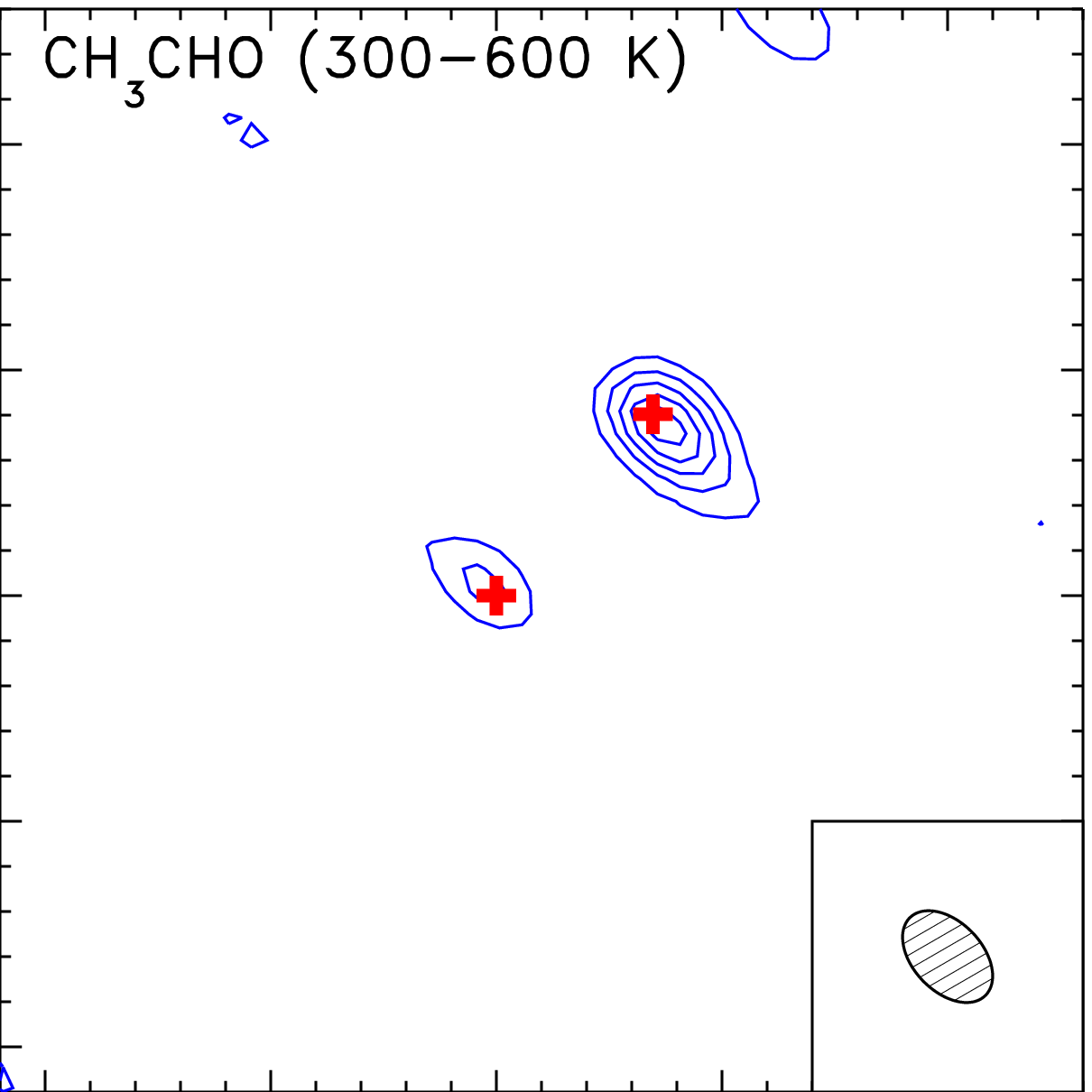}}
\resizebox{0.245\hsize}{!}{\includegraphics{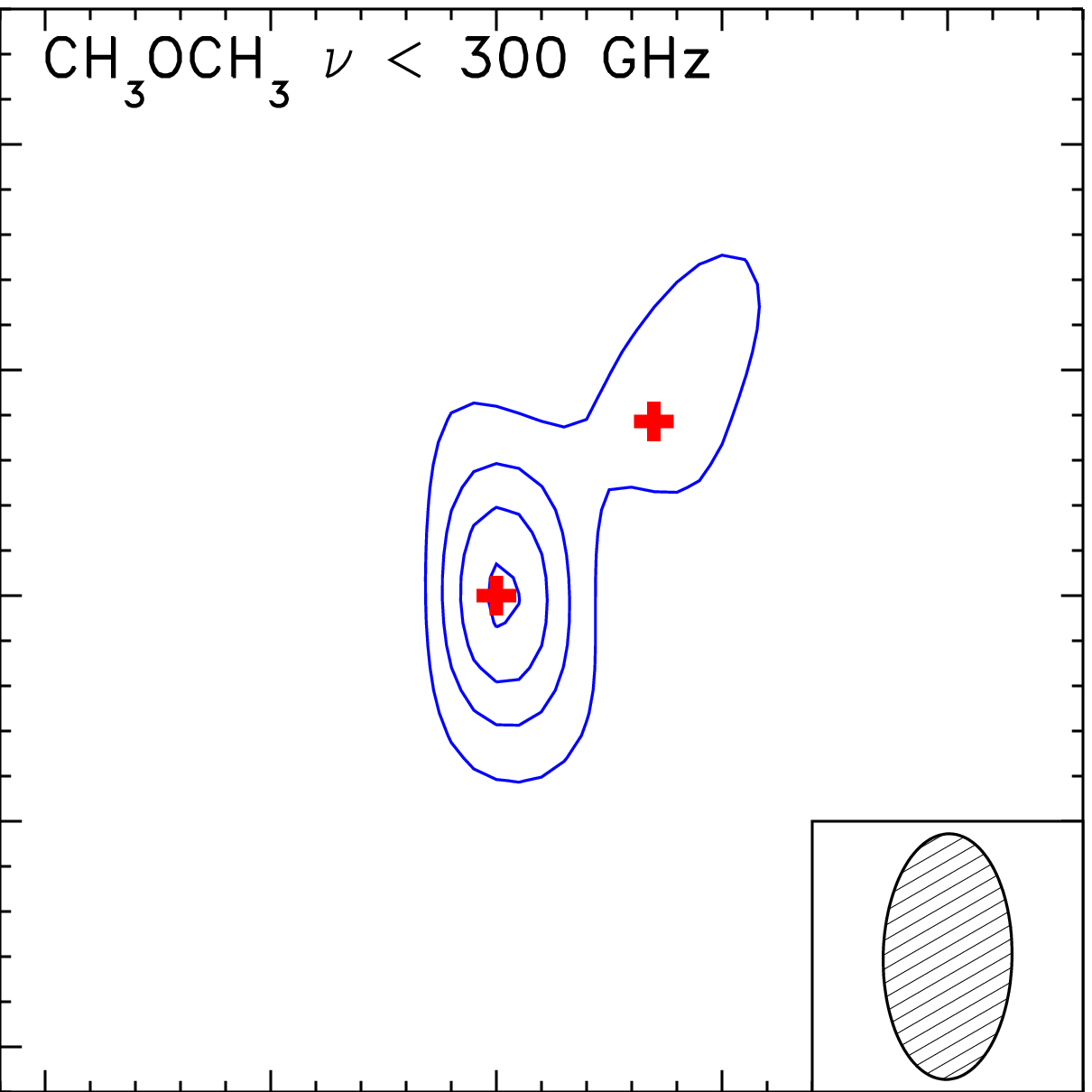}}
\resizebox{0.245\hsize}{!}{\includegraphics{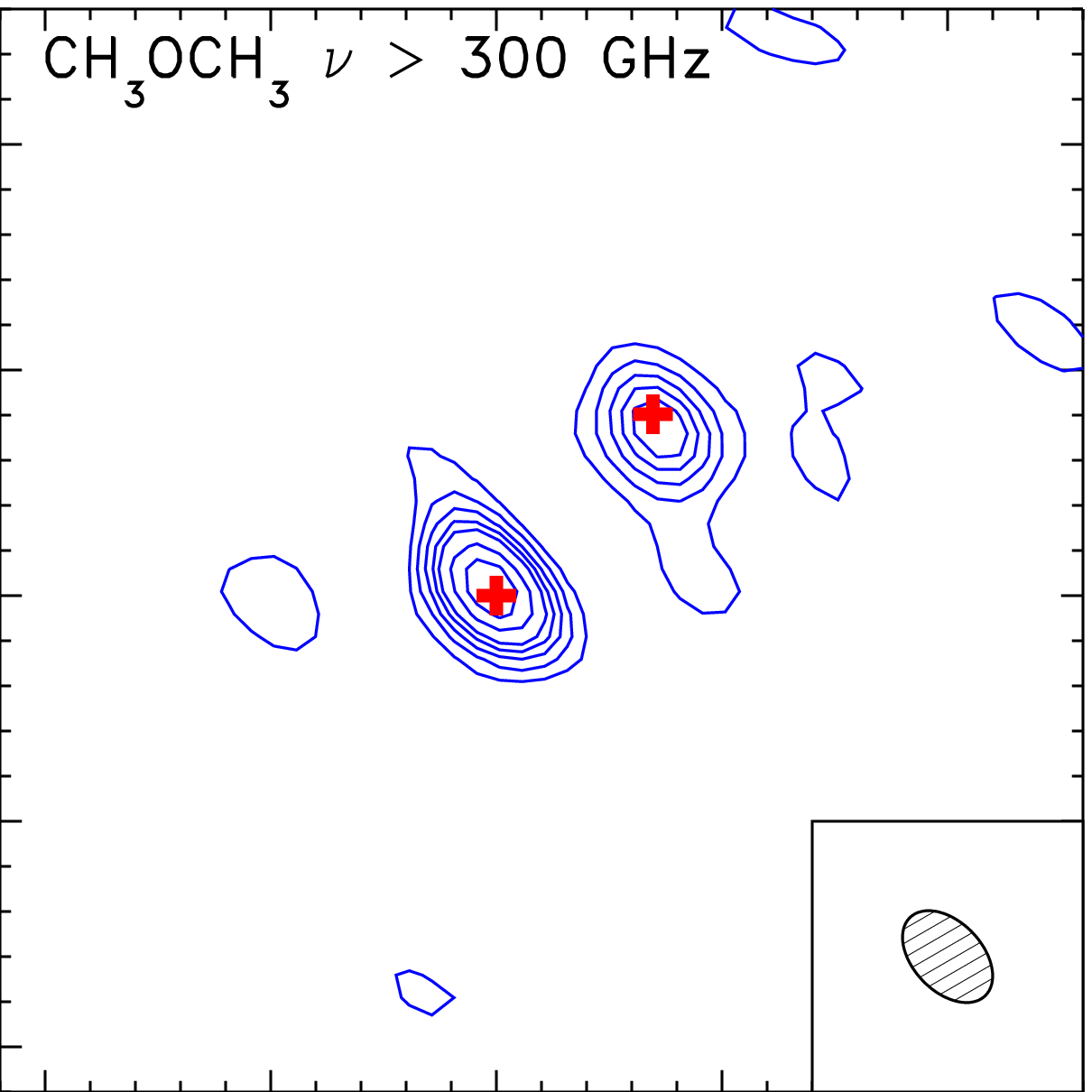}}
\resizebox{0.245\hsize}{!}{\includegraphics{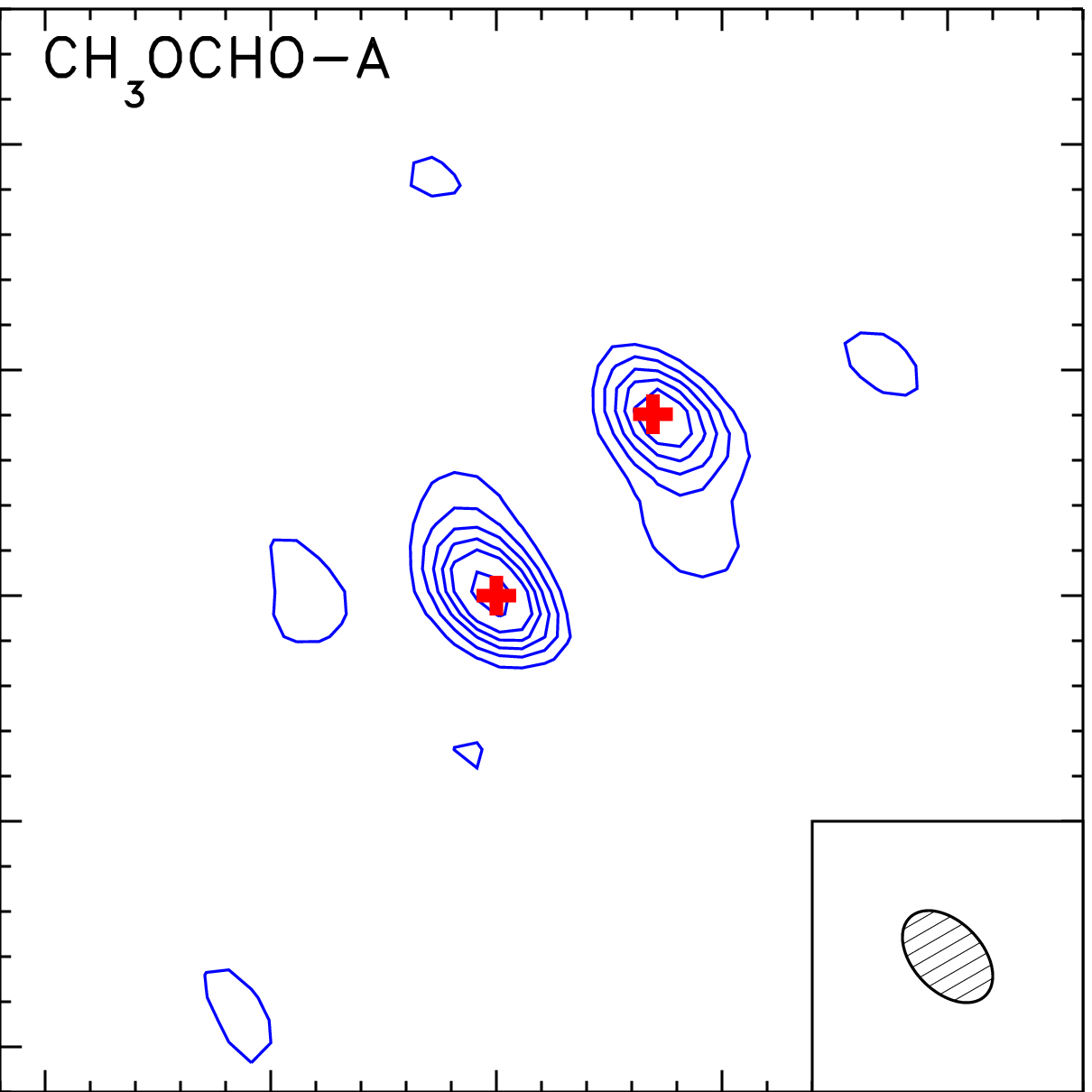}}
\resizebox{0.245\hsize}{!}{\includegraphics{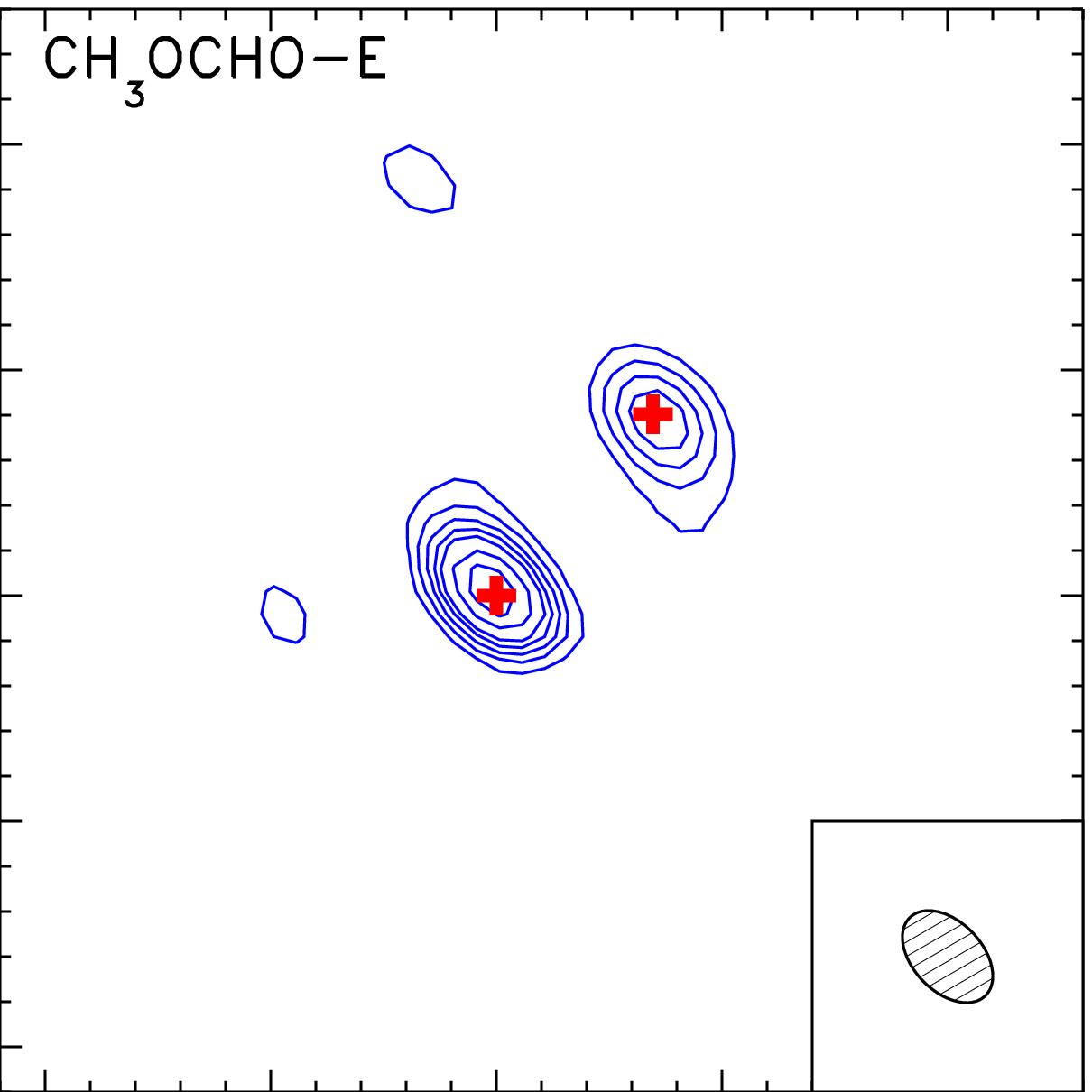}}

\phantom{xxx}
\caption{As in Fig.~\ref{image_first} for the O-bearing organic molecules.}
\end{minipage}
\end{figure*}
\clearpage
\clearpage
\begin{figure*}
\begin{minipage}[!h]{0.15\linewidth}\phantom{xxx}\end{minipage}
\hspace{0.5cm}
\begin{minipage}[!h]{0.85\linewidth}
\resizebox{0.245\hsize}{!}{\includegraphics{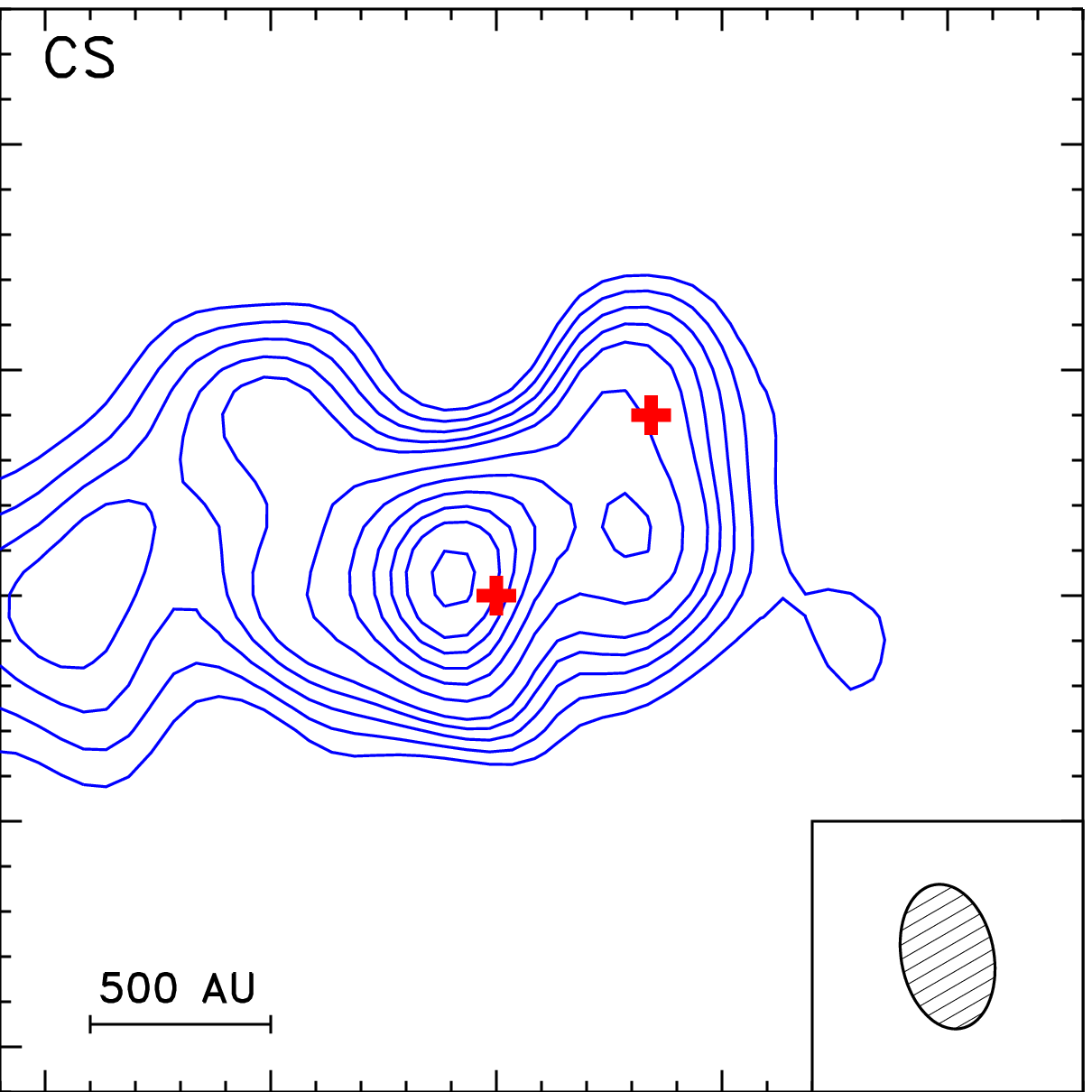}}
\resizebox{0.245\hsize}{!}{\includegraphics{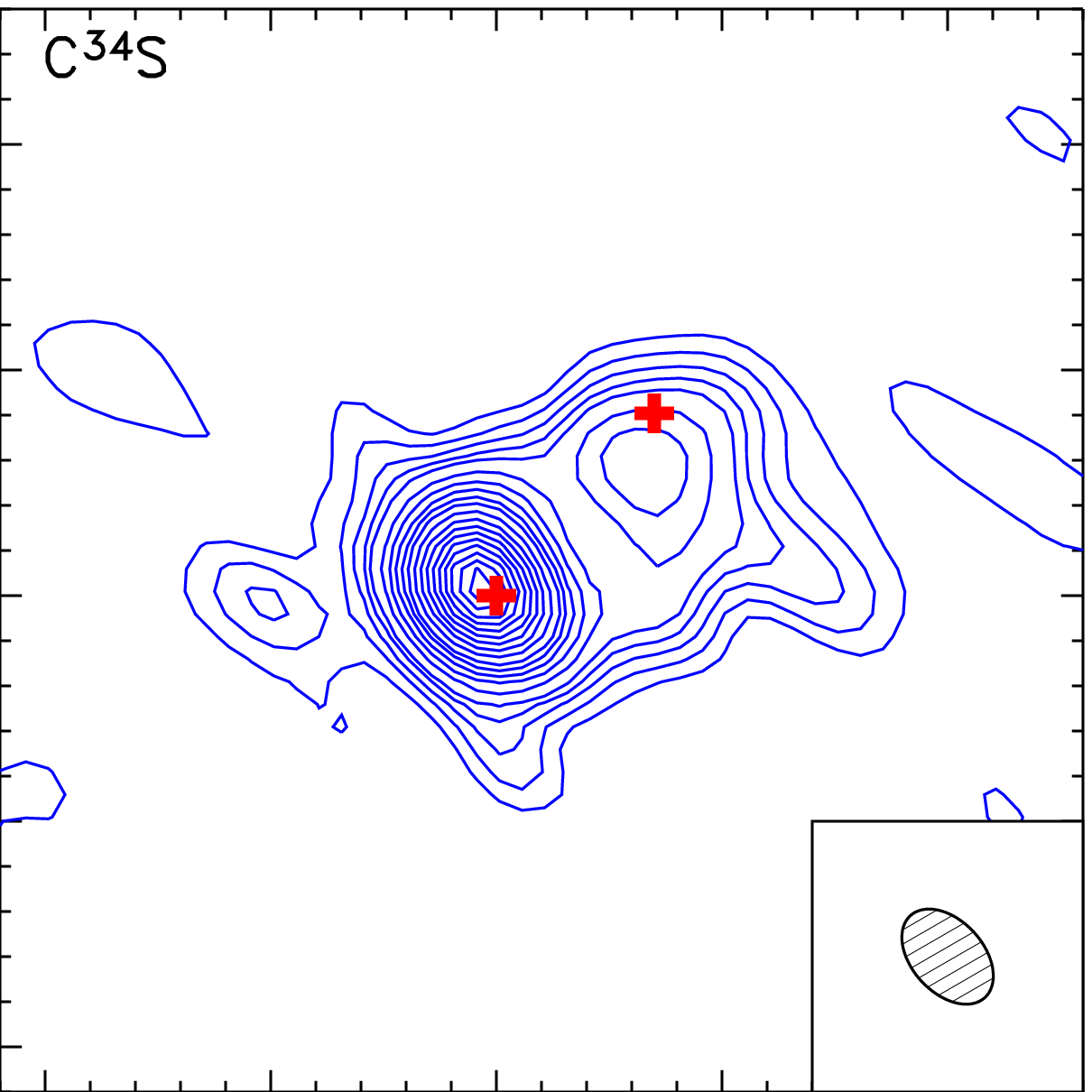}}
\resizebox{0.245\hsize}{!}{\includegraphics{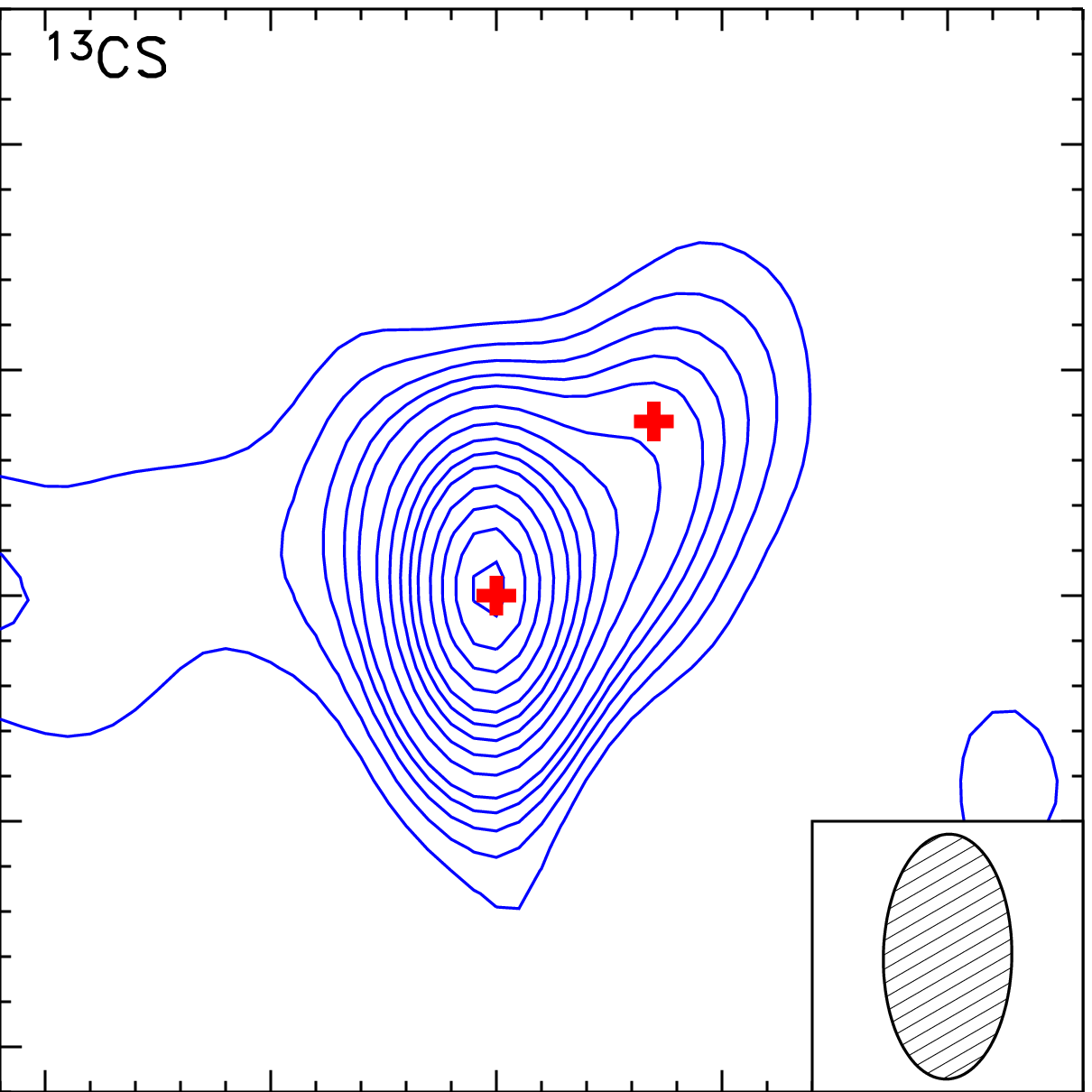}}
\resizebox{0.245\hsize}{!}{\includegraphics{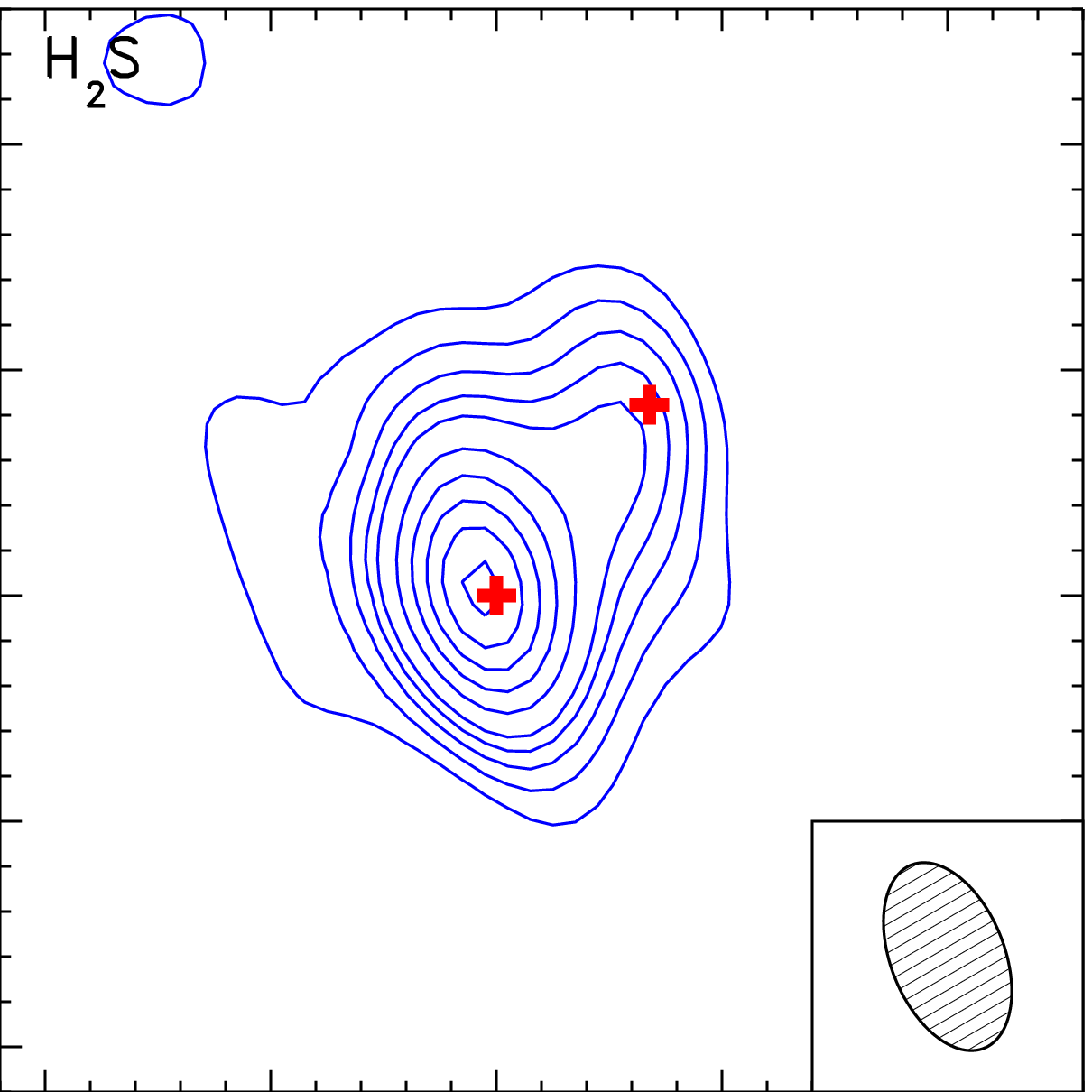}}
\resizebox{0.245\hsize}{!}{\includegraphics{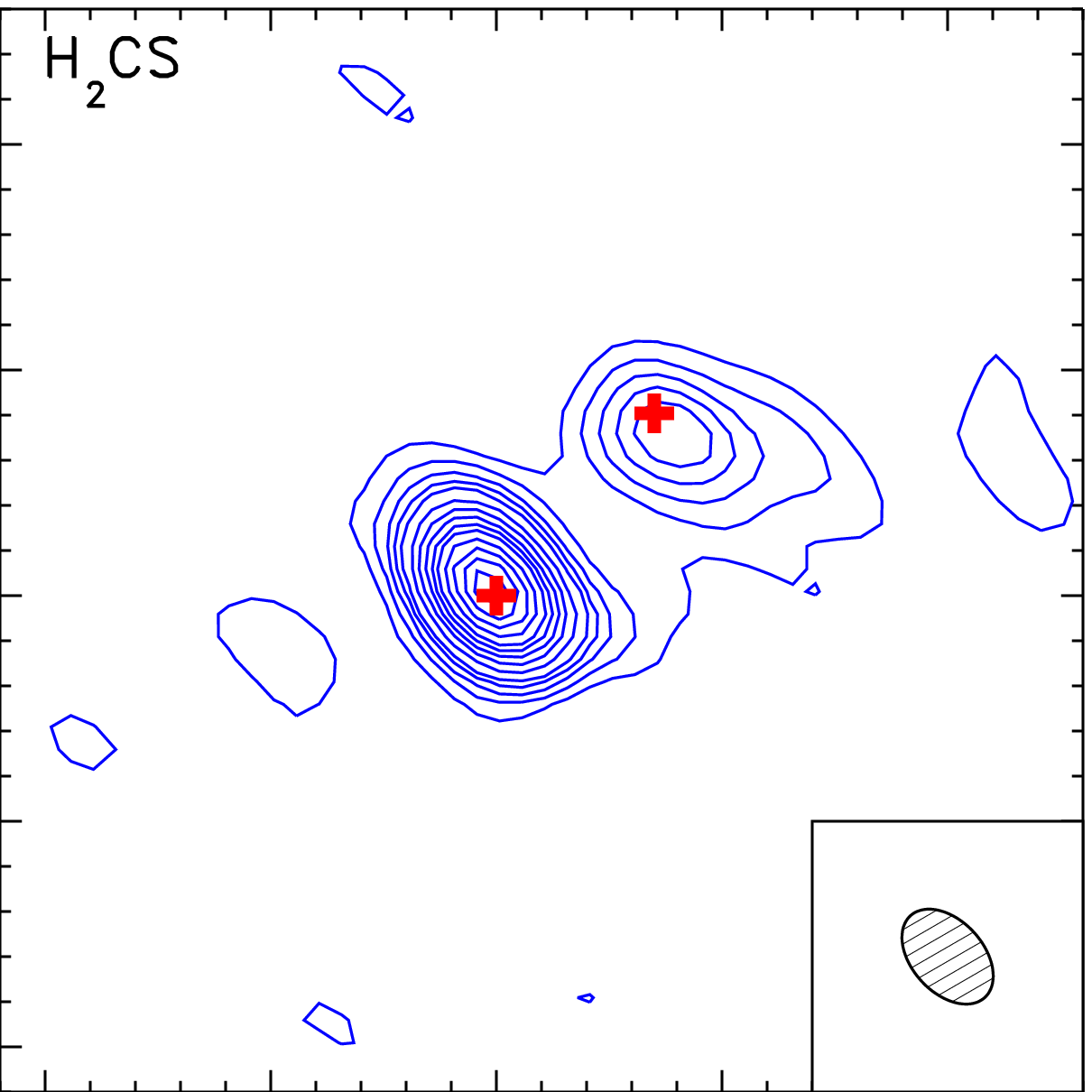}}
\resizebox{0.245\hsize}{!}{\includegraphics{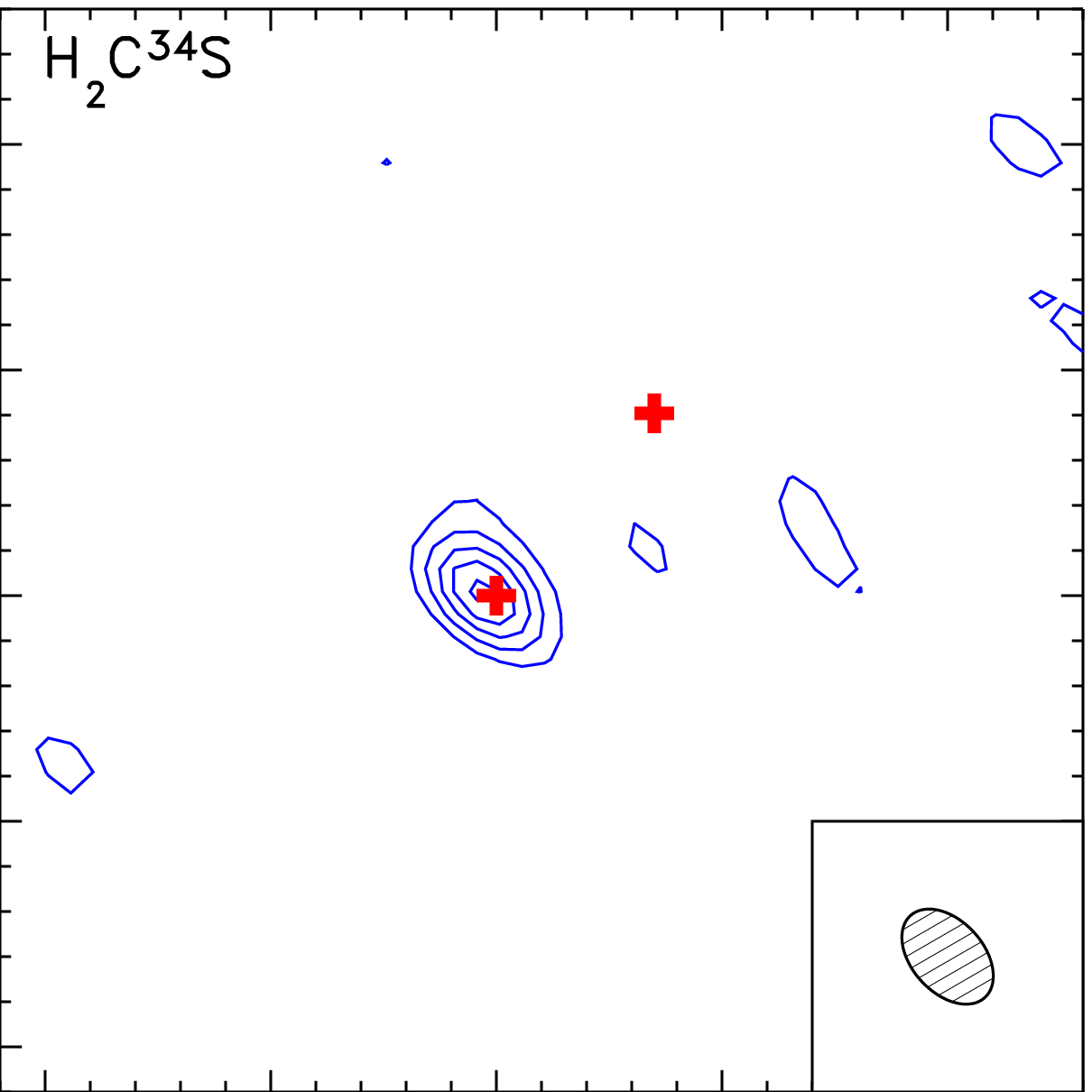}}
\resizebox{0.245\hsize}{!}{\includegraphics{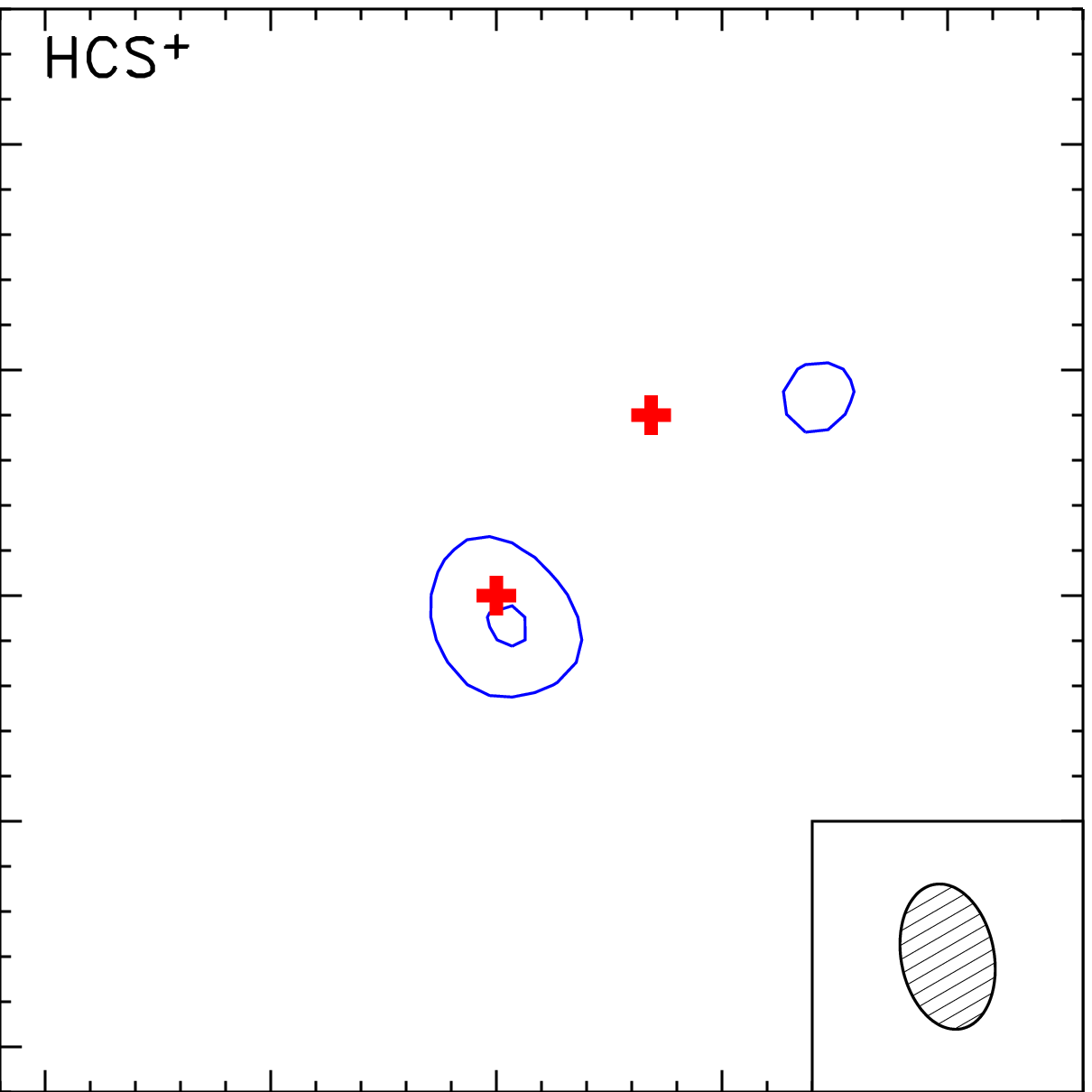}}
\begin{minipage}{0.245\hsize}\hfill\vspace{0.1in}\end{minipage}
\resizebox{0.245\hsize}{!}{\includegraphics{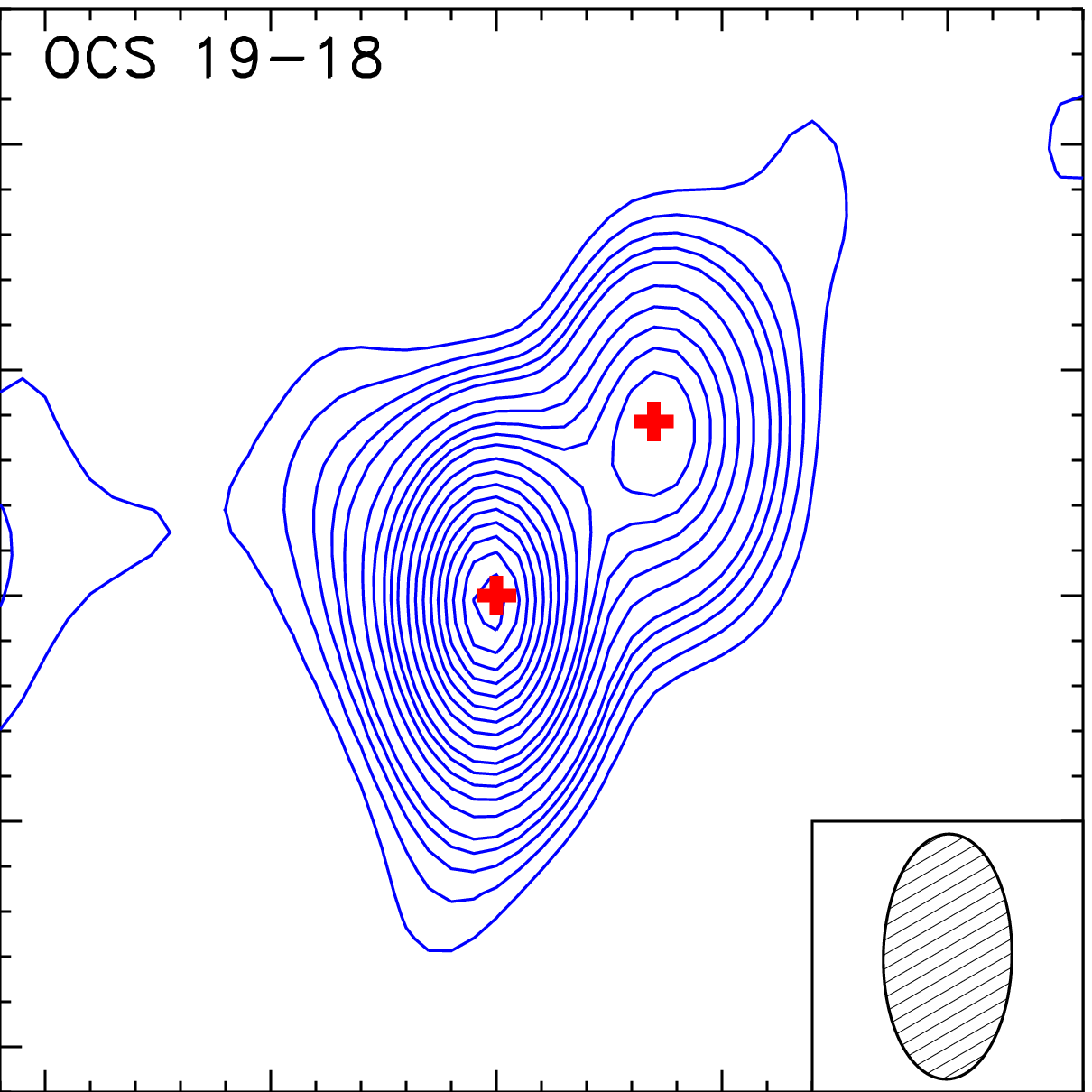}}
\resizebox{0.245\hsize}{!}{\includegraphics{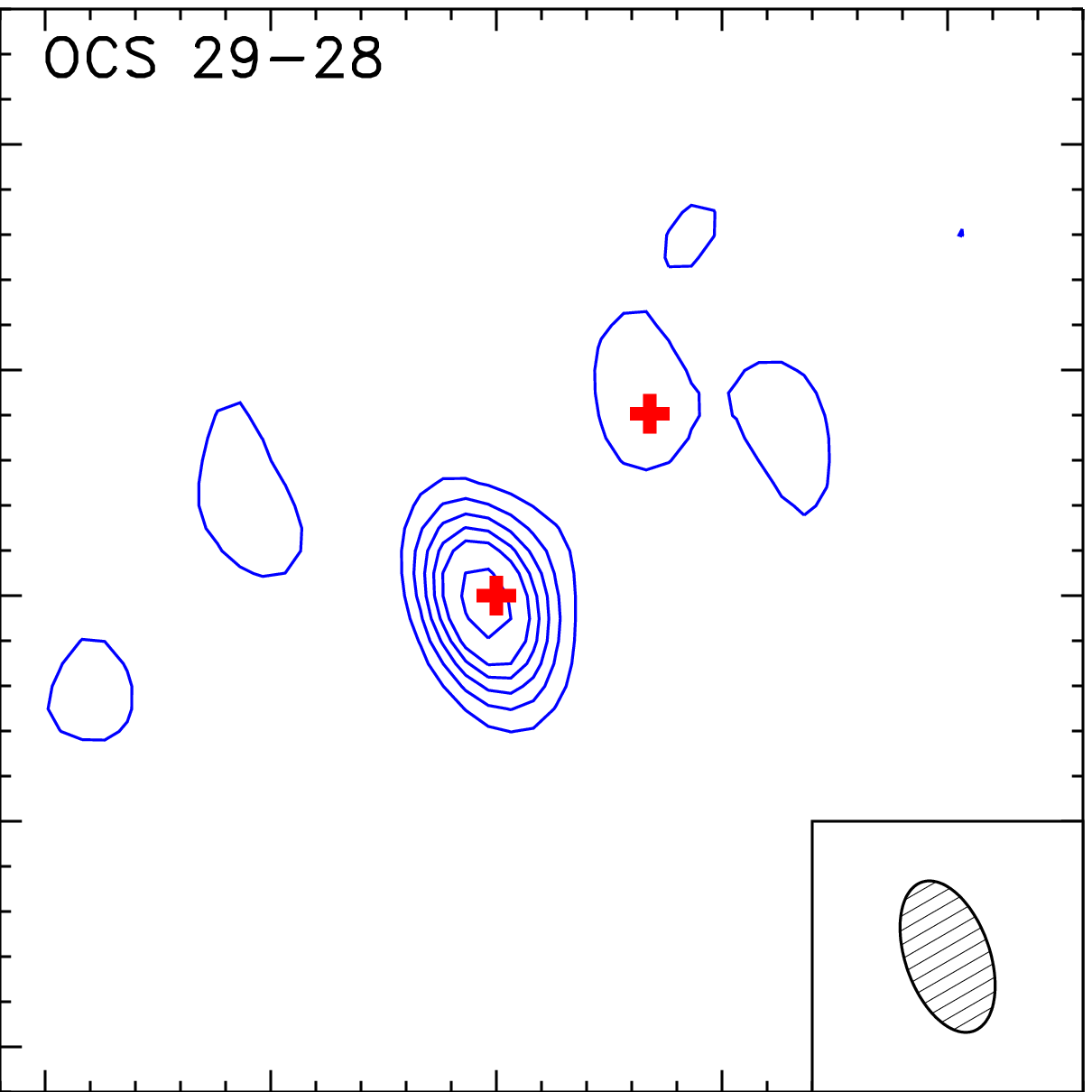}}
\resizebox{0.245\hsize}{!}{\includegraphics{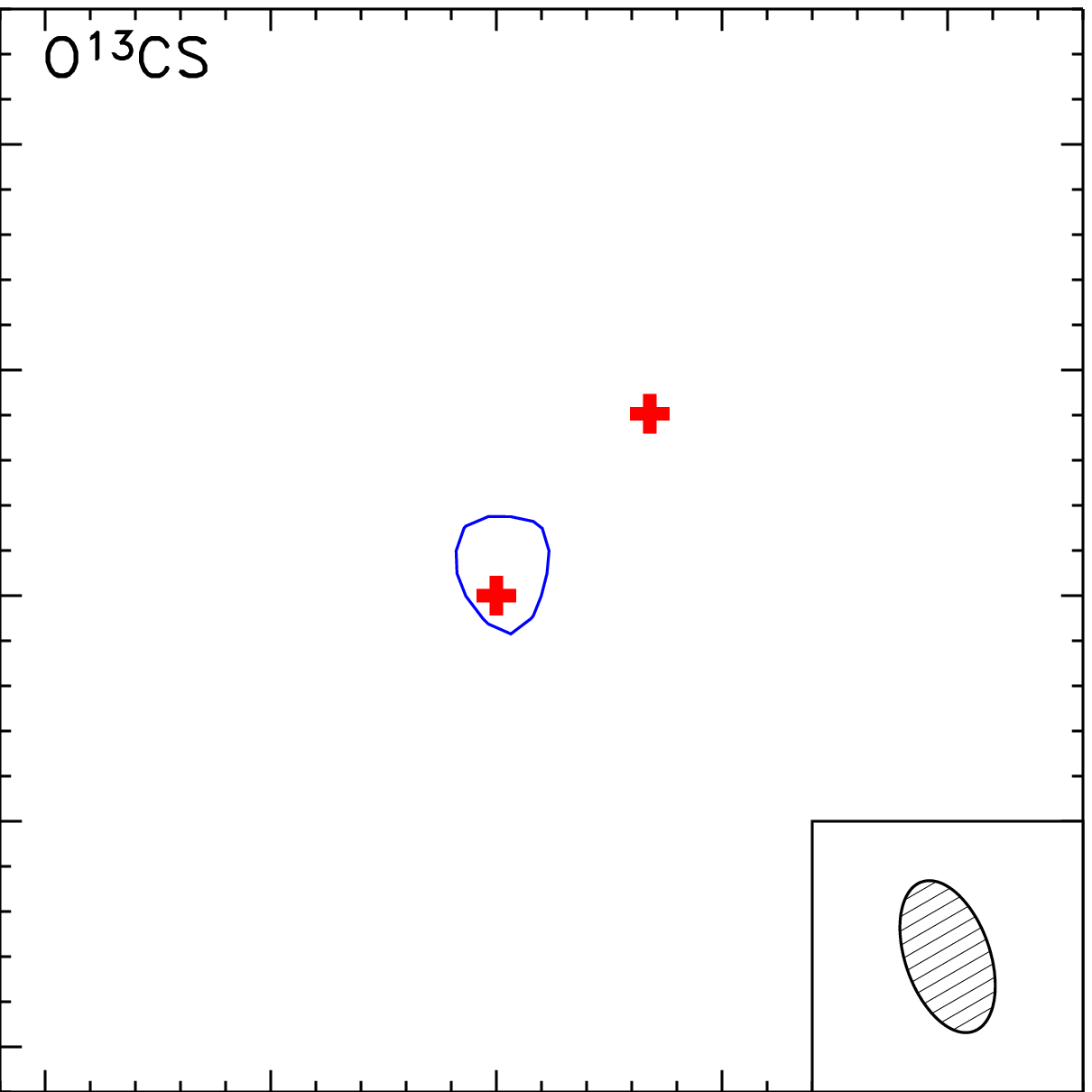}}
\begin{minipage}{0.245\hsize}\hfill\vspace{0.1in}\end{minipage}
\resizebox{0.245\hsize}{!}{\includegraphics{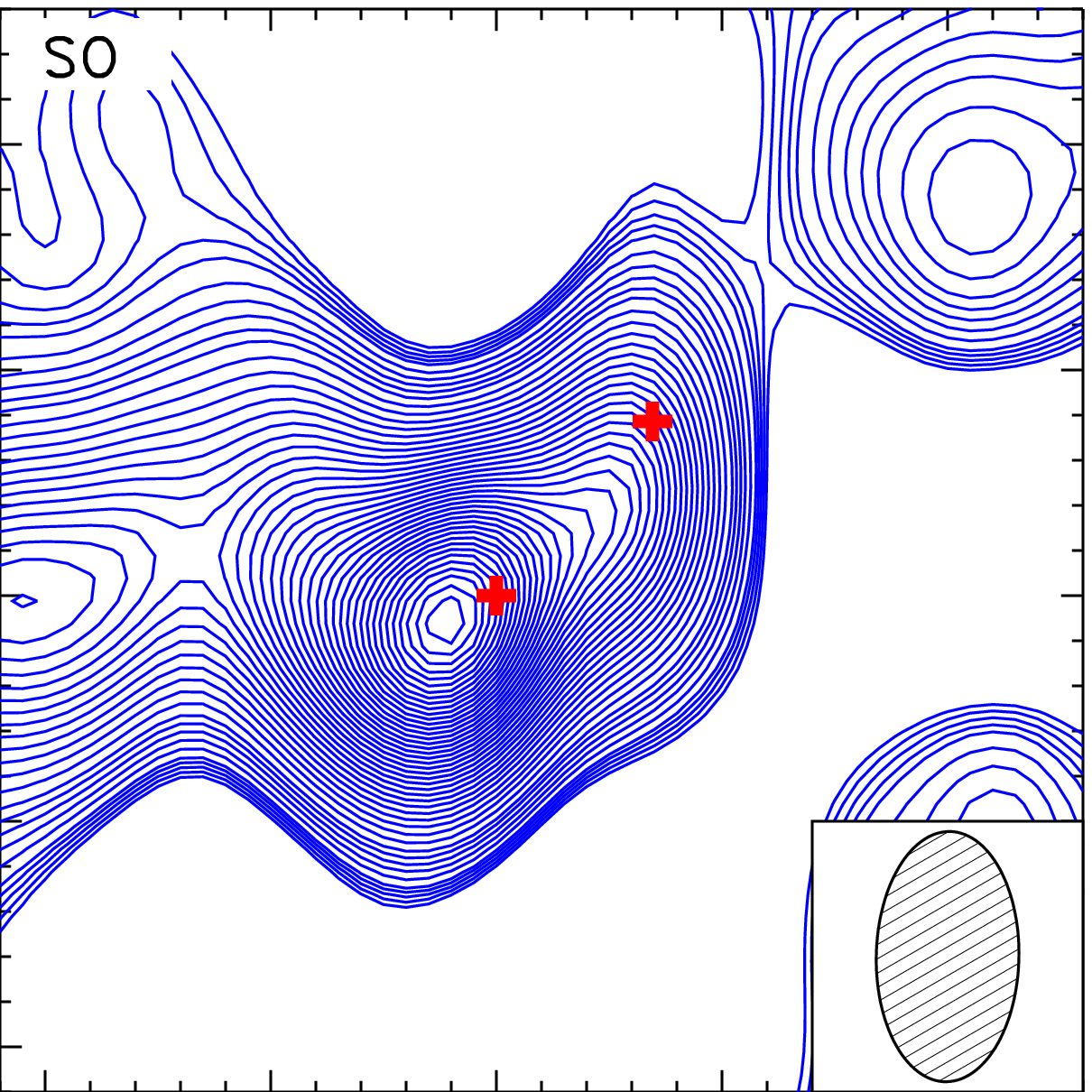}}
\resizebox{0.245\hsize}{!}{\includegraphics{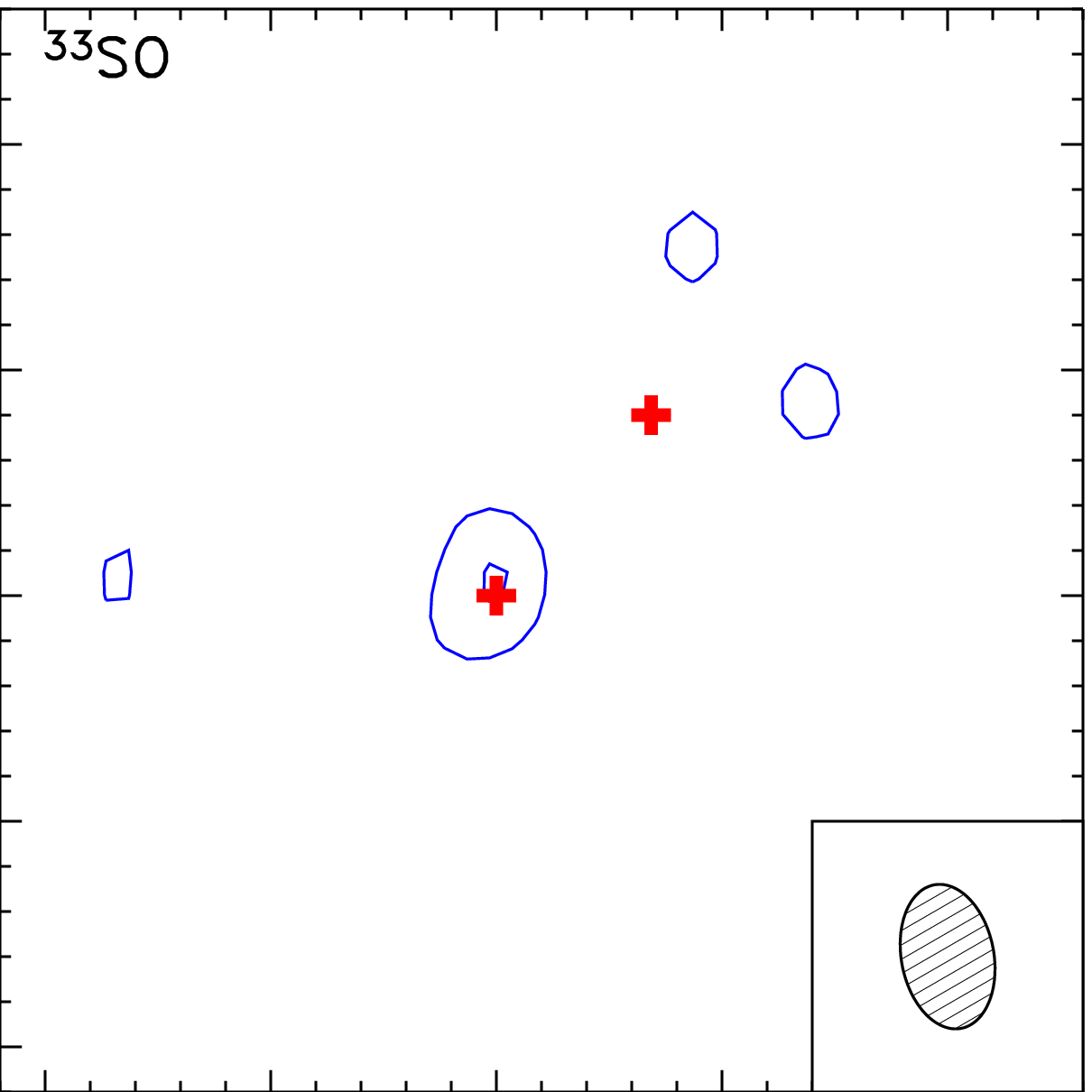}}
\resizebox{0.245\hsize}{!}{\includegraphics{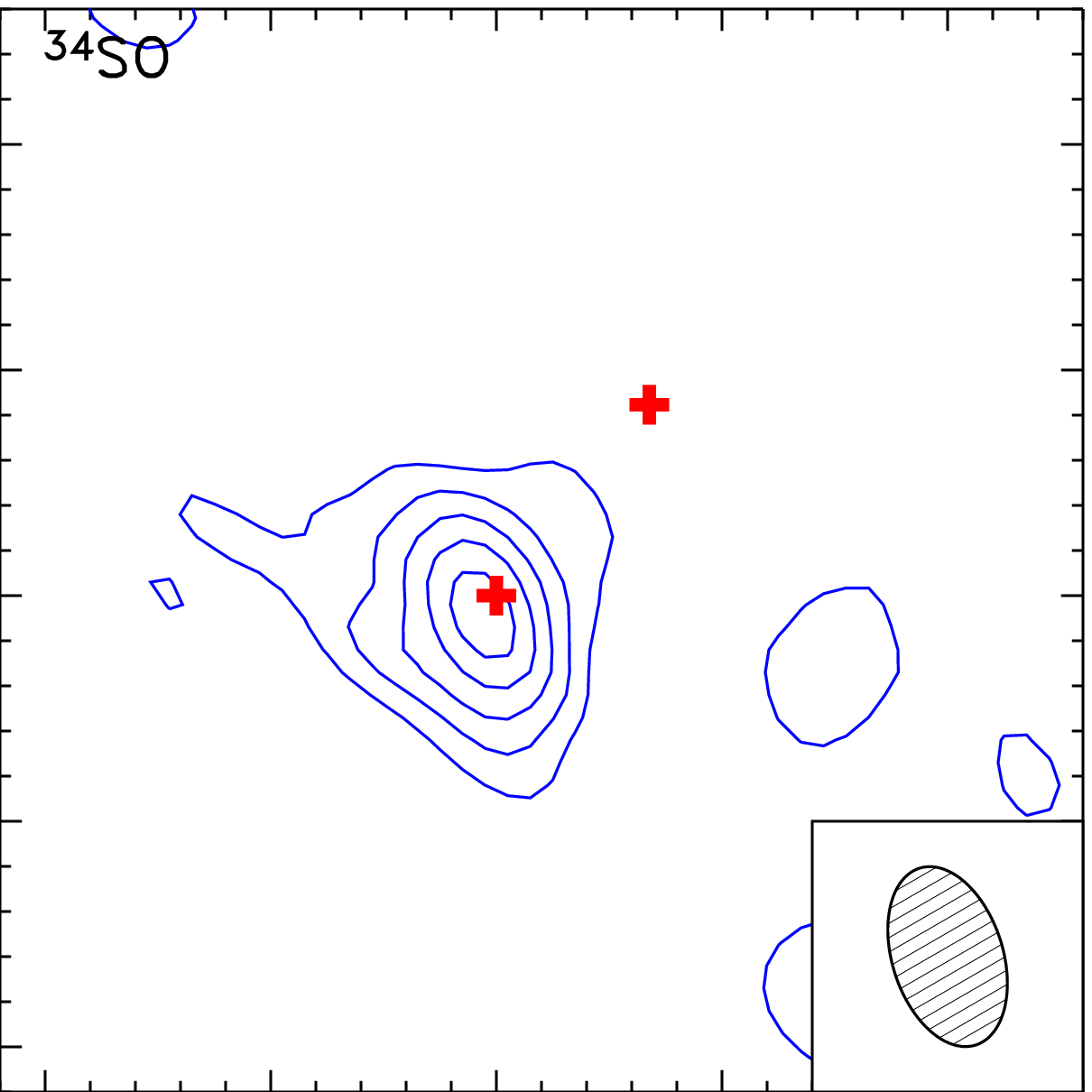}}
\begin{minipage}{0.245\hsize}\hfill\vspace{0.1in}\end{minipage}
\resizebox{0.245\hsize}{!}{\includegraphics{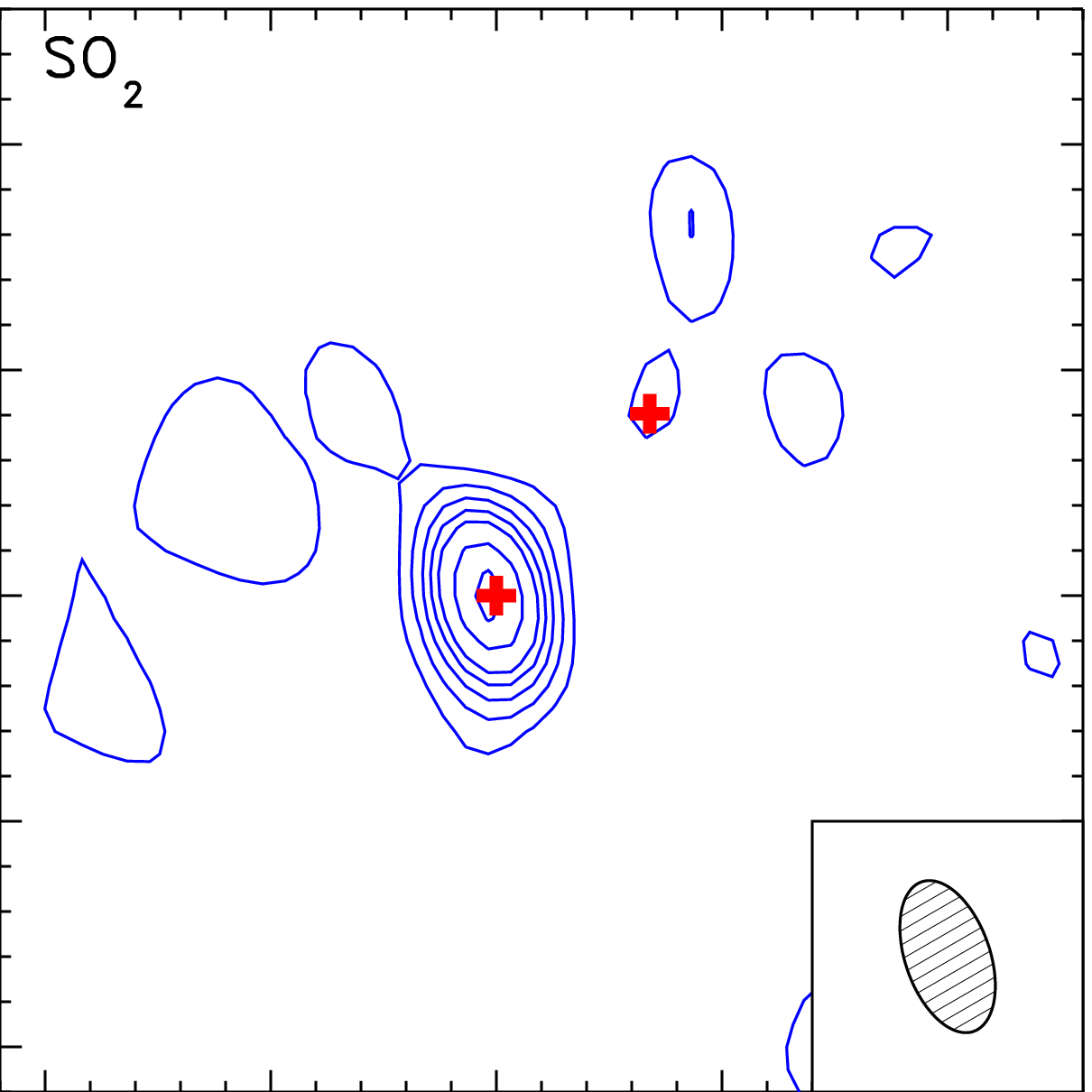}}
\resizebox{0.245\hsize}{!}{\includegraphics{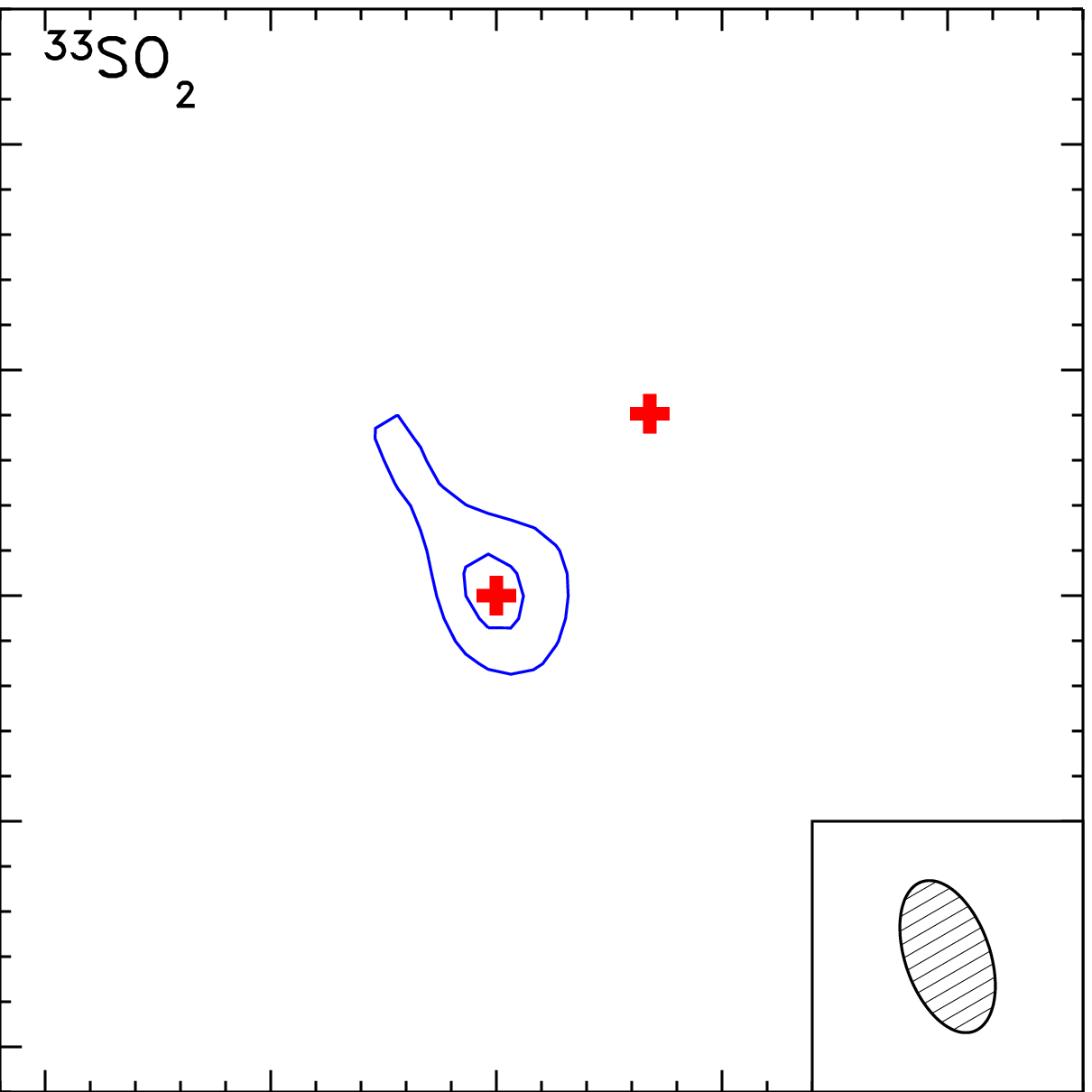}}
\resizebox{0.245\hsize}{!}{\includegraphics{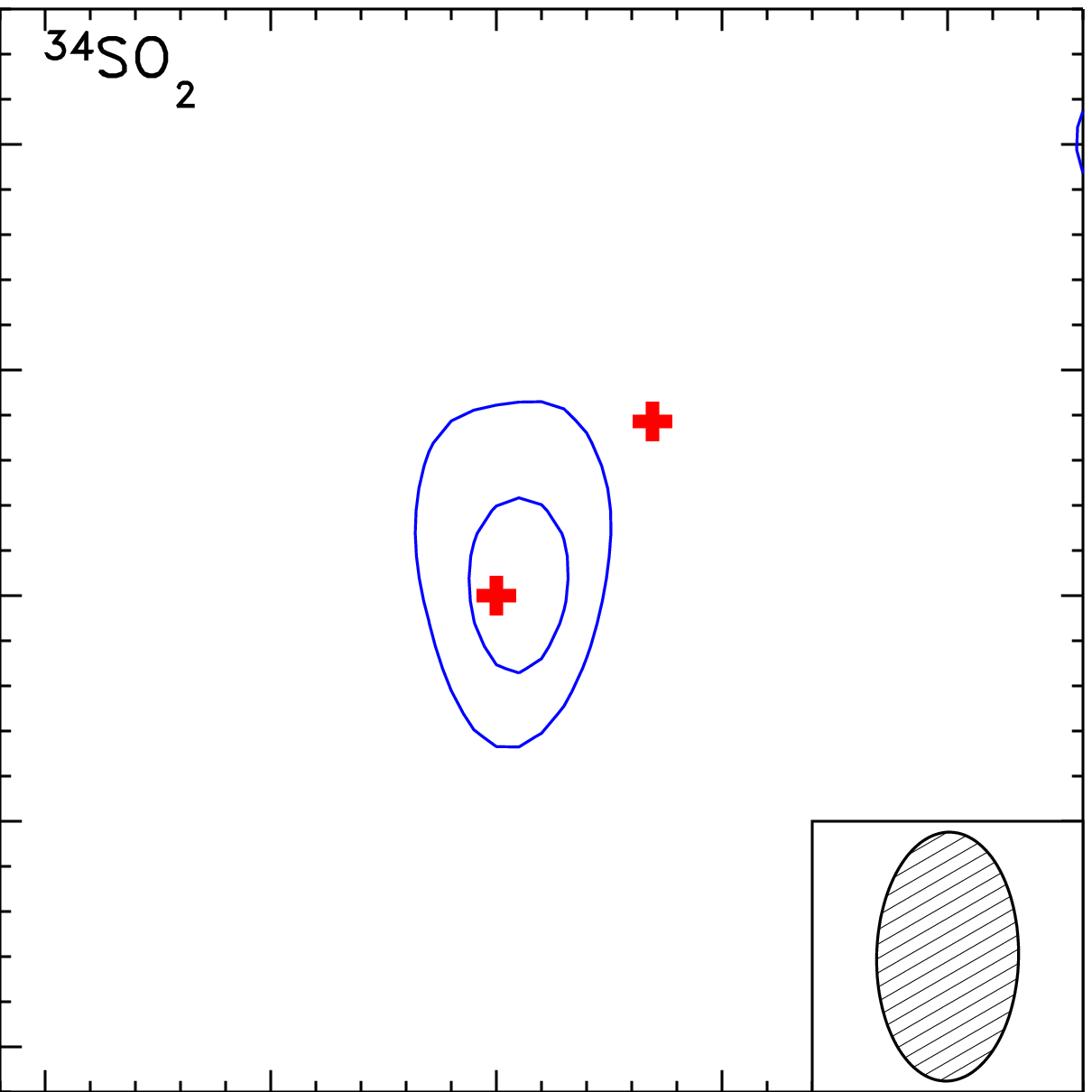}}
\resizebox{0.245\hsize}{!}{\includegraphics{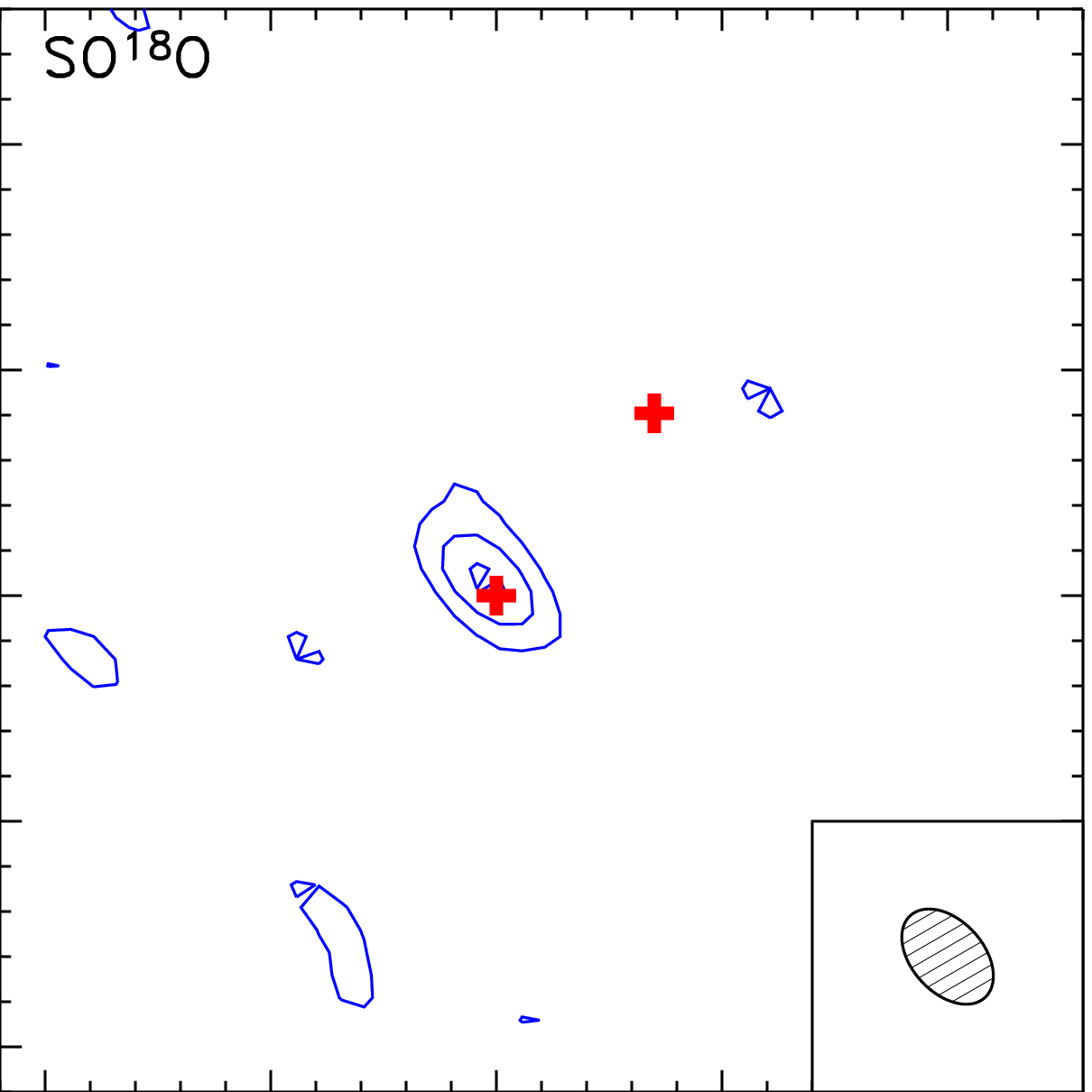}}

\phantom{xxx}
\caption{As in Fig.~\ref{image_first} for the S-bearing molecules.}
\end{minipage}
\end{figure*}
\clearpage
\begin{figure*}
\begin{minipage}[!h]{0.15\linewidth}\phantom{xxx}\end{minipage}
\hspace{0.5cm}
\begin{minipage}[!h]{0.85\linewidth}
\resizebox{0.245\hsize}{!}{\includegraphics{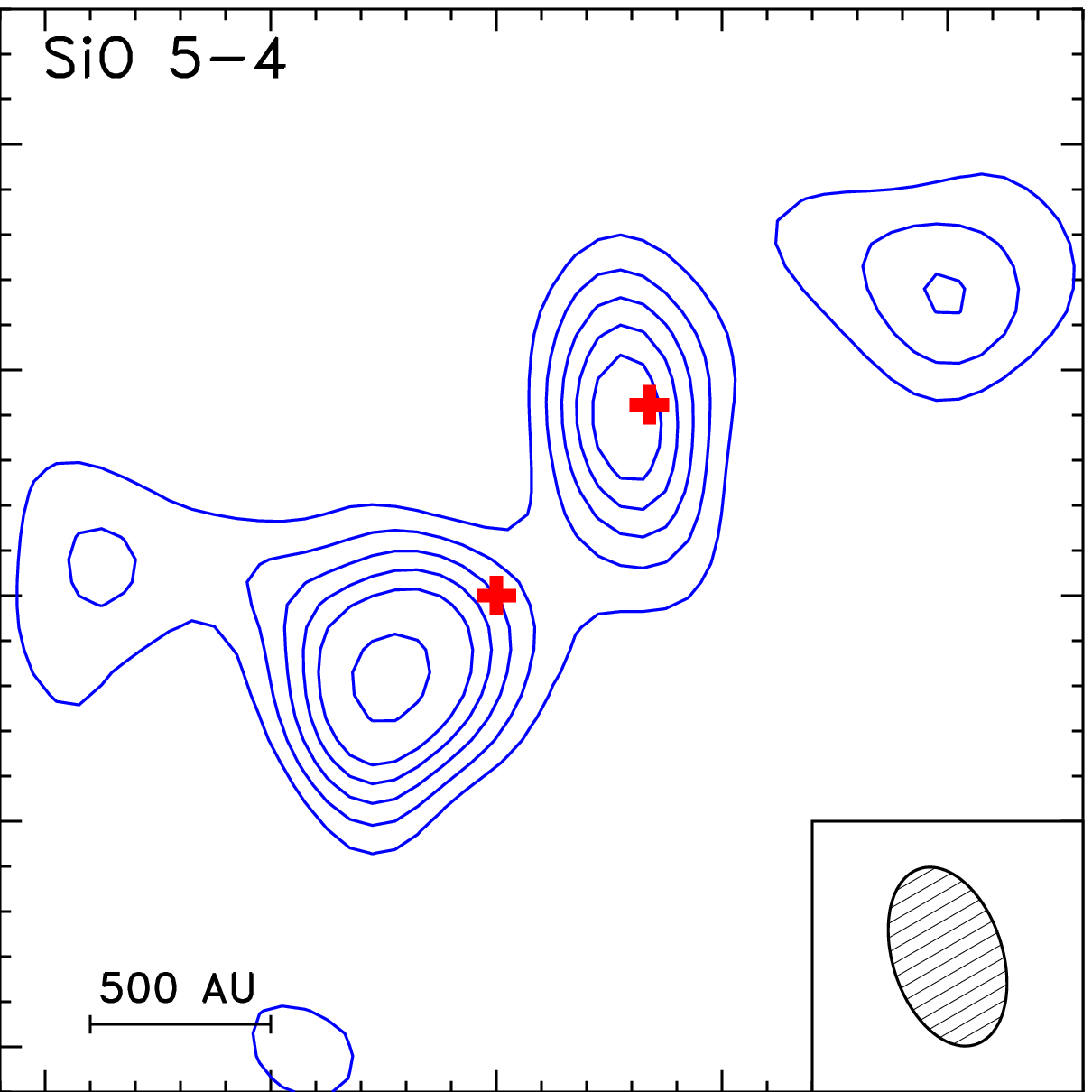}}
\resizebox{0.245\hsize}{!}{\includegraphics{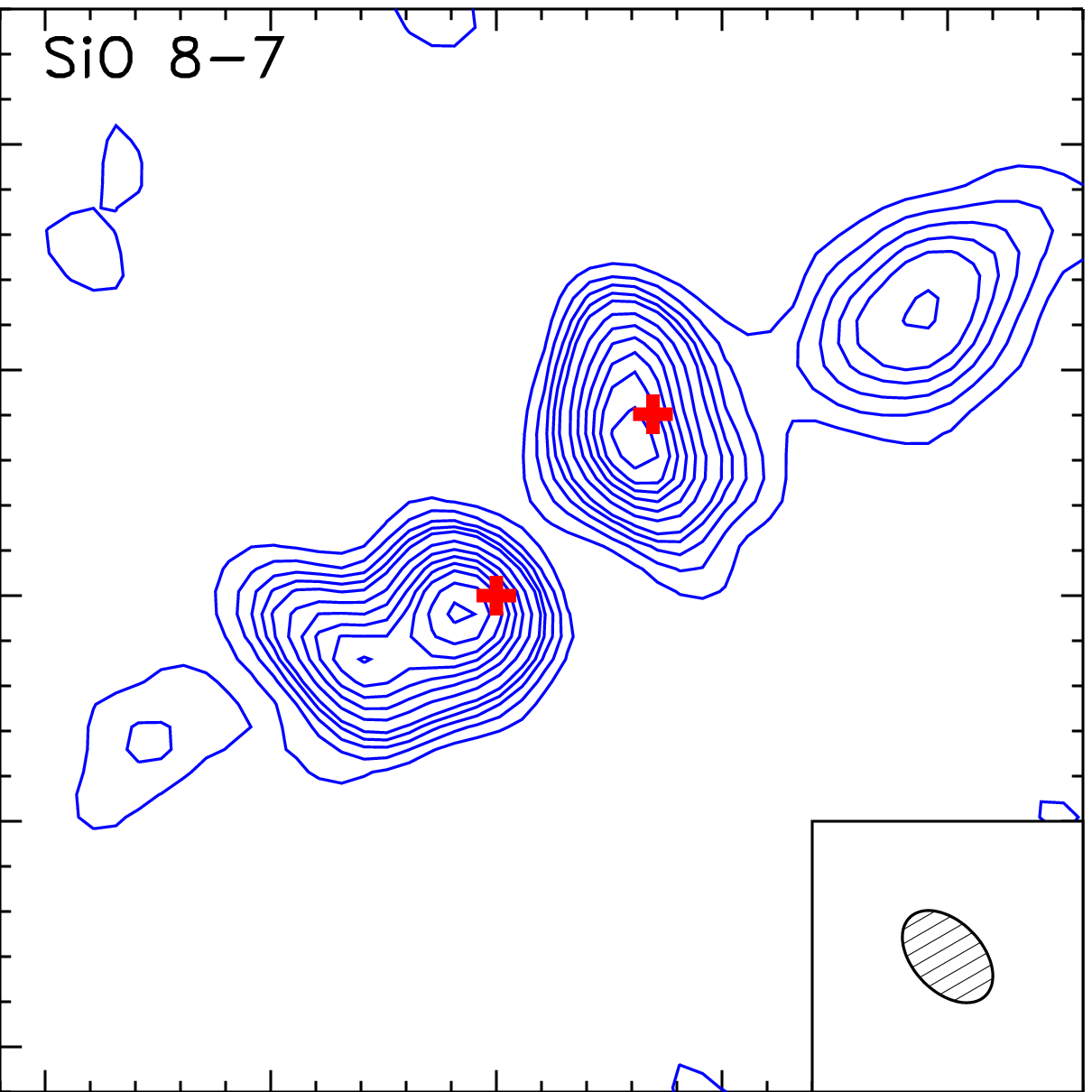}}
\resizebox{0.245\hsize}{!}{\includegraphics{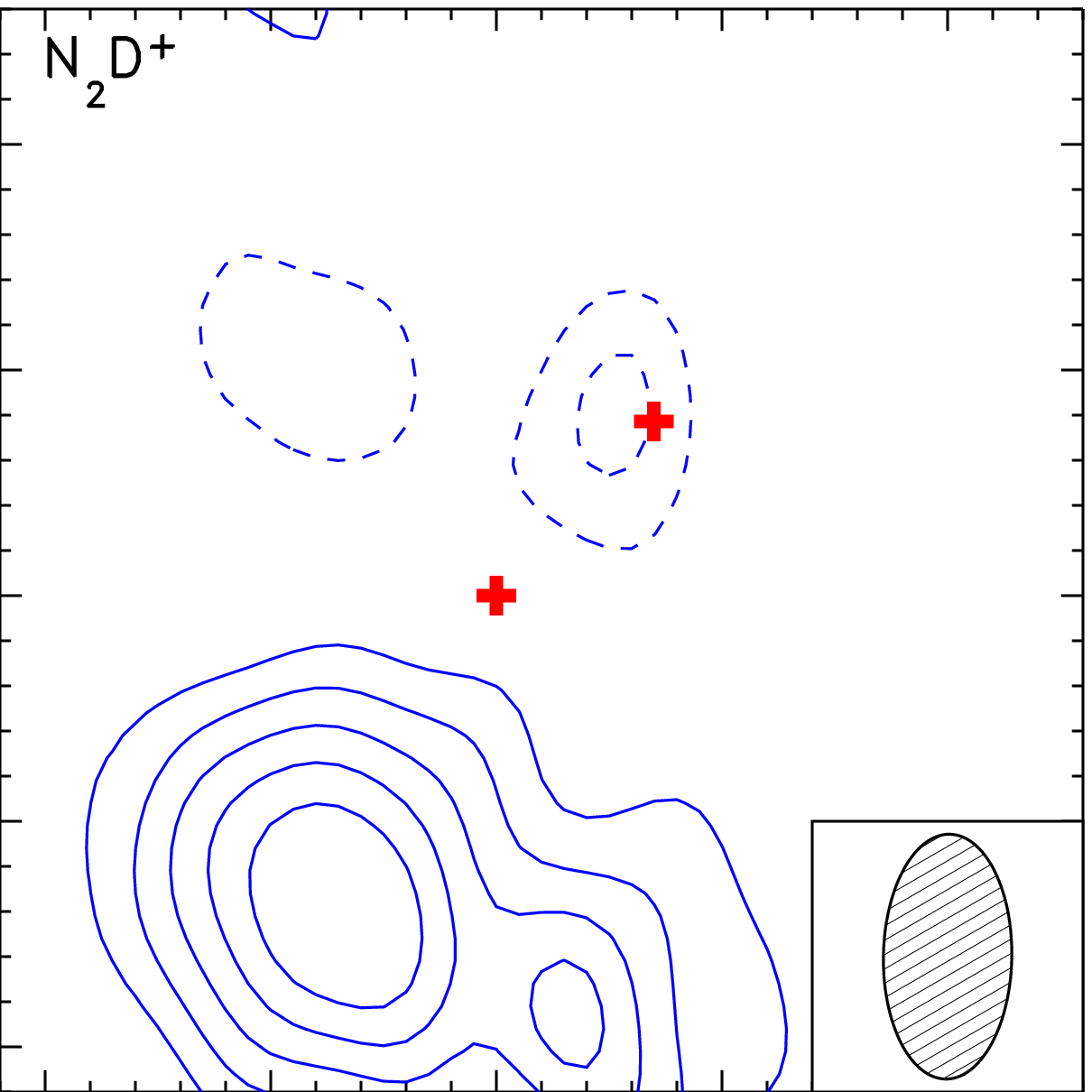}}
\resizebox{0.245\hsize}{!}{\includegraphics{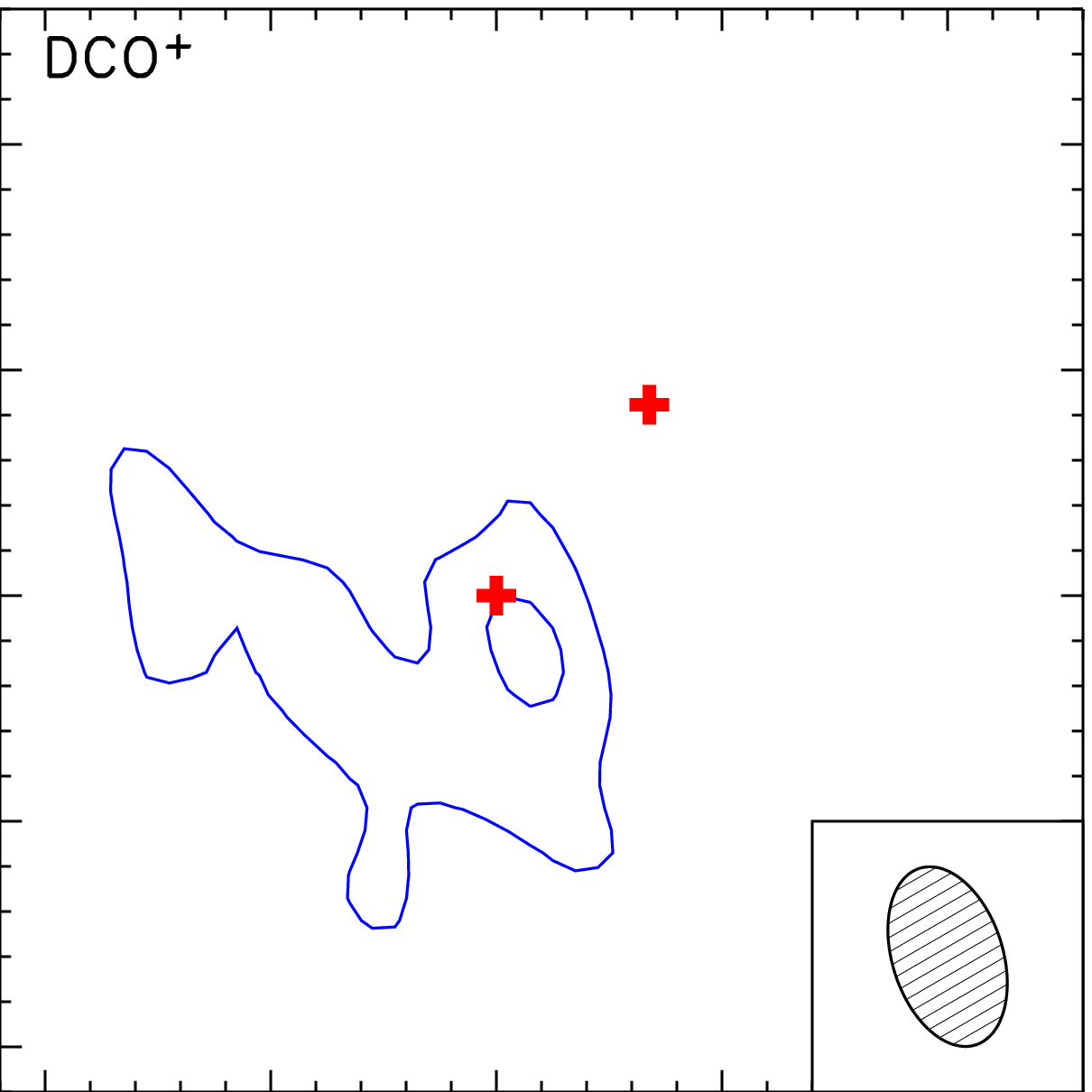}}
\resizebox{0.245\hsize}{!}{\includegraphics{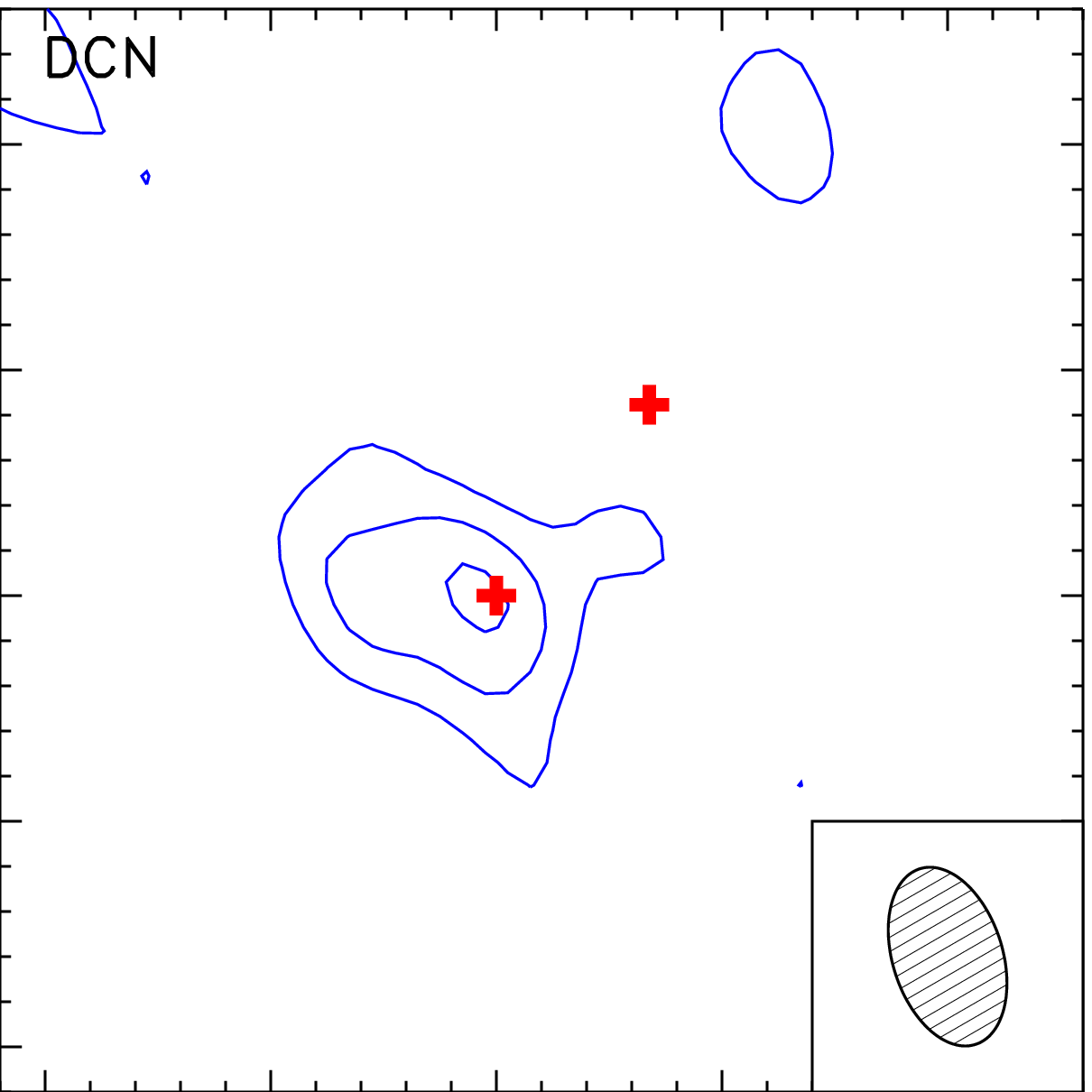}}
\resizebox{0.245\hsize}{!}{\includegraphics{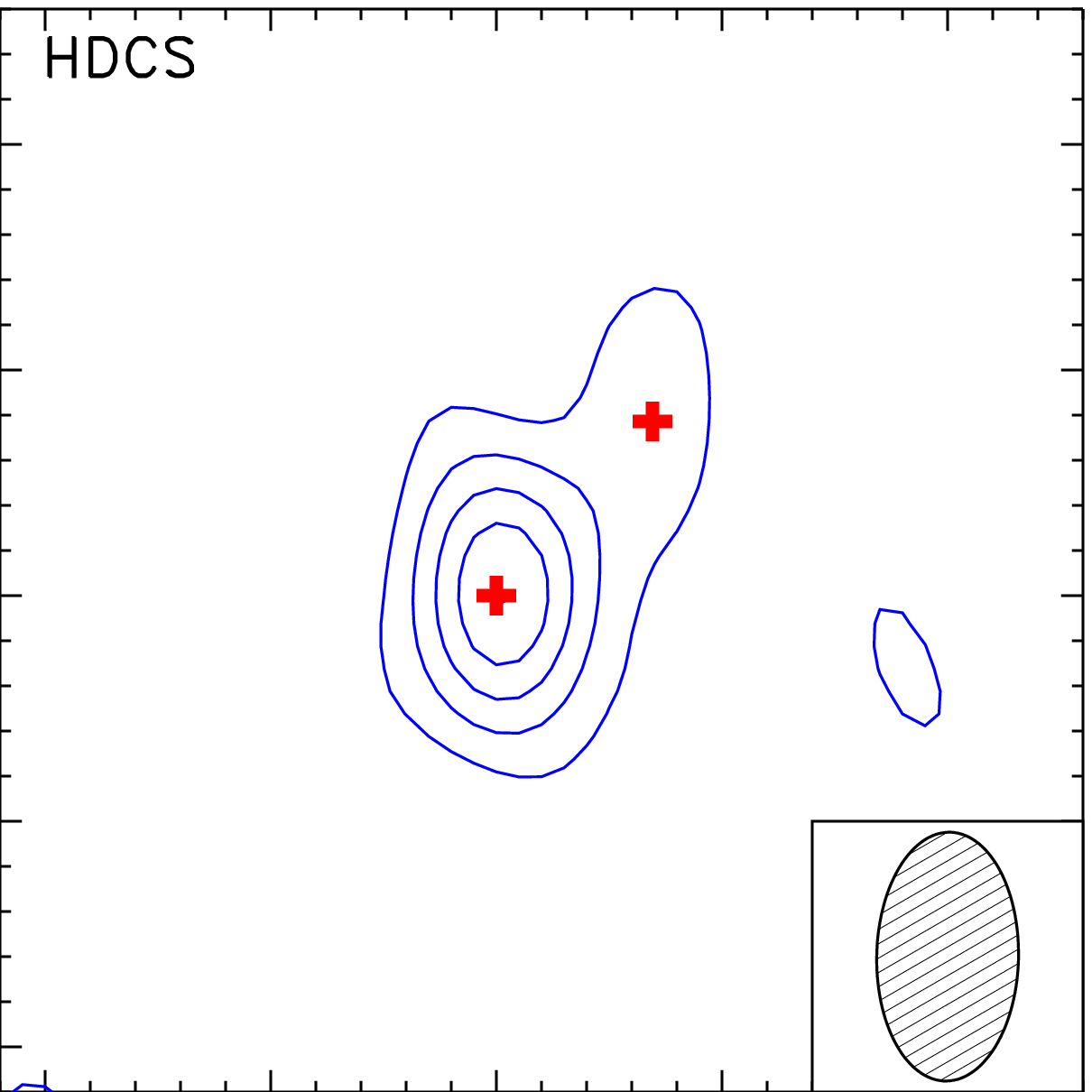}}
\resizebox{0.245\hsize}{!}{\includegraphics{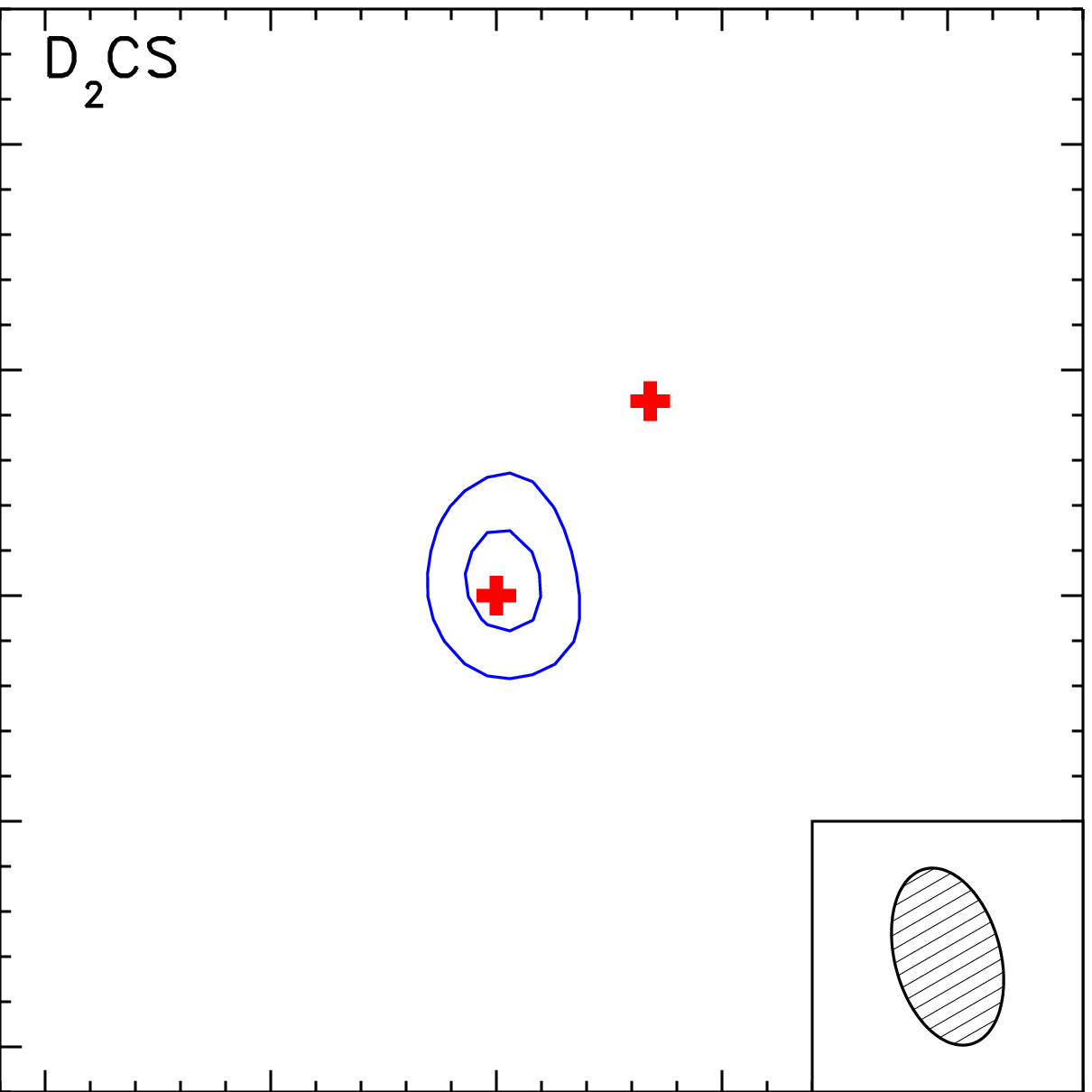}}
\resizebox{0.245\hsize}{!}{\includegraphics{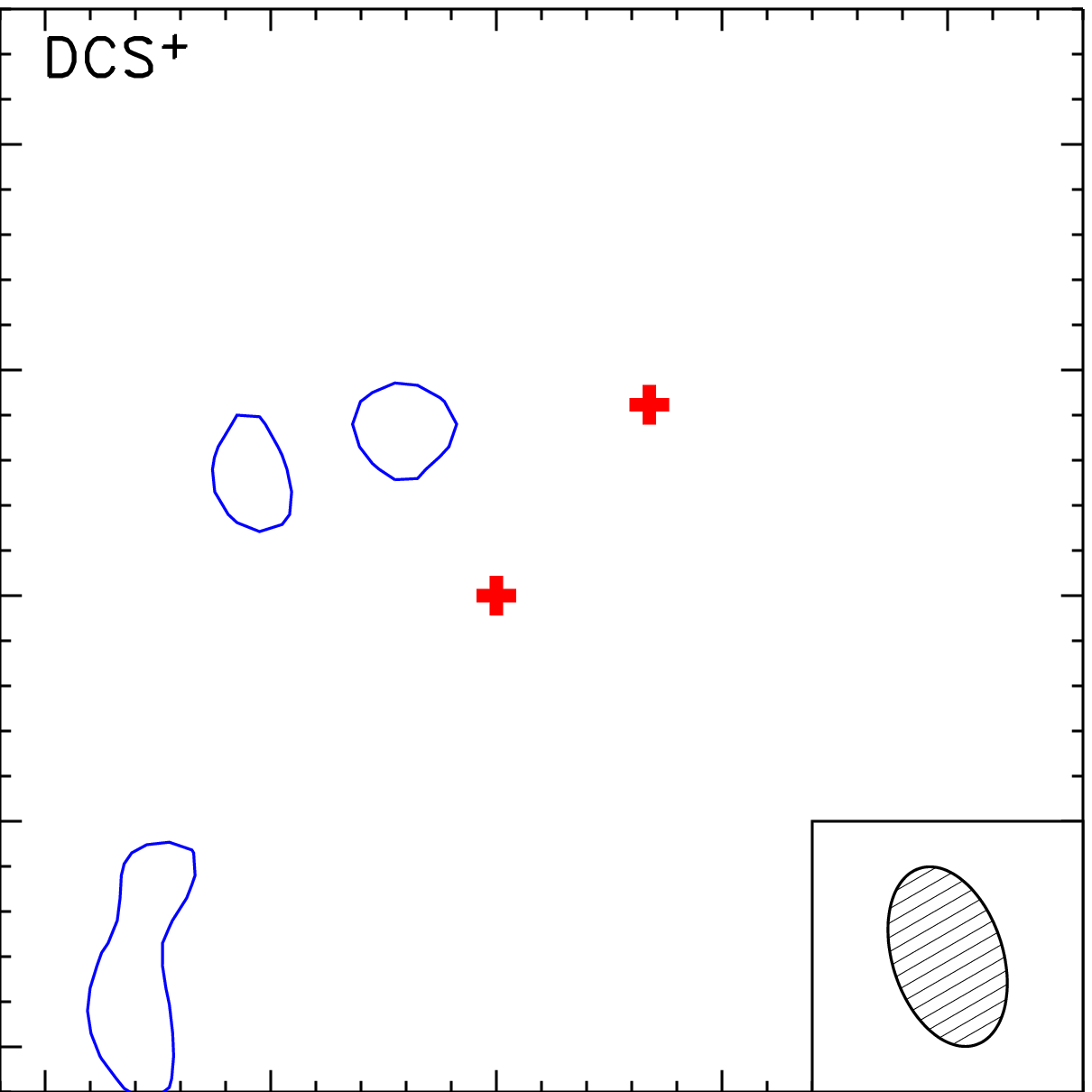}}
\resizebox{0.245\hsize}{!}{\includegraphics{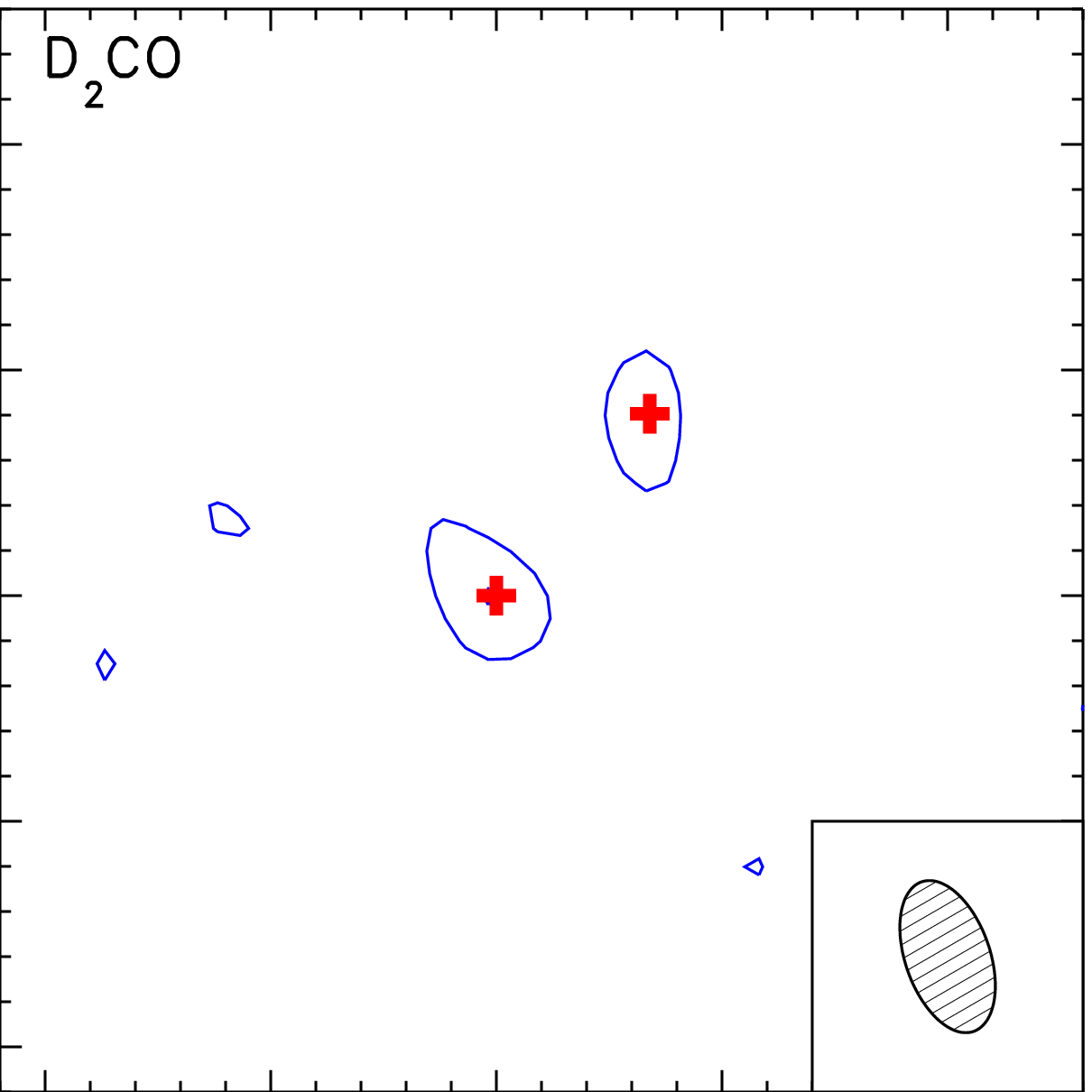}}
\resizebox{0.245\hsize}{!}{\includegraphics{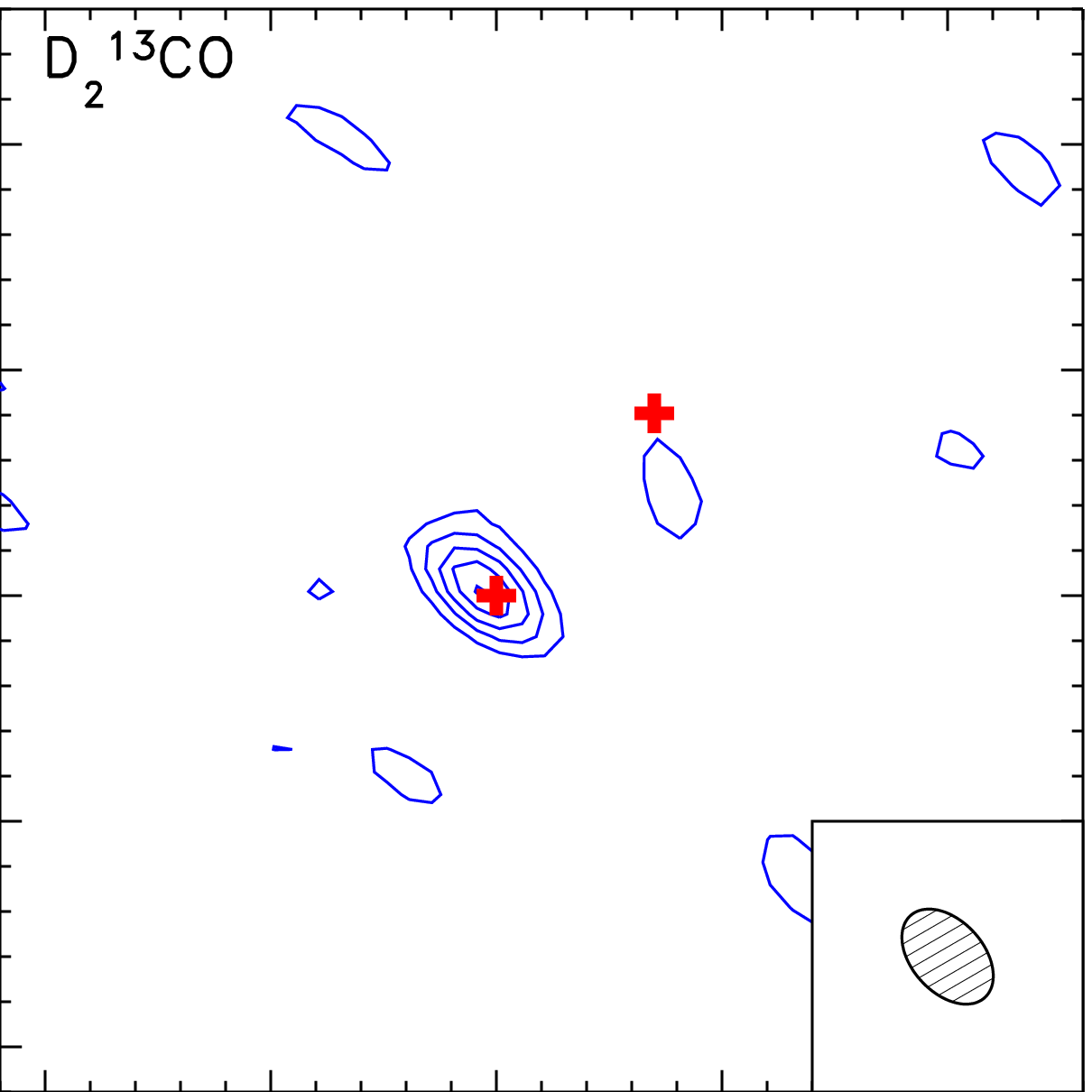}}
\resizebox{0.245\hsize}{!}{\includegraphics{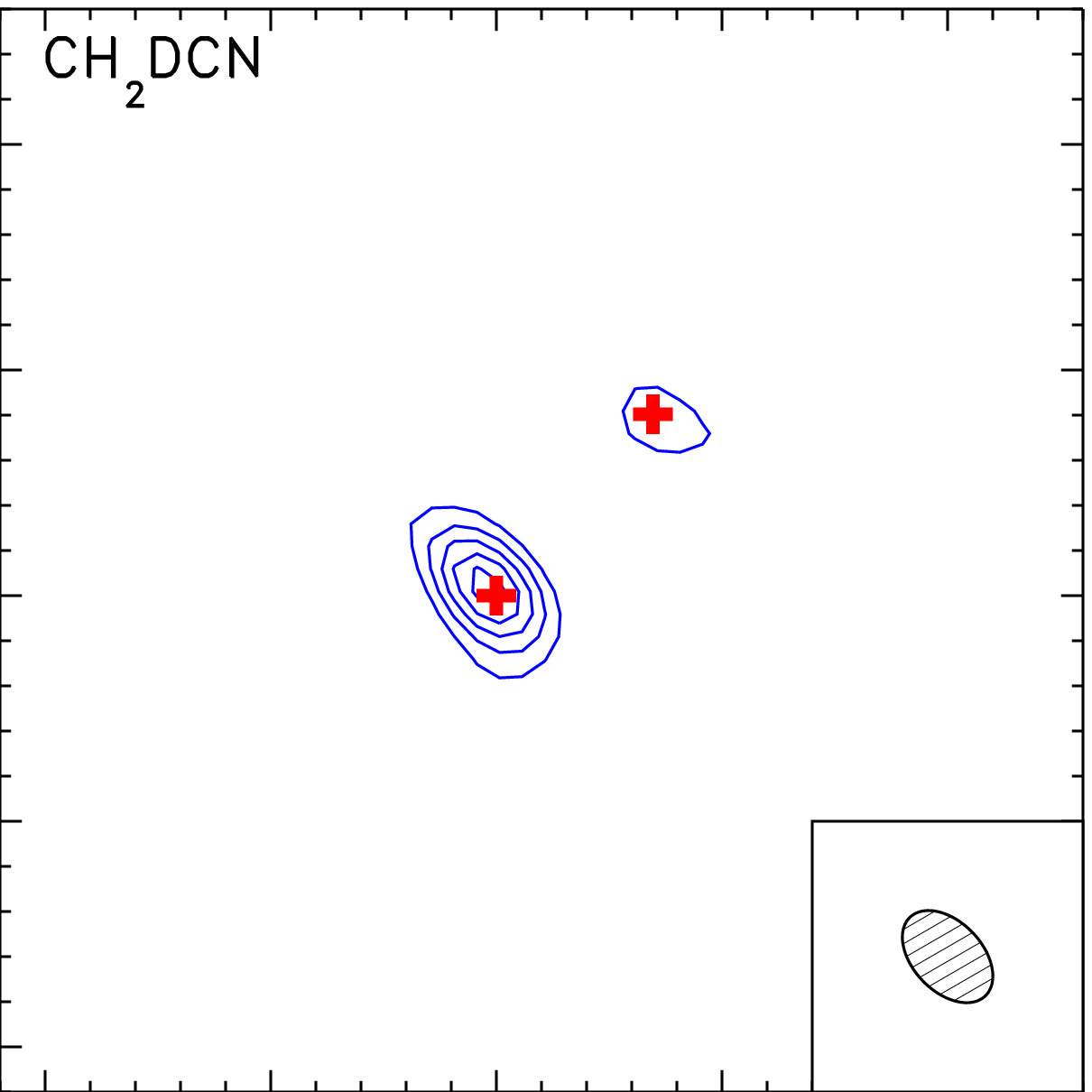}}
\resizebox{0.245\hsize}{!}{\includegraphics{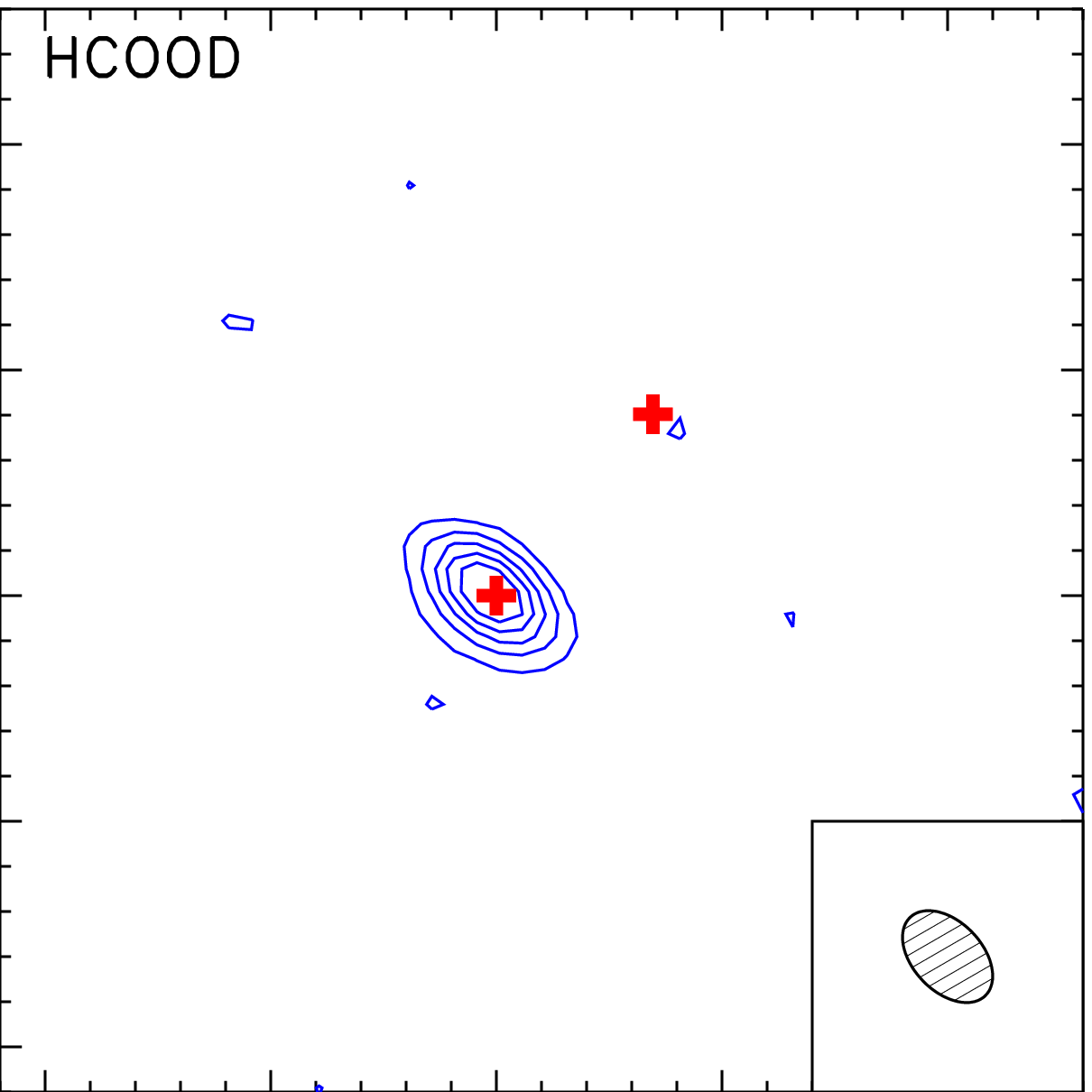}}
\resizebox{0.245\hsize}{!}{\includegraphics{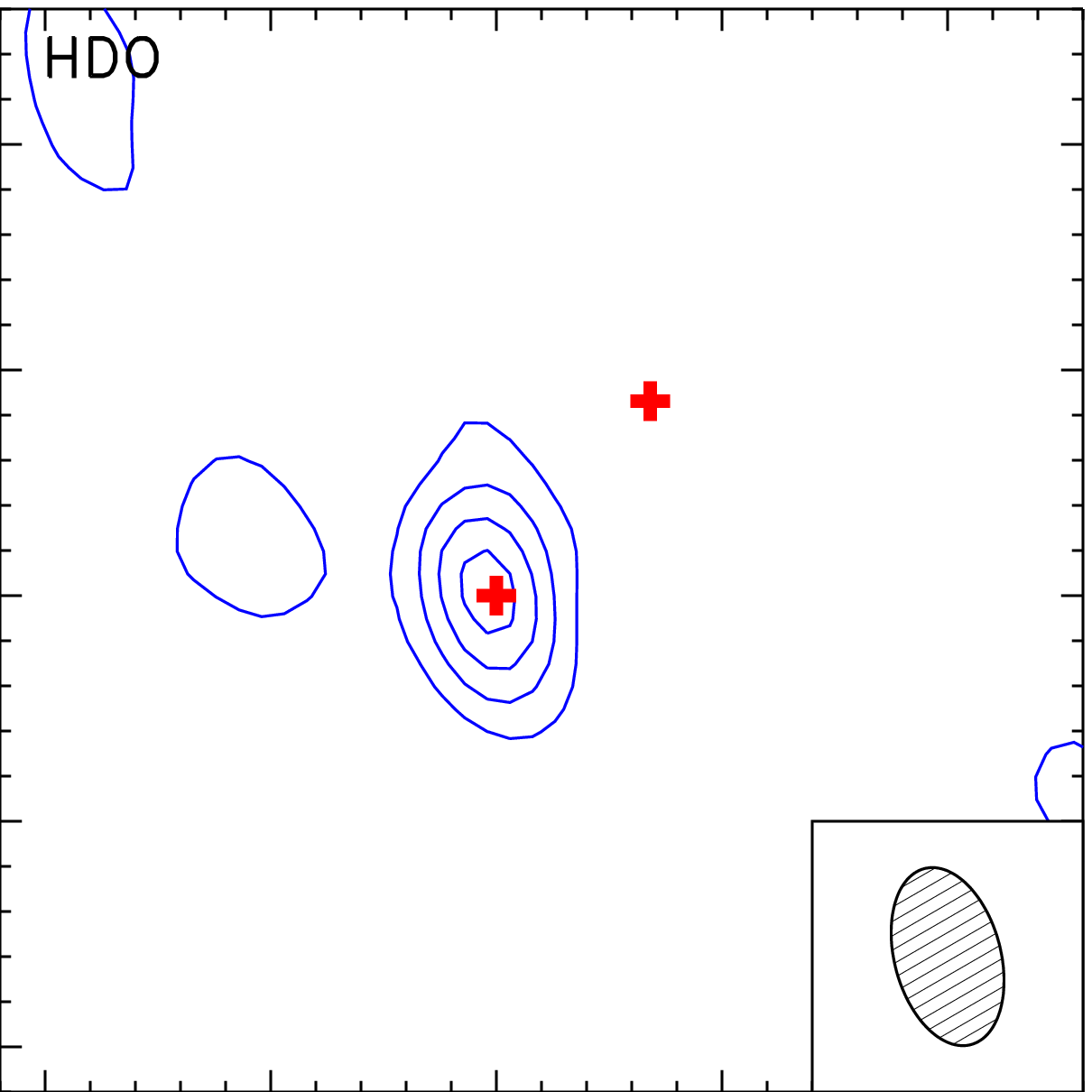}}

\phantom{xxx}
\caption{As in Fig.~\ref{image_first} for SiO and the deuterated
  molecules. For the N$_2$D$^+$ panel the observed ``absorption''
    feature is illustrated by the dashed contours corresponding to
    negative contour levels.}\label{image_last}
\end{minipage}
\end{figure*}
\clearpage

\section{Results and discussion}\label{discussion}
In this section we discuss the general features of the detected
emission before going into a few more specific topics in
detail. Generally the integrated line emission maps of IRAS~16293-2422
can be divided into groups:
\begin{enumerate}
\item Maps with significant extended emission encompassing both
  sources but without clear peaks or with peaks offset from the
  continuum sources. Examples: CO, SiO and H$_2$CO $5_{1,5}-4_{1,4}$.
\item Maps with significant extended emission peaking on one or both
  continuum sources. Examples: $^{13}$CO, C$^{18}$O, C$^{17}$O, CS and
  isotopologues, SO, H$_2$CO / H$_2^{13}$CO $3_{1,2}-2_{1,1}$,
  H$_2$C$^{18}$O.
\item Maps with localized peaks at the continuum sources, with
  IRAS~16293A being significantly stronger than IRAS~16293B (the latter
  in a few cases even absent). Examples: nitrogen- and sulfur-bearing
  species (including HC$_3$N, CH$_3$CN, HNCO, isotopologues of SO,
  SO$_2$, OCS among others), most deuterium-bearing species, CH$_3$OH
  and its $^{13}$C isotopologue.
\item Maps with localized peaks at the continuum sources, with the two
  peaks being approximately similar in strength. Examples: 
  CH$_3$OCH$_3$, CH$_3$OCHO, D$_2$CO.
\item Maps with localized peaks at the continuum sources, with
  IRAS~16293B source being stronger than IRAS~16293A. Examples:
  CH$_3$CHO, CH$_2$CO.
\item Maps with extended faint emission not strongly correlated with
  the continuum peaks. Examples: N$_2$D$^+$, H$^{13}$CO$^+$, DCO$^+$.
\end{enumerate}

\begin{figure}
\resizebox{\hsize}{!}{\includegraphics{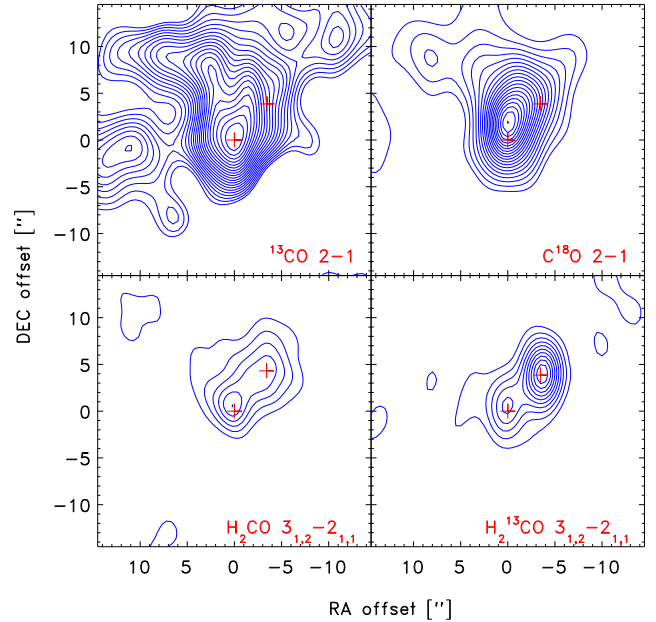}}
\caption{Comparison between $^{13}$CO 2--1 (upper left), C$^{18}$O
  2--1 (upper right), H$_2$CO $3_{1,2}-2_{1,1}$ (lower left) and
  H$_2^{13}$CO $3_{1,2}-2_{1,1}$ (lower right). For all species the
  emission is integrated from 1 to 5~\kms. The contours are shown in
  steps of 9~Jy~beam$^{-1}$~\kms\ for the CO isotopologues and
  3~Jy~beam$^{-1}$~\kms\ for the H$_2$CO isotopologues.}\label{h2co_co}
\end{figure}

These maps underline the complicated structure of the protostellar
system: the overlap between the emission in, e.g., the extended dense
gas tracers suggest that the two sources are embedded in one
larger-scale connected envelope, whose column density peaks close to
IRAS~16293A. A fainter local maximum in column density is present
toward IRAS~16293B, however. The absence of this secondary peak in,
e.g., the maps of $^{13}$CO and C$^{18}$O (Fig.~\ref{h2co_co}, upper
panels) suggest that these species are becoming optically thick on
scales close to the spatial scales probed by the interferometric
observations. A particular clear example of that is seen in the
emission from $3_{1,2}-2_{1,1}$ transitions of H$_2$CO and
H$_2^{13}$CO: for those two species an inversion is seen with the main
H$_2^{12}$CO isotopologue being stronger toward IRAS~16293A while the
fainter H$_2^{13}$CO isotopologue is stronger toward IRAS~16293B
(Fig.~\ref{h2co_co}, lower panels). The two transitions also appear
close to identical in strength in the maps -- further indicating a high
optical thickness of the main isotopologue and significant
resolved-out emission. Optical depth effects and general surface
brightness sensitivity may also explain the differences between for
example the main isotologue of SO and the fainter species ($^{33}$SO,
$^{34}$SO): the latter may also be extended similar to the main
isotopologue, but with too low surface brightnesses to be picked up by
the SMA observations. A few species show very strongly differentiated
emission between the two sources. The sulfur-species (SO$_2$ in
  particular) are strongly concentrated toward IRAS~16293A while a few
  of the complex organics (CH$_3$CHO in particular) appear most
  prominently toward IRAS~16293B. As the lines of the different
  species are not significantly different in excitation temperatures,
  these differences in their relative prominences point to different
  chemical structures of the two components in IRAS~16293-2422 --
  possibly reflecting differences in their physical evolution.

\subsection{Cold envelope chemistry}
Although a natural focus in the submillimeter interferometric
observations is on the higher excited lines, warmer gas and the
small-scale structure close to protostars themselves, the high
resolution offered by the interferometric observations provides
interesting insights into the chemistry in the colder envelope on
large scales. The survey includes a number of species that
predominantly are present in the cold 20--30~K gas and/or very
sensitive to changes in chemistry occurring at these temperatures.

\begin{figure}
\resizebox{\hsize}{!}{\includegraphics{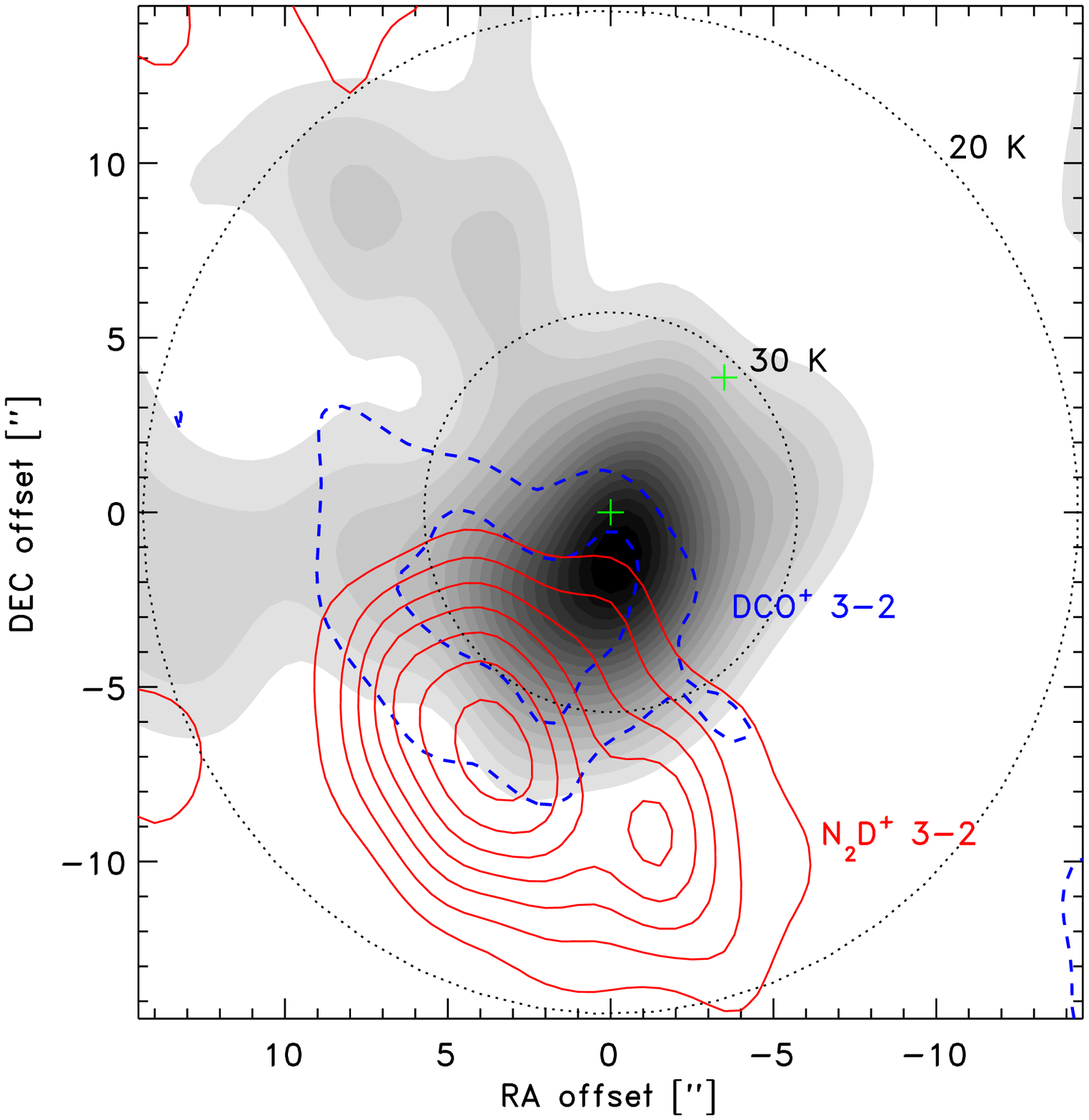}}
\caption{Comparison between the C$^{18}$O 2--1 (grey-scale), DCO$^+$
  3--2 (dashed blue contours) and N$_2$D$^+$ 3--2 toward the
  core. The lines are integrated from velocities of 3--7~\kms\
    (C$^{18}$O and N$_2$D$^+$) and 4--6~\kms\ (DCO$^+$). The dotted
  circles indicate the projected radii, where the dust temperature
  from radiative transfer calculations
  \citep{schoeier02,iras16293letter} has dropped to 30~K and 20~K,
  respectively -- assuming that the IRAS~16293A component is the sole
  source of luminosity and that the envelope is centered on this
  source.}\label{coldchem}
\end{figure}

One clear example is offered by a comparison between C$^{18}$O,
DCO$^+$ and N$_2$D$^+$ (Fig.~\ref{coldchem}). The emission from these
three species are all mainly associated with IRAS~16293A, and all show
significant emission extending over 5--10\arcsec\ scales.  C$^{18}$O
is the species located closest to the continuum peak.  DCO$^+$ is also
present there, but shows its maximum offset by 3--4\arcsec. The
N$_2$D$^+$ transition in contrast does not show any emission at the
continuum/C$^{18}$O peaks, but is offset in the same direction of
DCO$^+$ with its peak shifted even further. Of course, the
``absorption feature'' of N$_2$D$^+$ toward IRAS~16293B indicates that
this species is also present along line of sights toward the central
protostars, but simply resolved-out. It is likely that something
similar is the case for DCO$^+$ as well. Still, the maps reveal the
differences between the brightest spots in the emission in each of the
molecular species, reflecting that their underlying spatial
distributions also differ.

Fig.~\ref{coldchem} shows the temperature from self-consistent dust
radiative transfer models of IRAS~16293-2422
\citep{schoeier02,iras16293letter}. In those papers, the structure of
IRAS~16293-2422 is modeled as a single spherical protostellar dust
envelope heated by a central source of luminosity. Using the spectral
energy distribution and submillimeter continuum maps to constrain the
envelope density structure, the models then calculate the temperature
distribution as function of radius self-consistently. For this plot,
we locate the envelope at the position of IRAS~16293A and assign half
the total luminosity to this source. It is seen that the
differentiation between CO, DCO$^+$ and N$_2$D$^+$ is taking place at
scales corresponding to the radii of 20--30~K in the envelope. The
comparison to these spherical models is naturally a simplification
because of the binarity of the system: the dust and gas in the
northern-western part envelope is likely affected by the presence of
IRAS~16293B, as also suggested by the absence of the line emission
there.

\begin{figure}
\resizebox{\hsize}{!}{\includegraphics{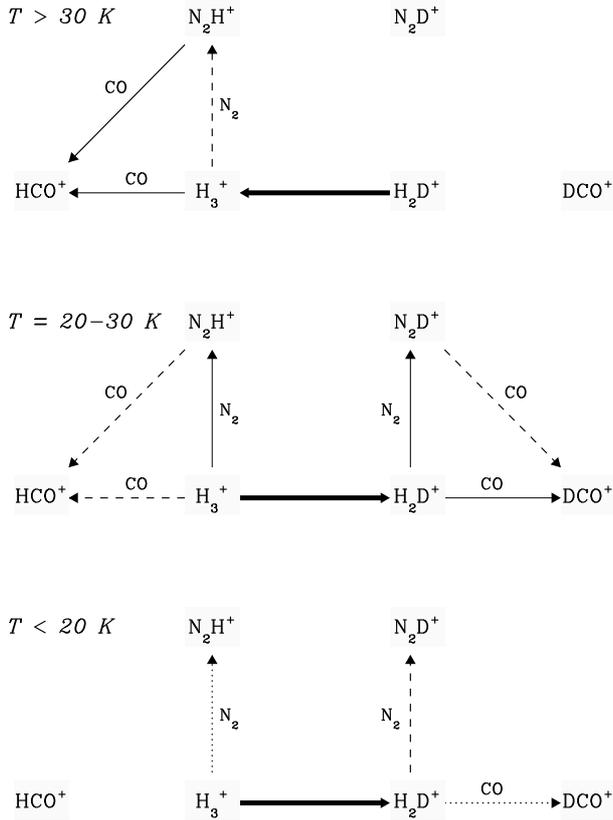}}
\caption{Schematic chemical network for HCO$^+$, DCO$^+$, N$_2$D$^+$
  and N$_2$H$^+$ in different regions of protostellar envelopes: at
  high temperatures ($T > 30$~K) CO is in the gas-phase and H$_3^+$ is
  dominating over H$_2$D$^+$ (the reaction in Eq.~1 proceeds to the
  left) enhancing HCO$^+$ and destroying N$_2$H$^+$. At intermediate
  temperatures ($T = 20-30$~K) CO depletion and H$_2$D$^+$ production
  starts to kick-in (the reaction in Eq.~1 proceeds to the right),
  enhancing N$_2$H$^+$ (less destruction through CO) and the
  deuterated species (DCO$^+$ in particular, through reactions between
  H$_3^+$ and CO). At low temperatures CO depletion is very
  significant and N$_2$D$^+$ is enhanced relative to the other species.}\label{d_network}
\end{figure}

The differentiation between C$^{18}$O, DCO$^+$ and N$_2$D$^+$ can be
understood in the context of the gas-phase chemistry of the cold
envelope gas (Fig.~\ref{d_network}): CO has been shown to be freezing
out significantly at low temperatures in the outer regions of
protostellar envelopes, desorbing off the dust grains once the
temperature increases to $\approx$~30~K
\citep[e.g.,][]{jorgensen02,coevollet}. This gives a primary peak of
the C$^{18}$O emission associated with the compact continuum emission
marking the location of the protostar itself as seen in
Fig.~\ref{coldchem}. At larger radii and low temperatures, the
freeze-out of CO gives a boost to DCO$^+$ and N$_2$D$^+$: DCO$^+$ is
tied to the CO abundance through the reaction ${\rm H}_2{\rm D}^+ +
{\rm CO} \rightarrow {\rm DCO}^+$. This takes place most efficiently
at temperatures lower than 30~K where the amounts of H$_2$D$^+$
increases due to the small energy difference in the reaction:
\begin{equation}
{\rm H}_3^+ + {\rm HD} \rightarrow {\rm H}_2{\rm D}^+ + {\rm H}_2
\end{equation}
\citep{roberts00a}. At temperatures higher than 30~K in contrast,
H$_2$D$^+$ is rapidly transformed into H$_3^+$ through reactions with
H$_2$ and then further incorporated into HCO$^+$ through reactions
with CO. This effect is similar for N$_2$D$^+$ -- except that it is
even further enhanced by the fact that any CO present in the gas will
work very efficiently in destroying N$_2$D$^+$, thus limiting it even
further (N$_2$H$^+$ is also enhanced in the colder, CO depleted, gas --
because of the lack of its primary destruction agent; see, e.g.,
\citealt{paperii}). One would therefore expect a sequence in the
presence of these three species with CO being present most closely to
the center, DCO$^+$ following this at slightly lower temperatures and
N$_2$D$^+$ at the lowest temperatures (Fig.~\ref{d_network}). This scenario is in qualitative agreement with what is seen in the
  SMA data with the species peaking at the distances expected from the
  temperature of the protostellar envelope given its temperature
  profile in a simple spherical model (Fig.~\ref{coldchem}).

\subsection{The importance of the outflows in IRAS~16293-2422}
The outflow activity in IRAS~16293-2422 has been the topic of many
discussions in literature -- both in terms of the larger scales probed
by single-dish observations and on smaller scales trying to identify
what outflows are driven by each of the components. CO line emission
toward IRAS~16293-2422 show a characteristic quadrupolar structure
\citep{walker88}, which could reflect its binary
nature. \cite{stark04} suggested that this quadrupolar morphology
could be interpreted as being a superposition of an older outflow in
the East-West direction driven by IRAS~16293B and a younger outflow in
the Northeast/Southwest direction driven by IRAS~16293A. This led
these authors to suggest that the IRAS~16293B in fact was the older of
these two sources -- possibly a low luminosity T Tauri star. This view
has, however, been challenged by high resolution CO observations
showing that the East-West outflow is unambiguously associated with
IRAS~16293A \citep[e.g.,][]{yeh08}. On the smallest scales in the
system, it remains an important question if the outflows play a role
in regulating the temperature and density structure: through the high
angular resolution CO 2--1 and 3--2 maps \cite{yeh08} for example
showed the existence of two bright spots in their outflow maps offset
by about 1$''$ from the two protostars and toward IRAS16293A in
particular, \cite{chandler05} showed that excitation transitions of a
number of other species were offset likely due to the impact of this
outflow (in particular, their Fig.~21).

Besides CO often used tracers of outflow emission are the transitions
of SiO (in particular, $J=$2--1, 5--4 and 8--7) thought to be the
result of silicon in atomic form sputtering off dust grains and
reacting with OH in the gas-phase producing SiO in the process
\citep[e.g.,][]{schilke97}. The larger scale environment of
IRAS~16293-2422 has been mapped in the line of SiO 2--1 with
single-dish telescopes \citep{hirano01,castets01} and used for
discussions of the relation between the SiO emission and the
protostellar outflows.

\cite{ceccarelli00a} discussed multi-transition single-dish
observations of the SiO emission at the location of the central
protostellar binary and analyzed the emission in the context of models
of a spherical collapsing envelope. They demonstrated that the SiO
emission could be explained within this model if the SiO abundance
increased from a low (``molecular cloud'') value of $4\times 10^{-12}$
to a ``warm'' value of $1.5\times 10^{-8}$. They noted that in this
model, the bulk of the observed SiO 5--4 emission arises at a distance
less than 150~AU from the central source where the infall velocity is
large, about 2.8~\kms\ in their model, consistent with the large line
width seen in the single-dish observations.

The observations presented here show that this model does not provide
an adequate description of the emission morphology of these SiO
transitions. In fact, both the $J=5-4$ and $8-7$ transitions show
emission significantly extended on scales comparable to typical
single-dish beams extending over the two protostellar sources with
localized peaks close to the two protostars. In IRAS~16293A, the SiO
emission peaks slightly east of the central protostar close to the
location of the bright CO spot reported by \cite{yeh08}, whereas the
emission toward IRAS~16293B is slightly extended around the position
of the central protostar and the CO bright spot associated with that
source -- but not uniquely associated with either. Generally, very few
low-mass protostars show significant SiO line emission
\citep[e.g.,][]{prosacpaper}, so the strong detection toward
IRAS~16293-2422 \citep[e.g.,][]{blake94} is notable in itself. SiO is
thought to be produced by sputtering of silicon of dust grains in
shocks, reacting with OH once in the gas-phase. This would suggest
that the shock activity is prominent both in the vicinity of
IRAS~16293A and IRAS~16293B -- despite the absence of a clear CO
outflow driven by the latter. It is therefore likely that the outflow
driven shock is affecting the physics and chemistry of the gas in the
immediate vicinity of IRAS~16293B. Higher resolution observations of
the kinematics of the gas at this position are needed to clarify the
importance of these shocks compared to the radiation by the embedded
protostar.
\begin{figure}
\resizebox{0.95\hsize}{!}{\includegraphics{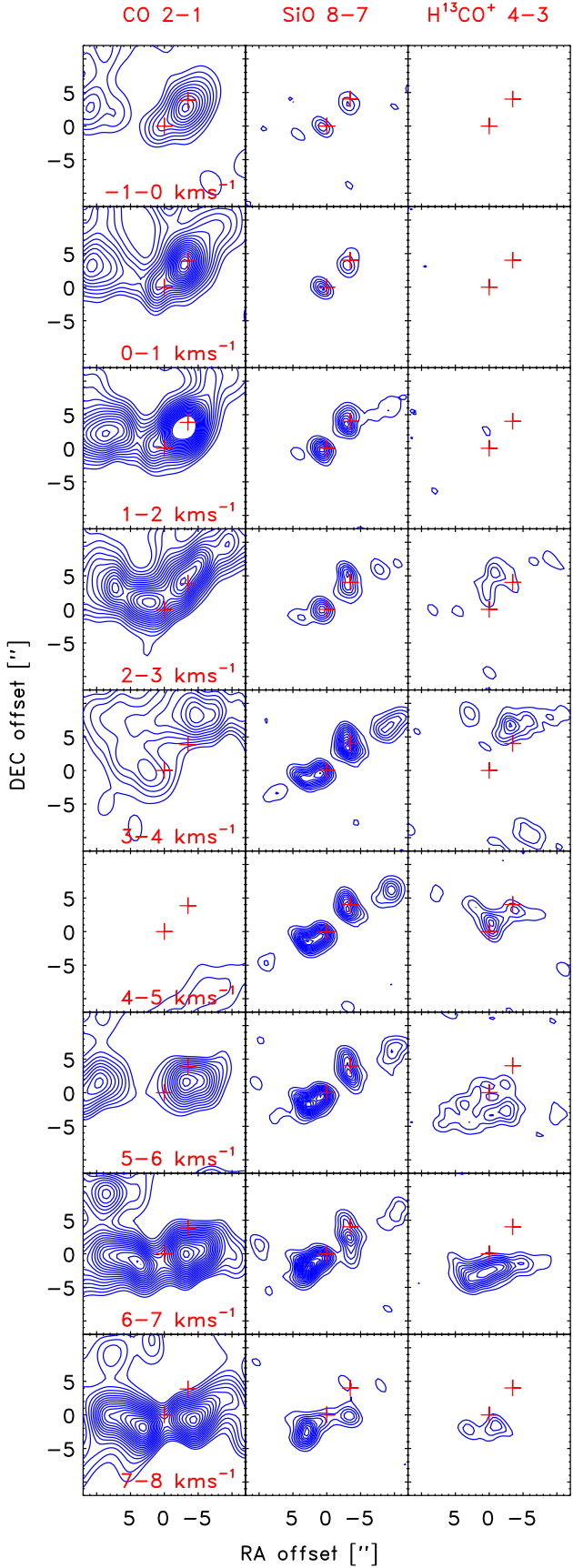}}
\caption{Channel maps for CO 2--1 (left column), SiO 8--7 (middle column) and
  H$^{13}$CO$^+$ 4--3 (right column). Contours are given in steps of
  20$\sigma$ for CO 2--1 and 3$\sigma$ for SiO 8--7 and H$^{13}$CO$^+$
  4--3.}\label{outflow_channel}
\end{figure}

\subsubsection{HDO emission}
Clearly the outflows in the IRAS~16293-2422 have an impact on the
emission on large scales. Some care therefore needs to be taken in
interpreting (unresolved) single-dish observations of the system -- in
particular, for discussing molecules tracing small-scale structures. A
few examples, of species probing gas affected by the outflow emission
are CS, CH$_3$OH and HDO. The two former species are known to be
tracers of dense envelope gas affected by outflow emission in
protostellar systems \citep[e.g.,][]{hotcoresample}, but the latter is
particular noteworthy: the SMA data encompass the HDO $3_{12}-2_{21}$
transition at 225.896~GHz, a relatively high excitation transition of
HDO, which potentially could reveal the presence and distribution of
(deuterated) water in this low-mass protostar.

\begin{figure}
\resizebox{\hsize}{!}{\includegraphics{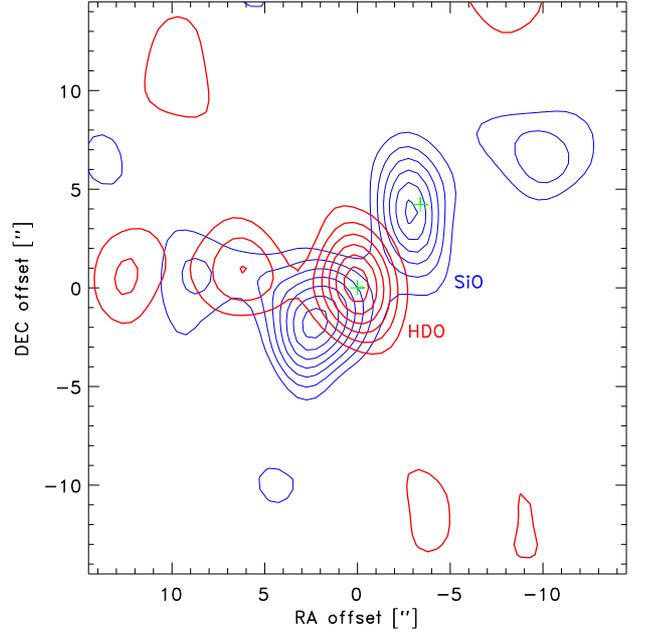}}
\caption{HDO (225.9~GHz; red) and SiO (217.1~GHz; blue) maps
  compared. Both
  datasets were integrated over the velocity range from $-$1 to
  6~\kms. The contours are given in steps of
  3$\sigma$.}\label{hdo_sio_compare}
\end{figure}

\begin{figure}
\resizebox{\hsize}{!}{\includegraphics{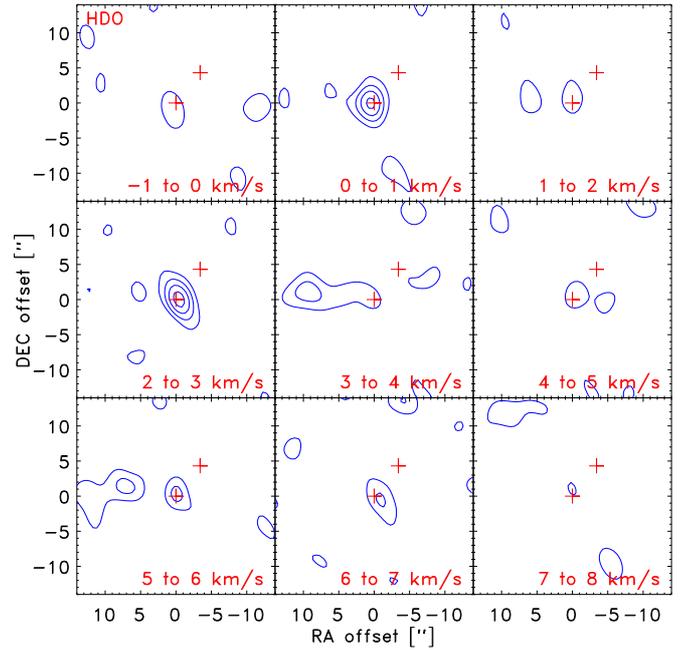}}
\caption{Channel maps (averaged over 2 channels / 1.1~\kms) of HDO
  (velocities given in upper right corner of each panel). The HDO data
  have had a taper of 3$''$ applied. }\label{hdo_channel}
\end{figure}

\begin{figure}
\resizebox{\hsize}{!}{\includegraphics{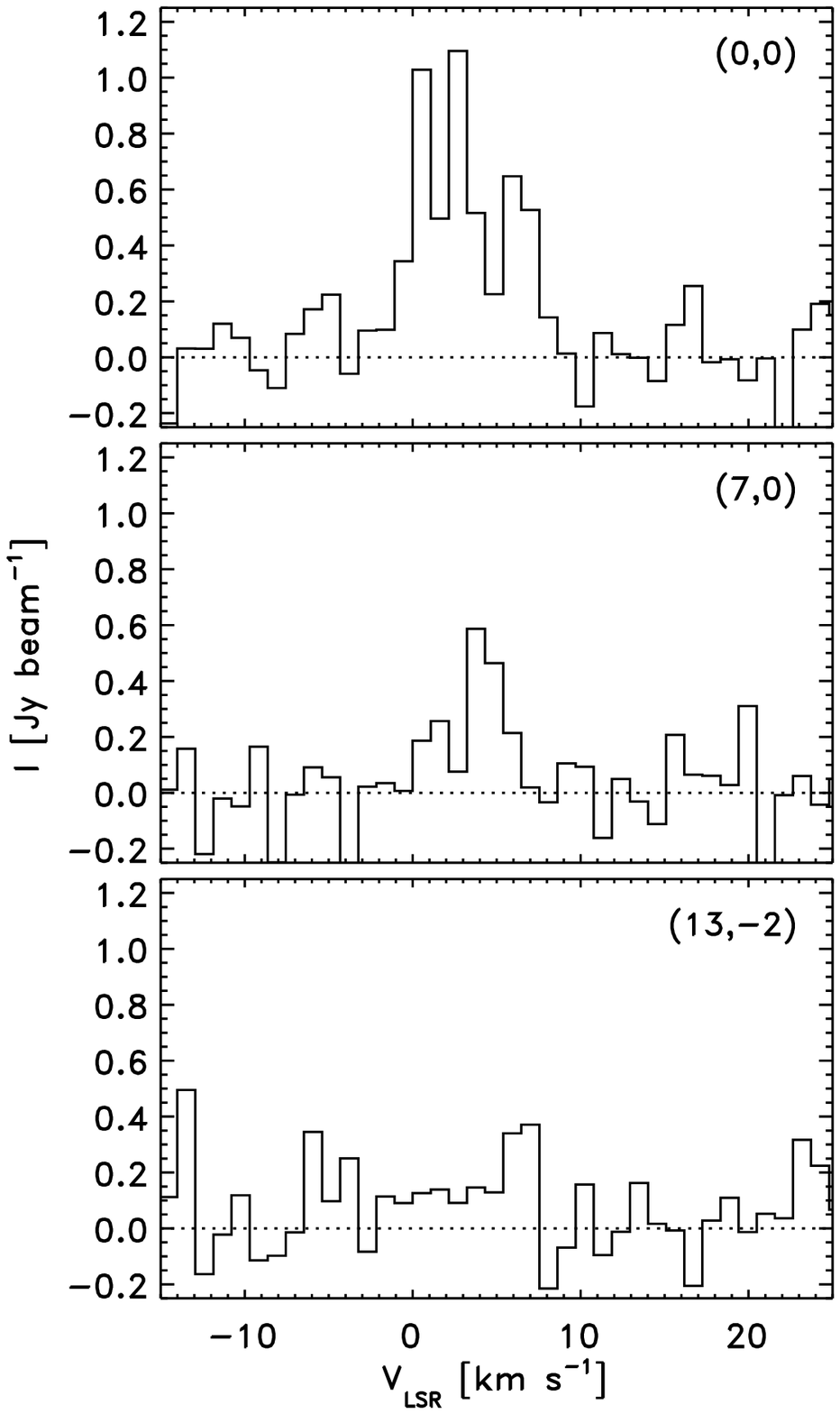}}
\caption{Representative spectra of HDO in the central beams from
    maps with a 3$''$ taper applied (see also Fig.~\ref{hdo_channel})
  toward the continuum peak of IRAS~16293A, the secondary peak and a
  position further offset toward the east where the emission is
  extended the offsets relative to the position of IRAS~16293A in
    arcseconds given in the upper right corner of each
    panel.}\label{hdo_spectra}
\end{figure}

Fig.~\ref{hdo_sio_compare} compares the integrated HDO emission to
that of SiO 5--4. The maps show two major peaks of the HDO line. One
is associated with I16293A and one offset at about (5$''$,0$''$).  A
third peak is seen at even larger distances -- but channel maps
(Fig.~\ref{hdo_channel}) suggests that this is an extension of the
secondary peak. Fig.~\ref{hdo_channel} shows the channel maps of the
HDO transition and Fig.~\ref{hdo_spectra} spectra toward these three
positions. The emission at continuum peak is seen most prominently at
the systemic velocity, $V_{\rm LSR}$, of the majority of the lines in
the data at about 4~\kms. The second peak east of the IRAS~16293A
continuum position is redshifted relative to this but still extending
around the systemic velocity. The morphology of the HDO emission
appear related although not directly correlated to that of SiO:
generally the HDO emission peaks appear in ``valleys'' of the
red-shifted lobe of the SiO outflow emission
(Fig.~\ref{hdo_sio_compare}).

\section{Summary}\label{summary}
This paper has presented a large Submillimeter Array imaging survey of
the line emission from molecular species toward the deeply embedded,
Class~0, protostellar binary IRAS~16293-2422 down to 1.5--3\arcsec\
(190--380~AU) resolution scales. We have identified the molecular line
emission toward each of the two components in the protostellar binary
and discuss some general features of the emission.

\begin{itemize}
\item 515 transitions of 54 species (including isotopologues) are
  identified. 90\% of the identified transitions have upper level
  energies lower than 550~K. A richness of both non-organic and
  organic molecules (of varying complexity) are seen toward both
  binary components.
\item Significantly narrower lines are observed toward IRAS~16293B
  than IRAS~16293A in the system (1.9~\kms\ for IRAS~16293B
  vs. 2.6~\kms\ for IRAS~16293A) and the lines toward IRAS~16293B are
  found to be blue-shifted by about 0.5~\kms\ compared to IRAS~16293A
  as well.
\item The molecular species show significantly different strengths
  relatively at the two components. IRAS~16293A in general has the
  stronger emission and shows many nitrogen and sulfur-bearing species
  weak or absent toward IRAS~16293B. The latter in contrast harbors a
  number of stronger transitions from oxygen-bearing complex organics
  -- in particular, CH$_3$CHO, which is very faint toward IRAS~16293A
  (see also \citealt{bisschop08}). There is no evidence for
  significantly different excitation conditions in the two sources,
  however.
\item Outflowing motions are clearly witnessed by the maps of CO (see
  also \citealt{yeh08}) and SiO. The latter shows extended emission
  toward IRAS~16293A but notably very compact emission near
  IRAS~16293B suggesting the presence of shocks in the immediate
  vicinity of this source.
\item A few specific examples are discussed. More extended emission
  from N$_2$D$^+$, DCO$^+$ and CO shows differences that can be
  attributed to cold gas-phase chemistry on large scales in the outer
  parts of the protostellar core. The emission from deuterated water
  (HDO) is only present toward IRAS~16293A with evidence for some
  water coming off dust grains in regions impacted by the protostellar
  outflow traced by SiO.
\end{itemize}

The data presented in this paper illustrate the potential of (and need
for) high angular resolution imaging when discussing the physics and
chemistry of nearby embedded low-mass protostars based on
submillimeter wavelength observations -- and caution against
over-interpretations based on lower resolution data in complex systems
such as IRAS~16293-2422. On the other hand, the wealth of information
in these and similar studies provide many independent, strong
constraints and could be used to shed light on some of the unanswered
questions concerning the physical and chemical structure and evolution
of low-mass protostars. Once the Atacama Large Millimeter Array (ALMA)
is fully operational with its large spectral bandwidth (up to 8~GHz in
each of two sidebands) and collecting area, similar types of
observations will be routinely done and thus provide a significant
boost to studies of the molecular astrophysics in star-forming
regions. The data presented in this paper will guide higher
sensitivity and resolution observations with ALMA, but also contain a
wealth of information in their own right. We therefore welcome
everyone to make use of these data and make them publicly available.

\acknowledgements We would like to thank the anonymous referee
  for a number of good suggestions helping to improve the presentation
  of this survey. This paper is based on data from the Submillimeter
Array: the Submillimeter Array is a joint project between the
Smithsonian Astrophysical Observatory and the Academia Sinica
Institute of Astronomy and Astrophysics and is funded by the
Smithsonian Institution and the Academia Sinica. It is a pleasure to
thank everybody involved with the Submillimeter Array for the
continued development of this observatory. The research of JKJ was
supported by a Junior Group Leader Fellowship from the Lundbeck
foundation. Research at Centre for Star and Planet Formation is funded
by the Danish National Research Foundation and the University of
Copenhagen's programme of excellence. Quang Nguyen Luong's research at
Bonn University was supported by a M.Sc. fellowship from Bonn
International Graduate School of Physics and Astronomy (BIGS) and
Argelander Institut f\"{u}r Astronomie.

\clearpage

\Online

\onecolumn
\clearpage

\begin{appendix}
\section{Compilation of all identified emission lines}

\begin{longtable}{lllrrrr}
\caption{Compilation of all identified emission lines.}\label{detectlines}\\
\hline\hline
{Frequency} & {Molecule} & {Transition} &
  {log($\mu^2$S)\tablefootmark{a}} & {$E_u$}
    [K]\tablefootmark{b} & {Flag\tablefootmark{c}} & {Plot\tablefootmark{d}} \\
\hline
\endfirsthead
\caption{continued.}\\
\hline\hline
{Frequency} & {Molecule} & {Transition} &
  {log($\mu^2$S)\tablefootmark{a}} & {$E_u$}
    [K]\tablefootmark{b} & {Flag\tablefootmark{c}} & {Plot\tablefootmark{d}} \\
\hline
\endhead
\hline
\endfoot
215.8399200 & $^{34}$SO                 & $ 5_{ 6 }  - 4_{ 5 }  $                  &   1.15  &  34.38     & A & +  \\
216.1097290 & CH$_{3}$OCHO-E            & $19_{ 2,18}-18_{ 2,17}$                  &   1.69  & 109.34     & B & -  \\
216.1125822 & DCO$^{+}$                 & $ 3        - 2        $                  &   1.66  &  20.74     & A & +  \\
216.1154858 & CH$_{3}$OCHO-A            & $19_{ 2,18}-18_{ 2,17}$                  &   1.69  & 109.32     & B & -  \\
216.2065780 & DCS$^{+}$                 & $ 6        - 5        $                  &   1.36  &  36.32     & A & +  \\
216.2947260 & CH$_{3}$CHO               & $11_{ 1,10, 5} - 10_{ 1, 9, 5}$          &   2.14  & 269.62     & B & -  \\
216.5686510 & H$_{2}$CO                 & $ 9_{ 1, 8}- 9_{ 1, 9}$                  &   0.54  & 173.99     & 1 & +  \\
216.5819360 & CH$_{3}$CHO               & $11_{ 1,10, 2} - 10_{ 1, 9, 2}$          &   2.14  &  64.87     & B & -  \\
\end{longtable}
\tablefoot{{This table is also available in electronic form at http://www.nbi.dk/$\sim$jeskj/sma-iras16293.html
\tablefoottext{a}{Catalog line strength for transition. ``$\ast$'' indicates multiple hyperfine
  components from the catalog entry added together.}
\tablefoottext{b}{Energy of upper level for transition.}
\tablefoottext{c}{Flag describing identification of  line: ``1''
  indicates transition detected toward both components, ``A''
  transition only detected to IRAS~16293A and ``B'' transition only
  detected toward IRAS~16293B. Transitions that are close in frequency
  / blended are grouped with the vertical lines in this column.}
\tablefoottext{d}{``+'' indicates transitions included in the maps in Fig.~\ref{image_first}--\ref{image_last} for specific species.}
}}

\end{appendix}

\end{document}